\documentclass[a4paper,fleqn,usenatbib]{mnras}

\usepackage{newtxtext,newtxmath}

\usepackage[T1]{fontenc}
\usepackage{ae,aecompl}

\usepackage{array}
\usepackage{eqnarray}
\usepackage{graphicx}	
\usepackage{amssymb}	
\usepackage{amsmath}	
\usepackage{multicol}        
\usepackage{bm}		
\usepackage{pdflscape}	
\usepackage{threeparttable}

\def\kms{\mbox{km\,s$^{\rm -1}$}}

\defcitealias{Mackey13}{Ma13}
\defcitealias{Kamann18}{K18}

\title[Masses and $M/L$ Ratios of Star Clusters. I.]{Dynamical Masses and Mass-to-light Ratios of Resolved Massive Star Clusters. I. NGC~419 and NGC~1846\thanks{This paper includes data gathered with the 6.5-meter {\it Magellan} Telescopes located at Las Campanas Observatory, Chile.}}

\author[Y.-Y. Song et al.]{Ying-Yi Song,$^{1}$\thanks{E-mail: yysong@umich.edu, songyingyi@gmail.com}\href{https://orcid.org/0000-0002-6270-8851}{\includegraphics[scale=0.6]{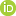}}
Mario Mateo,$^{1}$
A. D. Mackey,$^{2}$\href{http://orcid.org/0000-0002-6529-8093}{\includegraphics[scale=0.6]{orcid.png}}
Edward W. Olszewski,$^{3}$
\newauthor Ian U. Roederer,$^{1,4}$\href{https://orcid.org/0000-0001-5107-8930}{\includegraphics[scale=0.6]{orcid.png}}
Matthew G. Walker,$^{5}$\href{https://orcid.org/0000-0003-2496-1925}{\includegraphics[scale=0.6]{orcid.png}}
and John I. Bailey, III$^{6}$\href{http://orcid.org/0000-0002-4272-263X}{\includegraphics[scale=0.6]{orcid.png}}
\\
$^{1}$Department of Astronomy, University of Michigan, 1085 S. University Avenue, Ann Arbor, MI 48109, USA\\
$^{2}$Research School of Astronomy and Astrophysics, Australian National University, Canberra, ACT 2611, Australia\\
$^{3}$Steward Observatory, The University of Arizona, 933 N Cherry Avenue, Tucson, AZ 85721, USA \\
$^{4}$Joint Institute for Nuclear Astrophysics---Center for the Evolution of the Elements (JINA-CEE), USA\\
$^{5}$McWilliams Center for Cosmology, Department of Physics, Carnegie Mellon University, 5000 Forbes Avenue, Pittsburgh, PA 15213, USA\\
$^{6}$Department of Physics, University of California Santa Barbara, Santa Barbara, CA 93016, USA
}

\date{Accepted XXX. Received YYY; in original form ZZZ}

\pubyear{2019}

\begin{document}
\label{firstpage}
\pagerange{\pageref{firstpage}--\pageref{lastpage}}
\maketitle

\begin{abstract}
As an introduction of a kinematic survey of Magellanic Cloud (MC) star clusters, we report on the dynamical masses and mass-to-light ($M/L$) ratios of NGC~419 (SMC) and NGC~1846 (LMC).  We have obtained more than one hundred high-resolution stellar spectra in and around each cluster using the multi-object spectrograph M2FS on the {\it Magellan}/Clay Telescope. Line-of-sight velocities and positions of the stars observed in each cluster were used as input to an expectation-maximization algorithm used to estimate cluster membership probabilities, resulting in samples of 46 and 52 likely members ($P_{M}\geq 50$\%) in NGC~419 and NGC~1846, respectively. This process employed single-mass King models constrained by the structural parameters of the clusters and provided self-consistent dynamical mass estimates for both clusters. Our best-fit results show that NGC~419 has a projected central velocity dispersion of $2.44^{+0.37}_{-0.21}$ \kms, corresponding to a total mass of $7.6^{+2.5}_{-1.3}\times10^4\ {\rm M}_{\sun}$ and $V$-band $M/L$ ratio of $0.22^{+0.08}_{-0.05}$ in solar units.  For NGC~1846, the corresponding results are $2.04^{+0.28}_{-0.24}$ \kms,  $5.4^{+1.5}_{-1.4}\times10^4\ {\rm M}_{\sun}$ and $0.32^{+0.11}_{-0.11}$.  The mean metallicities of NGC~419 and NGC~1846 are found to be $\rm [Fe/H]=-0.84\pm0.19$ and $-0.70\pm0.08$, respectively, based on the spectra of likely cluster members. We find marginal statistical evidence of rotation in both clusters, though in neither cluster does rotation alter our mass estimates significantly.  We critically compare our findings with those of previous kinematic studies of these two clusters in order to evaluate the consistency of our observational results and analytic tools.
\end{abstract}

\begin{keywords}
galaxies: star clusters -- Magellanic Clouds -- stars: kinematics and dynamics -- techniques: spectroscopic 
\end{keywords}


\section{Introduction}
\label{sec:intro}
The mass-to-light ($M/L$) ratio is frequently used to translate the luminosity of a stellar system to a common baryonic mass scale.
As a result, $M/L$ ratios serve as fundamental tools to trace the baryons distributed within galaxies over cosmic time. 
In most instances, $M/L$ ratios are applied to composite stellar systems that have complex mixtures of ages, metallicities and star formation histories.
Population synthesis models used to estimate $M/L$ ratios for such systems rely crucially on stellar evolutionary grids that attempt, often parametrically, to account for the complex star formation and chemical evolutionary histories of galaxies \citep[e.g.,][]{Bruzual03, Maraston05, Vazdekis10, Conroy09, Conroy10}. 

One way to test the reliability of model-dependent $M/L$ ratios is to determine dynamical masses of simple stellar populations (SSP).
Star clusters are particularly well-suited for such measurements since they are generally regarded as prototypical SSPs \citep[though see e.g.][for reviews]{Gratton12, Bastian18}.
Clusters also span a wide range of metallicity and age, hence making them useful also as clean tests of how $M/L$ ratios vary with these fundamental population parameters. Using clusters for $M/L$ ratio determinations also offers simplicity. In the most basic terms, measuring the $M/L$ ratio of star clusters involves mass determinations via direct kinematic measurements and straightforward dynamical analyses, and obtaining luminosities from independent photometric observations.

There are two broad approaches commonly used to obtain the kinematic data needed to estimate cluster masses, integrated-light spectroscopy and spectroscopy of individual stars within clusters to serve as dynamical tracers.  Integrated-light spectroscopy is best suited for star clusters that have very condensed, unresolved cores and/or that are so distant that they cannot be readily resolved into their constituent stars.  An influential early example of this approach---though applied to mostly resolved systems---is the study of \citet{Illingworth76b} who obtained scanned integrated-light spectra of ten Southern Galactic globular clusters (GCs). \citet{Mandushev91} employed a similar approach to obtain $M/L$ ratios of 32 Galactic GCs.  Local Group (LG) clusters have also proven to be popular targets for  integrated-light kinematic spectroscopy.
\citet{Zaritsky12, Zaritsky13, Zaritsky14} reported $M/L$ ratios obtained using integrated-light spectroscopy of a sample of 29 clusters from four different LG 
galaxies, including the Milky Way, the Large and Small Magellanic Clouds (MCs), and the Fornax dwarf spheroidal galaxy.  
\citet{Larsen02} measured $M/L$ ratios of four suspected intermediate-age M33 GCs, while 
\citet{Strader09, Strader11} produced a large sample of $M/L$ ratios for 163 M31 GCs that comprised new and previously-published results.
Some studies have probed beyond the LG to obtain measurements of the internal kinematics of clusters (e.g.
\citealt{Martini04} who studied 14 GCs associated with NGC 5128).

The other approach of measuring internal cluster kinematics via observations of individual stars is best suited for comparatively nearby, well-resolved systems.  This method has become significantly more practical in recent years with the development of wide-field multi-object spectrographs (MOSs) and comparatively wide-field IFUs.   For example, \citet{Lane10b} used the AAOmega spectrograph to derive the $M/L$ ratios in 10 halo GCs.  More recently, \citet{Kimmig15} published a new catalog of $M/L$ ratios for 25 Galactic GCs based solely on MOS measurements of individual stars. Their data, as is the case for most MOS results,  are particularly uniform in kinematic  precision.  The large number of GCs that have been observed in this manner make it possible to examine trends of $M/L$ as a function of mass and metallicity, but, notably, not in age. Some examples of this approach applied beyond the Galaxy include \citet{Feast80}, \citet{Lupton89}, \citet{Mateo91}, \citet{Fischer92a, Fischer92b, Fischer93}, \citet{Suntzeff92}, \citet{Ferraro06}, \citet{Mackey13} and \citet{Kamann18}. 

The most challenging aspect of cluster kinematic studies, regardless of approach, is the comparatively small velocity dispersions---ranging from 1 to 15 \kms---of these systems.  Resolving such dispersions demands moderate to high spectral resolution and excellent instrumental stability.  Integrated-light spectroscopy can generally succeed in clusters only when the instrumental resolution is precisely measured---and sufficiently stable---to extract these comparatively small cluster dispersions reliably.  In many of the studies cited above, the instrumental resolution ranged from 10 to 50 \kms, meaning that the clusters with dispersions $\sim 5$ \kms\ or smaller  inflated line profiles by at most 10\% and often less than 1\% compared to the line spread function of the spectrographs.  Measuring kinematics using resolved-star spectroscopy also requires moderate to high resolution, though determining line centers---as opposed to line widths---is comparatively much more precise at any given S/N level.  The more acute problem here is that moderately large samples are needed to beat down stochastic errors.  MOS help in this regard, but they often can target only limited numbers of members in a given cluster, especially for more distant systems.

Some useful compilations of kinematics studies of star clusters include \citet{Pryor93} who tabulated central velocity dispersions and $M/L$ ratios of 56 Galactic GCs with integrated-light and individual-star spectroscopy, the latter often from heroic studies before MOS were available.  \citet{McLaughlin05} derived dynamical properties in a consistent manner for 38 Galactic GCs and 19 LG galaxies (16 in the MCs) from published single-star spectroscopic studies.  The latter paper also lists photometric structural data for some 46 additional MC clusters that do not have any kinematic measurements.

This paper represents the first in a series in which we aim to measure in a consistent manner dynamical masses and $M/L$ ratios of resolved massive star clusters in the Magellanic Clouds using individual-star spectroscopy.  The key motivations behind this work are (a) to obtain data from which reliable masses can be determined for massive clusters that span a large range in metallicity, and most importantly, age, (b) to exploit the availability of an MOS---Michigan/{\it Magellan} Fiber System \citep[M2FS,][]{Mateo12}---that is capable of targeting clustered fields efficiently and obtaining spectra with individual velocity precisions well below the expected internal dispersions of the clusters.  The first aim will allow us to critically compare $M/L$ models with SSPs over as broad a range in the metallicity-age plane as feasible, while the second provides us with the unprecedented means of routinely and systematically obtaining precise masses in MC clusters with central dispersions as small as 1 \kms.  As of this writing, we have obtained high-quality data for over 20 MC clusters that span the range from 50 Myr to 13 Gyr in age, and from $-2.2$ to $+0.0$ in [Fe/H].
The ultimate aim of this work is to critically compare theoretical $M/L$ estimates with what we glean from this cluster sample.
The more specific aims of this first paper are to introduce key features of this survey and to assess the quality of our results via comparisons with previous observations of two clusters---NGC~419 in the SMC and NGC~1846 in the LMC---that are in common with our cluster sample (\citealt{Kamann18}, hereafter \citetalias{Kamann18}; \citealt{Mackey13}, hereafter \citetalias{Mackey13}).

The paper is organized as follows. In \autoref{sec:data} we describe the  target selection, observational and data reduction procedures as they apply to NGC~419 and NGC~1846 and that we will adopt for all clusters in our survey.
\autoref{sec:analysis} starts with a generalized description of the Bayesian method we used to derive velocity and physical parameters from M2FS spectra.  We continue to explain our determination of dynamical properties such as systemic velocities and central velocity dispersions, applying the techniques, as feasible, to published and our own new data for NGC~419 and NGC~1846.  In \autoref{sec:analysis} we also discusses the accuracy and limitations of our analyses. In \autoref{sec:mass} we report our derived masses and $M/L$ ratios and their rotation signatures for these two clusters. \autoref{sec:comp} compares our results critically with those from previous studies for these two clusters.
In \autoref{sec:summary} we conclude with a summary of our methods and results, and provide a brief outline of the content and scope of subsequent papers for this survey.

\section{Data Overview}
\label{sec:data}
\subsection{Cluster Candidates}
\label{sec:cluster_select}
The clusters identified for use in the present papers and for all future papers of our study were chosen from catalogs of MC clusters with good-quality age and metallicity estimates and for which we could expect to obtain samples of about 50 stellar members.
The latter requirement---imposed to ensure that we can obtain statistically reliable mass estimates for the clusters---restricted us in practice to candidates more luminous than an absolute V-band magnitude of $-6$.   

\begin{figure}
   \centering
   \includegraphics[width=0.45\textwidth]{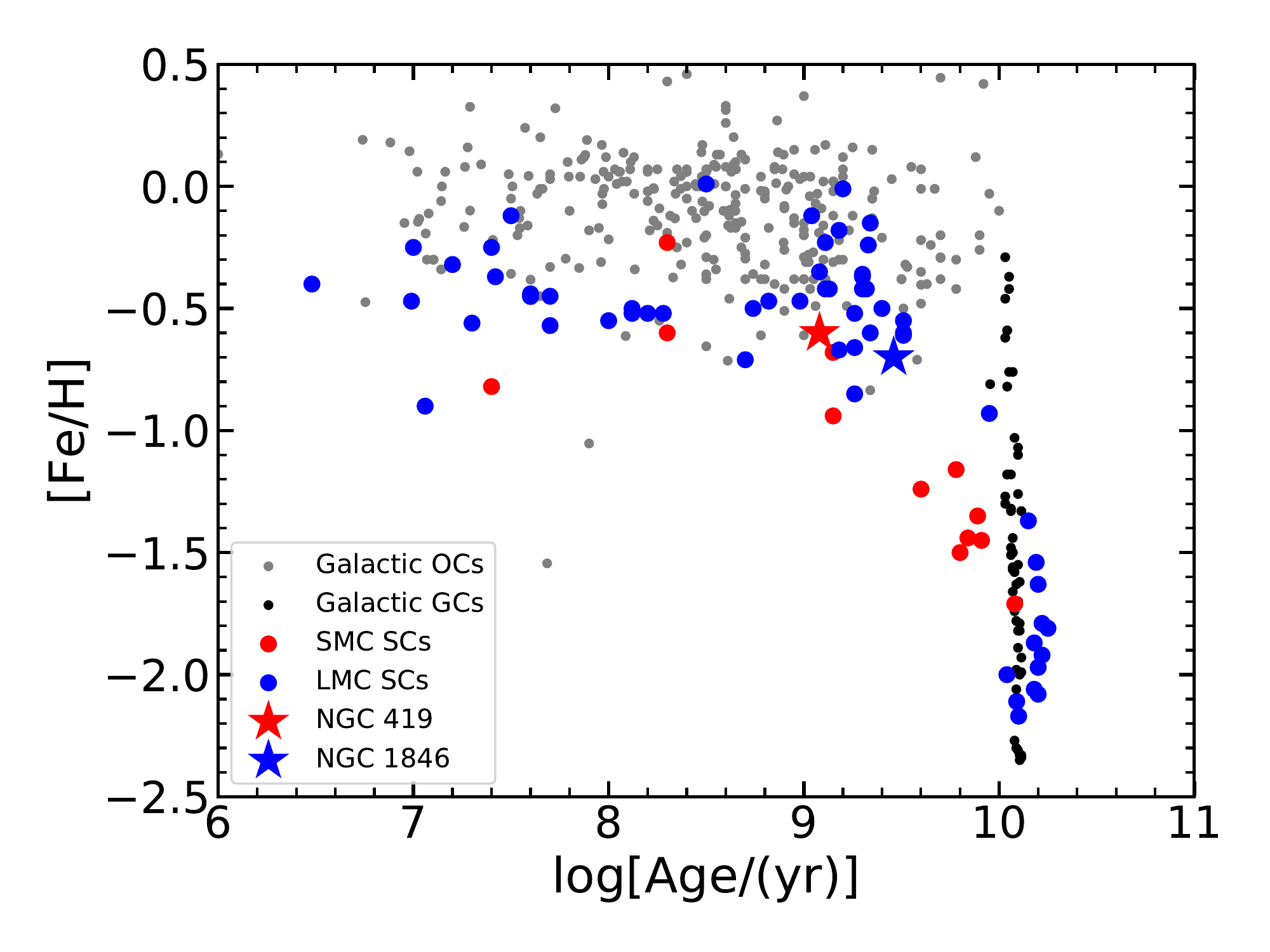}
   \caption{A plot of metallicity vs. $\log{\rm (Age)}$ for star clusters (SCs) in the Milky Way and the Magellanic Clouds. Blue points correspond to the SCs in the LMC \citep{Pessev06, Pessev08}, red points correspond to the SCs in the SMC \citep{Pessev06, Pessev08}, black points correspond to the Galactic GCs \citep{VandenBerg13}, and gray points correspond to the Galactic open clusters \citep{Dias02}. The two clusters studied in this paper, NGC~419 and NGC~1846, are marked as red and blue star-shaped symbols, respectively.}
   \label{fig:amr}
\end{figure}

A further selection on our sample was to identify clusters with good integrated-light or star-count analyses.  These data are essential for constraining the structural parameters needed to carry out the dynamical analysis of the clusters.  Integrated photometry of our candidates was taken from the catalogs of \citet{McLaughlin05} and \citet{Goudfrooij06}, while  
radial profiles based on photometric and/or star-count measurements are taken from \citet{McLaughlin05}, \citet{Glatt09} and \citet{Goudfrooij09, Goudfrooij11b, Goudfrooij14}.   
Good age and metallicity estimates are also required of clusters in our sample.   Thus, we identified systems with adequate stellar photometry suitable to provide precise cluster age and (photometric) metallicity estimates from their color-magnitude diagrams (CMDs).   
We gave preference to clusters with deep {\it Hubble Space Telescope} ({\it HST}) photometry, but good-quality ground-based photometry was acceptable.  
Finally, in the spirit of providing a broad range of SSPs to test modeled $M/L$ ratios, we chose clusters to span as broad a range in age and metallicity as practical within the Magellanic Clouds (see \autoref{fig:amr}).  
Although we include some ancient clusters (age $>10$ Gyr) in our sample, we have already noted that there are many dynamical studies and $M/L$ estimates of GCs in the literature \citep[e.g.][]{Larsen02, Lane10b, Strader11, Kimmig15}.  
At the other extreme, clusters younger than about 50 Myr were avoided as these may not yet have achieved dynamical equilibrium and hence would yield biased mass estimates.  
Our resulting sample from which we have drawn clusters to study in detail consists mostly of systems ranging in age from 50 Myr to 13 Gyr, and in [Fe/H] from $-2.2$ to $+0.0$ (see \autoref{fig:amr}). 

The adopted photometric and structural parameters of the two clusters of this introductory study---NGC~419 and NGC~1846---are listed in \autoref{tab:basic}. The aperture photometry for both clusters come from \citet{Goudfrooij06}, while structural profiles are taken from \citet{Goudfrooij09} for NGC~1846, and from a weighted average of previous results \citep{Glatt09, Goudfrooij14} for NGC~419. As listed in \autoref{tab:obs}, the centers of NGC~419 and NGC~1846 were taken from \citet{Glatt09} and \citetalias{Mackey13}, respectively.

For NGC~1846, \citet{Olszewski91} derived a spectroscopic metallicity of $\rm [Fe/H]=-0.7\pm0.2$. The metallicity of NGC~419 was estimated through iron and hydrogen spectral indices calibrated using SSPs by \citet[][]{dFP98} to be $\rm [Fe/H]=-0.60\pm0.21$. Both clusters possess so-called extended main-sequence turnoffs (eMSTOs) confirmed from the photometric analysis of {\it HST} imaging.  If taken as an internal age spread, these eMSTOs imply that the stellar populations of NGC~1846 span an age range of 1.6--1.9 Gyr \citep[e.g.,][]{Mackey07, Mackey08}, while for NGC~419 the implied age range is 1.2--1.6 Gyr \citep{Glatt08}.  In the latter case, the specified range in age appears to represent primarily the age uncertainty rather than clear evidence of a composite stellar population \citep{Martocchia17}. 

As noted in \autoref{sec:intro}, the specific selection of NGC~419 and NGC~1846 reflects the fact that both clusters have reasonably recent independent spectroscopic measurements of individual stars within and around the clusters \citepalias{Kamann18, Mackey13}.  This provides an opportunity for us to critically compare our kinematic results for these clusters with these earlier studies.   In the case of NGC~1846, which was observed in a manner similar to the present study but with a smaller sample, we can also compare our analysis by running the earlier data through our machinery to determine how well we recover previous results.  For the case of NGC~419, we have an opportunity to compare our analysis and findings with MUSE observations \citepalias{Kamann18} to determine how the immense multiplexing, but relatively low spectral resolution, of that instrument compares to our smaller sample of target spectra obtained at considerably higher resolution.  The focus here on the specific cases of NGC~419 and NGC~1846 also allows us to illustrate in a concrete manner some of the procedures common to all clusters in our $M/L$-ratio survey.

\subsection{Selection of Targets Within Cluster Fields}
\label{sec:star_select}

\begin{table*}
\caption{Archival Photometry and Structural Parameters.}
\label{tab:basic}
\begin{threeparttable}
\begin{tabular}{cccccccccc}
\hline
    {Cluster} & {$V_{\rm ap}$} & {Aper. $^{\rm a}$} & {$(m-M)_0$} &{$A_V$} &{$r_{0,\rm K62}$} &{$r_{t,\rm K62}$}  & {Ref. $^{\rm b}$} &{$r_{0,\rm K66}$} &{$r_{t, \rm K66}$}  \\
    {} & {(mag)} & {(arcsec)} & {(mag)} & {(mag)} & {(arcsec)} & {(arcsec)} & {} & {(arcsec)} & {(arcsec)} \\
\hline
NGC 419  & $10.30\pm0.16$ & 50 & $18.85\pm0.03$ & $0.15\pm0.02$ & $18.8\pm6.9$ & $129.7\pm47.6$ & 1, 2 & ... & ... \\ 
         & ... & ... & ... & ... & $15.22\pm1.78$ & $174.15\pm18.57$   & 3 & ... & ... \\ 
         & ... & ... & ... & ... & $12.98\pm1.47$ & $207.19\pm30.11$   & 3 & ... & ... \\ 
	 & ... & ... & ... & ... & $15.60\pm1.66$ & $275.91\pm53.94$  & 3 & ... & ... \\ 
         & $10.30\pm0.16$ & 50  & $18.85\pm0.03$ & $0.15\pm0.02$ & $14.5\pm0.9$ & $185.0\pm14.5$    & Avg. & $14.9\pm0.9$ & $238.2\pm18.7$ \\ 
NGC 1846 & $10.68\pm0.20$ & 50 & $18.42\pm0.03$ & $0.07\pm0.02$ & $26.0\pm1.6$ & $161.2\pm9.9$  & 1, 2, 4 & $26.9\pm1.7$ & $212.6\pm13.0$ \\ 
\hline
\end{tabular}
\begin{tablenotes}
   \item $^{\rm a}$ Radius of aperture used for integrated-light photometry. 
   \item $^{\rm b}$ References: (1) \citet{Goudfrooij06}; (2) \citet{Goudfrooij14}; (3) \citet{Glatt09}; (4) \citet{Goudfrooij09}.
\end{tablenotes}
\end{threeparttable}
\end{table*}

\begin{figure*}
   \centering
   \includegraphics[width=0.45\textwidth]{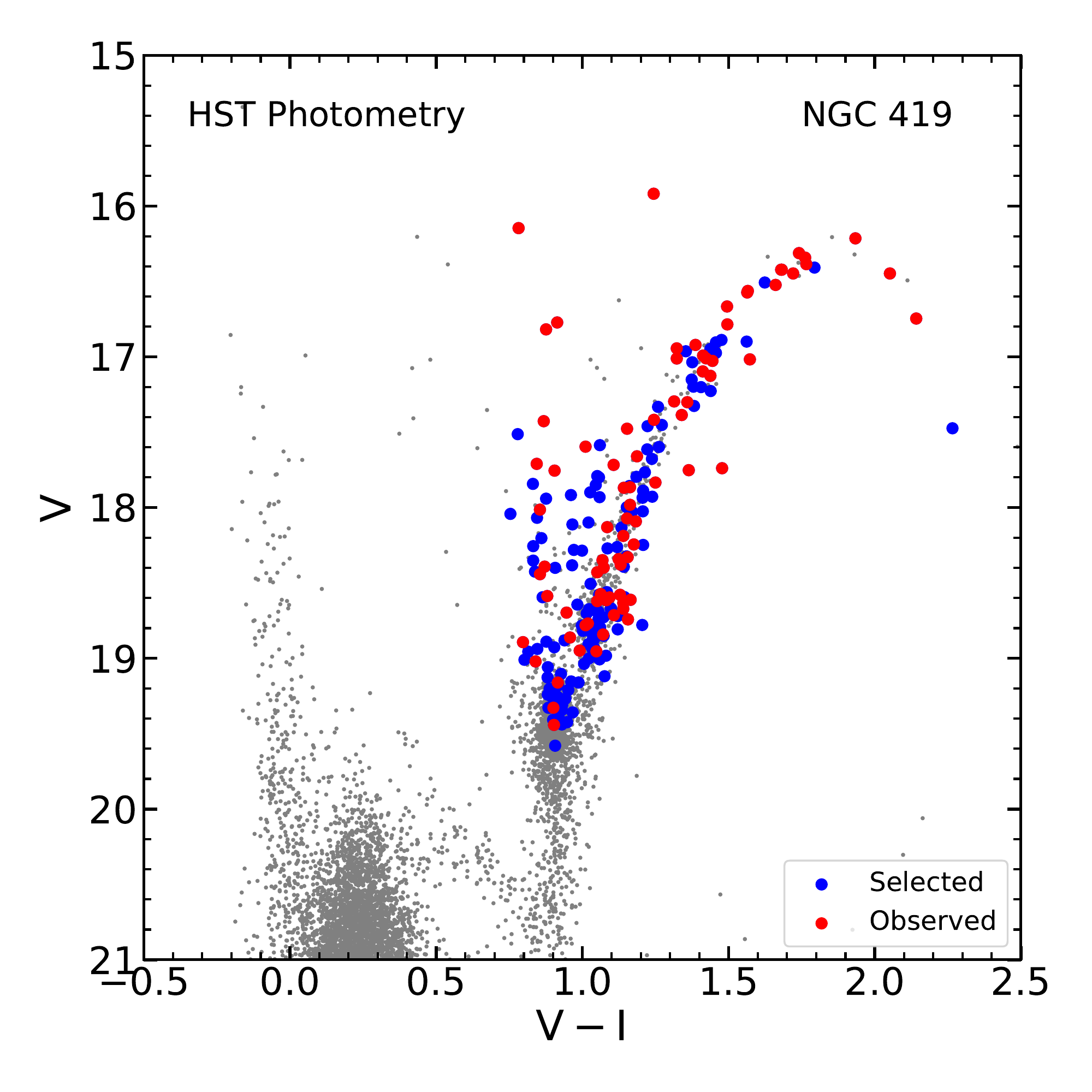}
   \includegraphics[width=0.45\textwidth]{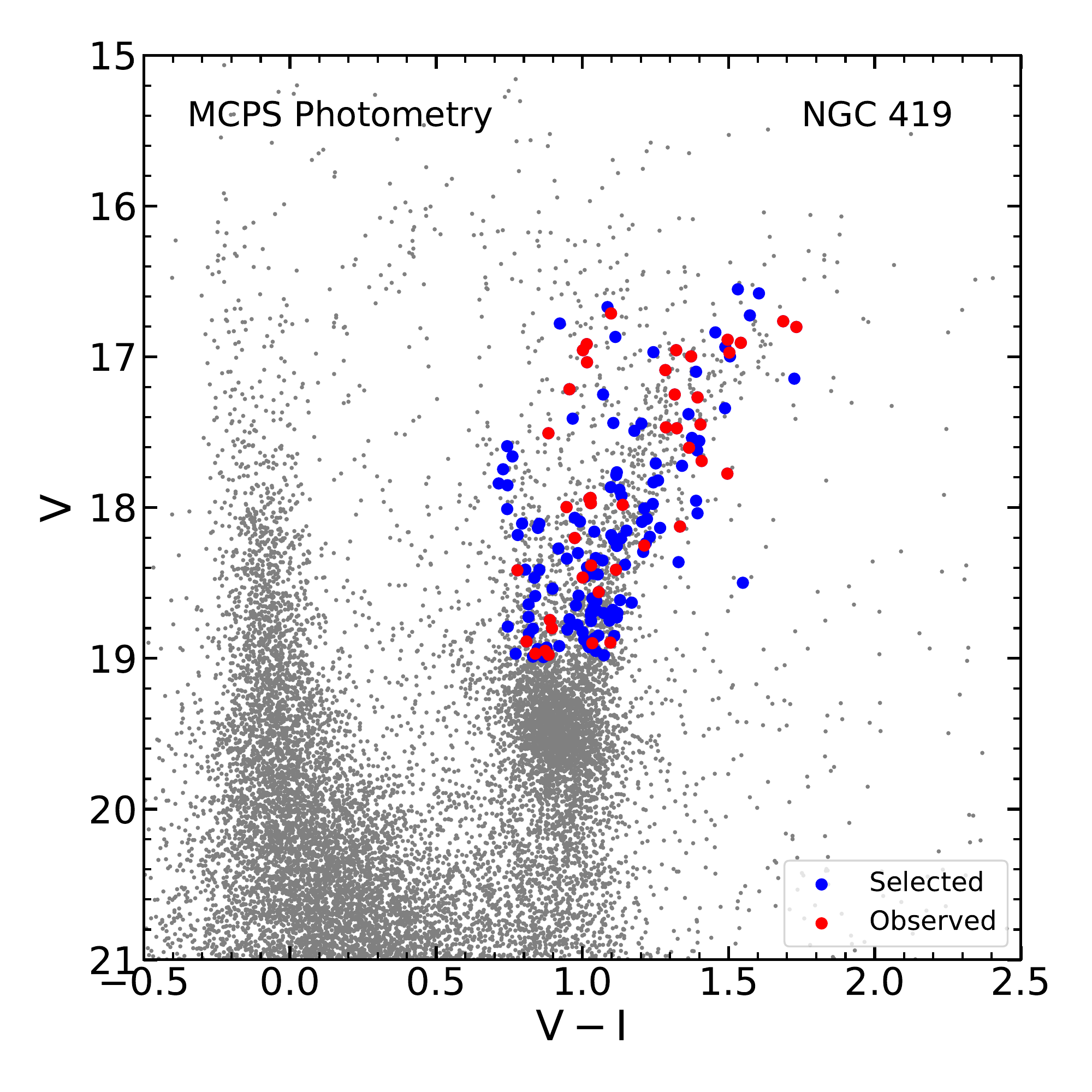}
   \includegraphics[width=0.45\textwidth]{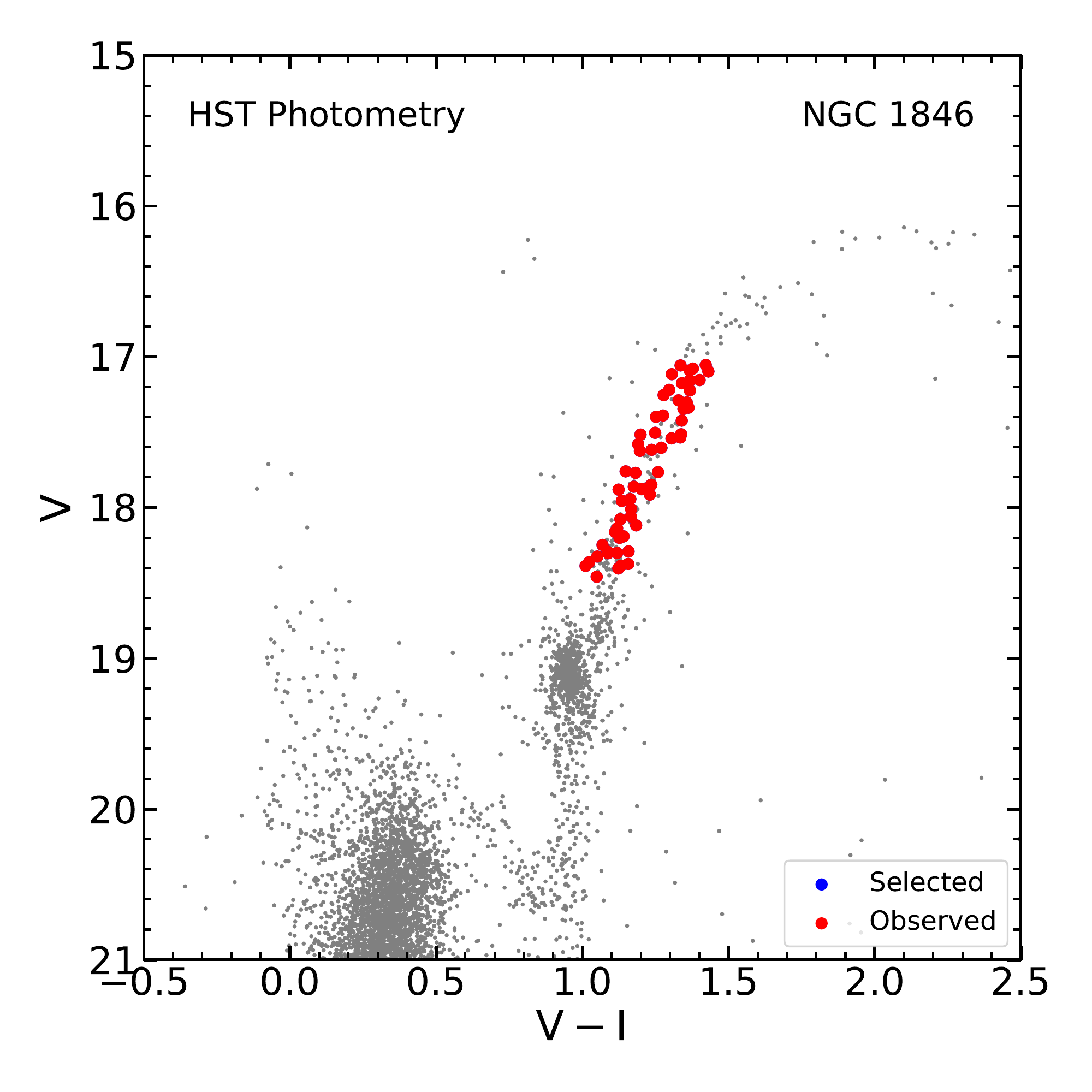}
   \includegraphics[width=0.45\textwidth]{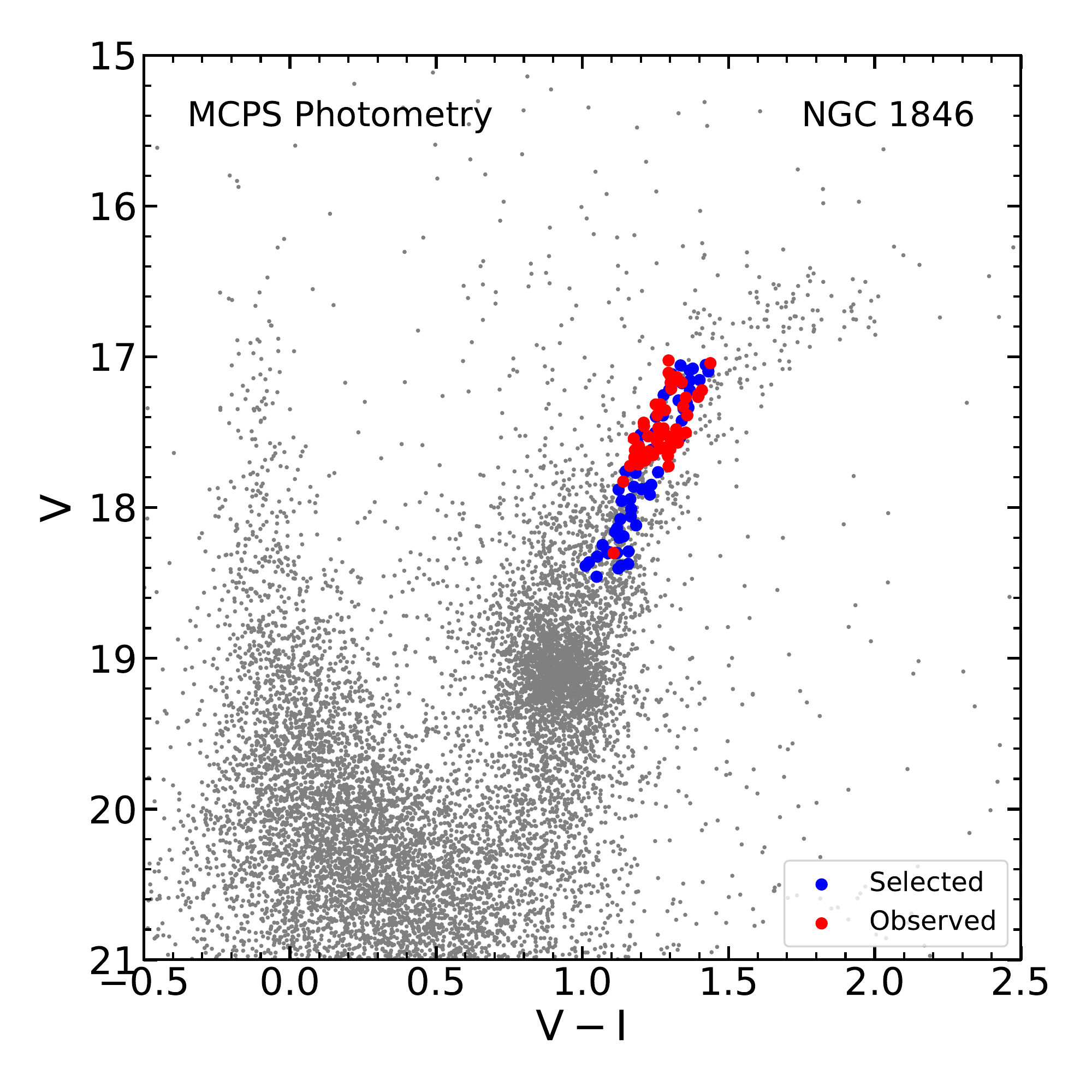}
   \caption{Color-magnitude diagrams for NGC~419 and NGC~1846 in the $(V-I,\,V)$ plane. The left panels based on {\it HST} photometry show the selection of potential targets in the cluster central regions. The right panels from the MCPS catalog  \citep{Zaritsky02, Zaritsky04} show the selection of potential targets in regions outside the tidal radii (see \autoref{sec:star_select}). In all panels, gray dots correspond to all the stellar objects from the respective photometric catalog.  Colored (blue and red) dots represent targets selected for potential M2FS observation. The red dots denote the objects observed in this study.}
   \label{fig:CMD}
\end{figure*}

For any given cluster in our study we typically selected a variety of specific types of targets for spectroscopic analysis.  The primary science targets are typically drawn from the red giant branch (RGB) of a cluster's CMD.  These targets are prioritized also by proximity to their respective cluster center.  Additional science targets beyond the formal tidal radii of clusters are also included to allow us to determine the kinematic and chemical distribution of their nearby field populations. These latter targets are typically identified from different photometric studies than candidates close to the cluster centers.  In all cases, we identified apparently isolated stars as potential spectroscopic targets.    We typically also identify `sky/background' positions both near the cluster centers and in their surrounding fields.   We discuss how we use these to determine backgrounds below in \autoref{sec:reduction}.

In the specific cases of the clusters that are the focus of this paper---NGC~419 and NGC~1846---we used the {\it HST} images from GO-10396 (NGC~419; PI: Gallagher) and from GO-10595 (NGC~1846; PI: Goudfrooij) to identify stars in the cluster centers. Both programs obtained relatively short exposures with the F555W and F814W filters (i.e., 40 s and 20 s for NGC~419, while 41 s and 9 s for NGC~1846).  We photometered all the images using the ACS modules from the \texttt{DOLPHOT} package \citep{dolphot}, and the output magnitudes were automatically transformed into the standard Johnson-Cousins $VI$ system. The targets were mainly selected from the RGB in the corresponding $(V-I,\ V)$ CMDs (see the left panels in \autoref{fig:CMD}). 
For regions around the clusters, we selected stars to characterize the superimposed field  populations on the basis of their position beyond the clusters' tidal radii (the `$r_t$' column in \autoref{tab:basic}), and their location in the same RGB region of the CMD from which candidates cluster members were identified (see the right panels in \autoref{fig:CMD}).  For both clusters, the field stars were drawn from the Magellanic Clouds Photometric Survey (MCPS) $UBVI$ catalog \citep{Zaritsky02, Zaritsky04}.  For all targets identified in clusters studied after the {\it Gaia} DR2 release \citep{Gaia2016, Gaia2018b}, we selected only stars identified in that catalog. 
To minimize the contamination of nearby stars, only apparently isolated stars were accepted into our final candidate sample. We regarded a star to be isolated when the integrated flux of all other stars in the corresponding photometric catalog ({\it HST} or {\it Gaia} DR2) located within 1 {\it arcsec} of the star adds up to $\leq$ 20\%\ of the candidate star's flux.

Positions for targets  selected and observed before the DR2 release were tied to the NOMAD astrometric system \citep{Zacharias:2004aa}.
We cross-matched any stars brighter than 17.5 mag in $V$-band from {\it HST} images or MCPS catalog with stars in the NOMAD catalog, and transformed the coordinates onto the NOMAD frame.
In some cases for {\it HST}-selected stars, rather large astrometric corrections of up to 1--2 {\it arcsec} were necessary.  For reference, the M2FS fiber apertures are 1.2 {\it arcsec} in diameter, and systematic precision of 0.25 {\it arcsec} is typically required. 

Individual background regions within the tidal radii of clusters were identified by eye from the F555W {\it HST} images when available. For the field regions beyond the clusters' tidal radii, background/sky positions were randomly chosen from the DSS red-band images and at least 2 {\it arcmin} from the clusters' centers. As noted below, we often supplemented these sky positions with an observational strategy to better determine the background contributions in the clusters and their surrounding fields. 

\subsection{Observations}
\label{sec:obs}

\begin{table*}
 \caption{Observations.}
 \label{tab:obs}
 \begin{tabular}{cccccccccc}
\hline
{Cluster} & {$\alpha_{2000}$} & {$\delta_{2000}$} & {Obs. Date} & {Exp. Type} & {Exp. Time} & {$N_{\rm HST}$} & {$N_{\rm MCPS}$} & {$N_{\rm sky}$} \\
 {} & {(hh:mm:ss)} & {(dd:mm:ss)} & {(UT)} & {} & {(s)} & {} & {} & {} \\
\hline
NGC 419  & 01:08:17.31 & -72:53:02.5 & 2017 Sep 21 & On target & $4\times2100$ & 77 & 40 & 4\\
      &  &  &  & Offset & $2\times480$ & 0 & 0 & 121\\
NGC 1846 & 05:07:33.66 & -67:27:40.7  & 2018 Feb 21 & On target & $4\times1200$ & 55 & 55 & 14 \\
\hline
\end{tabular}
\end{table*}

The spectral data used in this paper and throughout our $M/L$-ratio study have been obtained with the M2FS.  This instrument can deploy up to 256 optical fibers in a 29-{\it arcmin} diameter field using aluminum plugplates pre-drilled for each target.  As noted above, every fiber has a 1.2 {\it arcsec} entrance aperture.  However, the ferrules in which the fiber are mounted imposes a minimum separation between adjacent fibers of 13 {\it arcsec}.  The latter feature limits how densely we can pack fibers within the clusters' tidal radii; in practice, this limited the number of observable cluster candidate targets to about 100-140 per field, including sky positions and field stars.

The spectrograph in M2FS (MSpec)  consists of twin spectrographs (denoted `B' and `R'), each capable of observing 128 targets simultaneously over selected regions within the spectral range $3700$-$9500$~\AA. For this project, one arm of MSpec (usually the `B' arm) was configured to observe 128 targets in the so-called `HiRes' mode, in which an interference filter is used to isolate a single order of the cross dispersed spectrum spanning $5130$ to $5192$~\AA\ at an effective resolving power $\mathcal{R}\sim20,000$.   The second spectrograph (the `R' arm) produced spectra at similar resolution but used a much broader filter that passes 24 orders and covers the range  $3700$ to $5592$~\AA.  In this mode, up to five targets can observed simultaneously, though in practice, one fiber is usually assigned to a background/sky fiber.  
For this study we only use order~17 from the R-arm, corresponding to the single order on the B-arm.

Our NGC~419 observations had few suitable background/sky fibers, so we supplemented these by obtaining spectra of fields offset by 5 {\it arcsec} from the nominal field position.  These spectra were then used to estimate the background contribution in a manner described in detail in \autoref{sec:reduction}.  This process allowed us to sample the local backgrounds for every target throughout the cluster and in the corresponding field.  For NGC~1846 we relied on dedicated background/sky fibers (as described earlier in \autoref{sec:star_select}) to monitor the background.  The full background-subtraction process (including normalization) and a third approach used for other clusters is described in detail below in \autoref{sec:reduction}.  

\autoref{tab:obs} lists the full set of observations---including the offset exposures---used in this study. Though not detailed in the table, ThArNe arc calibration spectra were interleaved between science observations on a cadence of about 1 hour for the purpose of wavelength calibration. Additional calibration data (e.g. flats, aperture reference spectra, twilights, darks and biases) were obtained throughout the relevant M2FS runs when data for this paper and our overall $M/L$-ratio study were obtained.
  
\subsection{Data Reduction}
\label{sec:reduction}

\begin{figure*}
   \centering
    \includegraphics[width=0.45\textwidth]{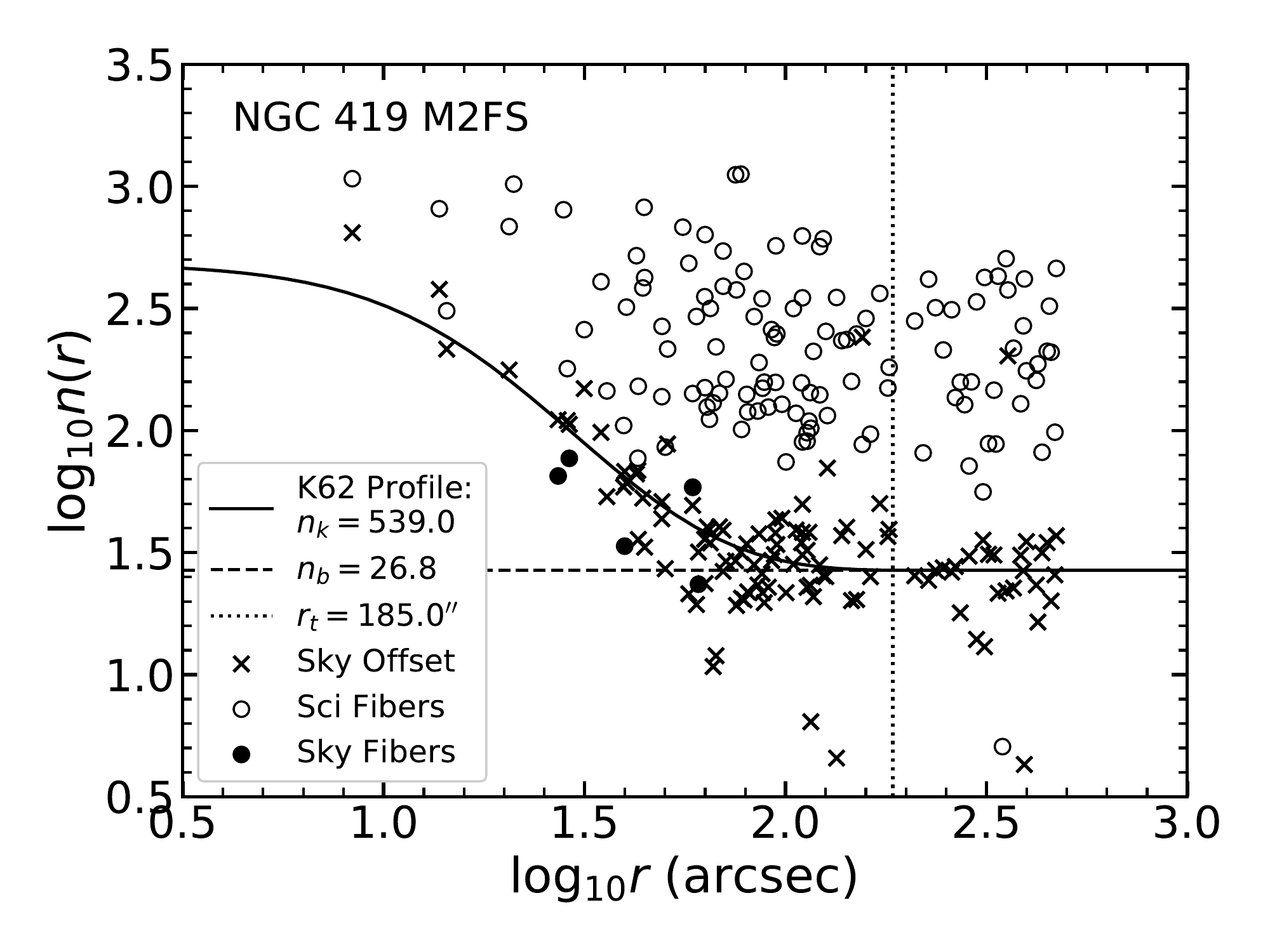}
    \includegraphics[width=0.45\textwidth]{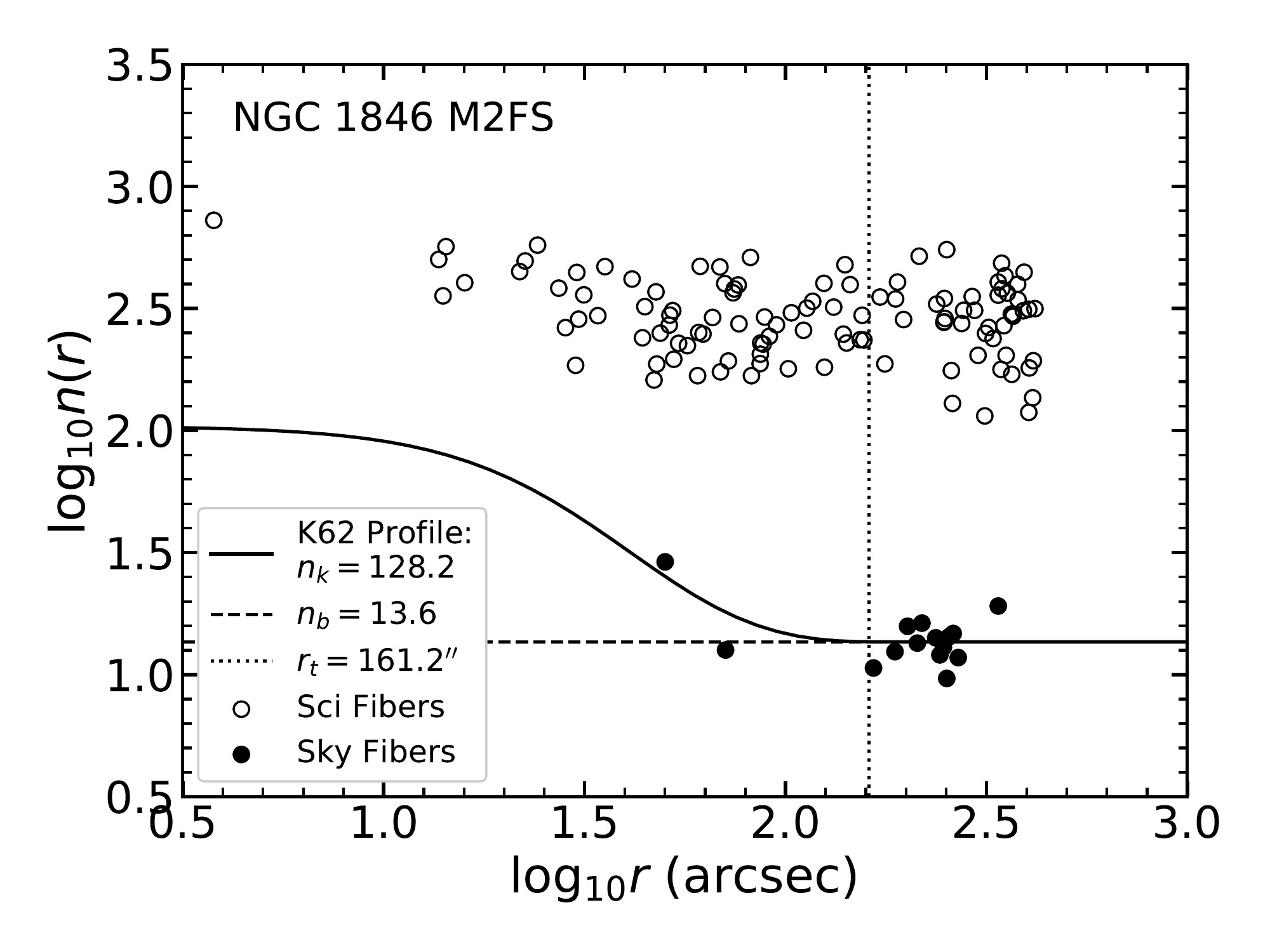}
   \caption{Plots of median counts in the spectra of stars in NGC~419 (left panel) and NGC~1846 (right panel) as a function of the stars' projected distances from their respective cluster centers. The adopted cluster centers are listed in \autoref{tab:obs}. The solid lines correspond to K62 profiles used in the sky subtraction process (see \autoref{sec:reduction}).  Open circles correspond to the target spectra. Closed circles correspond to dedicate sky spectra observed simultaneously with the targets. Crosses correspond to exposure-time-corrected median counts of sky spectra obtained through all the fibers while offsetting from on-target pointings (see \autoref{tab:obs} and \autoref{sec:obs}). The meanings of the symbols in the K62 profile---$n_k$, $n_b$ and $r_t$---are described in \autoref{sec:reduction}.}
   \label{fig:K62}
\end{figure*}

All data were processed using an M2FS pipeline based on IRAF.  The principal end products of this pipeline are the sky-subtracted spectra and their associated variances. Detailed reducing processes were thoroughly described in \citet{Walker15mn, Walker15apj}, and a brief description is available in \citet{Song17}.  To summarize, raw 2-D data obtained on both the `B' and `R' arms of MSpec were processed through overscan, bias and dark corrections, the latter using combined bias and dark calibration images obtained throughout M2FS runs.   For long science exposures, we then removed cosmic rays using the Laplacian filter algorithm developed by \citet{vanDokkum01}.  We then subtracted diffuse scattered light by fitting a moderate-order 2D surface to the inter-spectral regions in the images. Using templates that map the locations and wavelengths along for every fiber/target in an image, the pipeline then extracted and wavelength calibrated the spectra from each fiber.   All spectra were normalized to correct for fiber-to-fiber throughput variations using normalization factors obtained from twilight-sky exposures obtained with the fibers in the same spectral configuration used for the target observations.  All subsequent reduction steps, including final background subtraction (see below), employed these normalized 1-D spectra.    

The last reduction step is background/sky subtraction.  At its simplest, this process uses a master sky spectrum produced from all non-contaminated background/sky spectra. This master spectrum typically accounts well for the backgrounds experienced by targets in the outer parts of the clusters and in the surrounding field. A complication is that the internal background light from unresolved/faint cluster stars must also be considered for targets within the clusters' tidal radii. To properly subtract the backgrounds for stars near the cluster centers requires consideration of the variations in the clusters' light backgrounds with distance from the cluster centers.

As noted in \autoref{sec:star_select} and \autoref{sec:obs}, we obtained background spectra for NGC~419 from both dedicated background/sky fibers placed at preselected positions (see \autoref{sec:star_select}) as well as from offset exposures.  For NGC~1846, we only had background/sky measurements from fibers at pre-selected sky positions within the cluster and field regions.  In both cases, we normalized the background/sky measurements with a radial light profile (see below) plus a constant background that was required to pass through the summed counts from the relevant background/sky observations.  For NGC~419, the multiple background/sky data were suitable to quasi-independently constrain the background as a function of distance from the cluster center. We refer to this approach as `Method A' for background removal (see \autoref{fig:K62}, left panel).  For NGC~1846, we adopted the central surface brightness and cluster core radius (\autoref{tab:basic})  to fit the background/sky spectra (`Method B'; see  \autoref{fig:K62}, right panel).

In both Method A and B, we start by fitting the median counts of the offset spectra (crosses in \autoref{fig:K62}) with a cluster profile assumed to be of the form given by the  empirical density law described by \citet[][hereafter the K62 profile]{King62} 
\begin{equation}
n(r)=n_k\left[{1\over\sqrt{1+(r/r_0)^2}}-{1\over\sqrt{1+(r_t/r_0)^2}}\right]^2+n_b,
 \label{eq:K62}
\end{equation}
where $r_0$ is the King radius and $r_t$ is the tidal radius. Two constants, $n_k$ and $n_b$, account for the cluster's internal light (inside $r_t$) and field light (outside $r_t$), respectively. In our fitting process, $n_k$ is the only free parameter to be determined. This is because $n_b$ can be pre-determined as the average of the median counts of all offset spectra outside $r_t$, and we adopted $r_0$ and $r_t$ listed in \autoref{tab:basic}. The best-fit K62 profile for NGC~419---which uses Method A to determine $n_k$---is also plotted in the left panel of \autoref{fig:K62}.
 
For NGC~1846, no associated offset exposures were obtained and so we used Method `B' for background determination in this case.  Specifically, $n_b$ was determined by calculating the median of the 13 sky-fiber outside the cluster's tidal radius, $r_t$. 
The value of $n_k$ was determined by scaling from the value of $n_k$ for NGC~419 (see \autoref{fig:K62} for details).
The specific scaling factor for $n_k$ considered the ratio of the central surface brightnesses of the two clusters ($\Sigma_{V,0}$, which is constrained from aperture photometry given their K62 profiles using \autoref{eq:lum}) and the ratio of exposure times ($t_{\rm exp}$, see \autoref{tab:obs}).  For the general case, this procedure can be described by the equation 
\begin{equation}
{n_{k,\, \rm Target} \over n_{k,\,\rm Reference} }={\Sigma_{V,\,0,\,\rm Target} \over \Sigma_{V,\,0,\,\rm Reference}} \cdot {t_{\rm exp,\,Target} \over t_{\rm exp,\,Reference}},
 \label{eq:nk_methodB}
\end{equation}
where `Target' refers to the system in which the background is to be estimated relative to a `Reference' system; in the present case, NGC~1846 is the `Target' and NGC~419 serves as the `Reference'.  So long as the data for the Target and Reference were obtained in similar conditions, this procedure will scale the non-cluster background to reasonable precision.   If, for example, the photometric conditions for the Target/Reference observations are the same to 20\%, since the background typically constitutes no more than 20\% of the flux of individual stars, any errors in the background level should be at the $\pm 5$\% level.   The best-fit K62 profile obtained using this approach---Method `B'---for NGC~1846 is shown in the right panel of \autoref{fig:K62}. 

The actual background/sky subtraction was preformed as follows. First, all the sky spectra used for the $n_b$ estimation were averaged to serve as the master sky spectrum. Then for every science spectra a scaling factor was determined by the best-fit K62 profile based on the target's projected distance from the cluster center. Finally, we multiplied this factor by the master sky spectrum to determine the background/sky spectrum for each target and then subtracted this scaled spectrum from the respective target spectrum.  Throughout this process, we calculated the variances associated with all spectra to rigorously track the signal-to-noise (S/N) ratio for each pixel.  For every spectrum, we take the median over the full spectral range of the per-pixel S/N ratios to estimate the effective S/N ratio of a given spectrum. 

Note that for both Methods `A' and `B', we presume the background spectra do not depend on position relative to the cluster centers.  In reality, any off-target spectrum actually consists of contributions from telluric emission/scattered skylight and moonlight and unresolved/faint-star cluster light.  Telluric backgrounds will be mostly constant across a field during an exposure while the cluster background contributions will depend on the light profile of the cluster and is therefore position-dependent.  A careful accounting of the compound nature of the background in targets such as LMC clusters is typically beyond the scope of our data, since considerable sampling is required in the cluster centers.  The extent to which we can do this in a few cases will be explored in a future paper in which we analyze data from a much larger sample of MC star clusters.  For the current analysis of NGC~419 and NGC~1846, we have verified that background subtraction has little effect ($< 0.2$ \kms\ rms) on the kinematic results for individual stellar targets.  We therefore restrict ourselves here to background-subtraction Methods `A' and `B' for NGC~419 and NGC~1846, respectively.

\subsection{Previous Spectroscopic Samples for NGC~1846 and NGC~419}
\label{sec:prev_samples}
As noted in \autoref{sec:intro} and \autoref{sec:cluster_select}, prior to our study of NGC~1846, \citetalias{Mackey13} published a spectroscopic results for 105 stars drawn from the RGB region of the cluster's CMD and one planetary nebula Mo-17.  These data were obtained using the multi-object spectrograph FLAMES mounted at the ESO Very Large Telescope. 
With the full dataset of \citetalias{Mackey13}, we compared our analysis and results with theirs in \autoref{sec:comp_N1846}. As we shall see, these data also allowed us to expand the kinematic sample for NGC~1846 by combining it with our M2FS results.  We describe how we combined both datasets in \autoref{sec:combine_sample}.   For NGC~419, the full data for individual stars observed with MUSE \citepalias{Kamann18} are not available.  As a result, the comparison of the MUSE and our new M2FS results has been confined to higher-level conclusions.   A full discussion of those comparisons is also provided in \autoref{sec:comp}.

\section{Analysis}
\label{sec:analysis}
\subsection{Bayesian Fitting of M2FS Spectra}
\label{sec:spec_fit}

\begin{table*}
 \caption{Fits to the Twilight Spectra$^{\rm a}$.}
 \label{tab:twi}
 \begin{threeparttable}
 \begin{tabular}{ccccccccccc}
\hline
      Obs. Date & 
                  ${T_{\rm eff}}$ $^{\rm b}$ & 
                  $ \overline{v_{\rm los,raw}} $  & 
                  $ \overline{v_{\rm los,helio}} $  & 
                  $\sigma_{\overline{v_{\rm vlos}}}$ &          
                  $ \overline{\log{g}} $  & 
                  $\sigma_{\overline{\log{g}}}$ &
                  $\overline{\rm [Fe/H]}$ &  
                  $\sigma_{\overline{\rm [Fe/H]}}$ \\
      (UT) & \rm (K) & ($\rm km\,s^{-1}$) & ($\rm km\,s^{-1}$) & ($\rm km\,s^{-1}$) & \rm (dex) & \rm (dex) & \rm (dex) & \rm (dex) \\
\hline
      2017 Sep 21     & 5778 & -1.30 & -0.40 & 0.17 &  4.45 & 0.04 & -0.22 & 0.02 \\
      2018 Feb 16     & 5778 & +0.29 & -0.45 & 0.13 & 4.38 & 0.03 & -0.18 & 0.02 \\        
\hline
 \end{tabular}
  \begin{tablenotes}
\item $^{\rm a}$ {The solar values are $T_{\rm eff,\,\odot}=5778$ K, $\log{g_{\odot}}=4.44$ and ${\rm [Fe/H]_{\odot}}=0.0$.}
\item $^{\rm b}$ {The values of $T_{\rm eff}$ were fixed during our spectral fitting process (see \autoref{sec:spec_fit}).}
\end{tablenotes}
 \end{threeparttable}
\end{table*}

We have employed the Bayesian method introduced by \citet{Walker15mn, Walker15apj} to analyze the sky-subtracted spectra produced using the procedures described in the previous section.  The template spectra for these fits are taken from the SSPP library of stellar model atmospheres \citep{Lee08a, Lee08b, Walker15mn}.  The output of these fits are estimates of the line-of-sight (LOS) velocity ($v_{\rm los}$), surface gravity ($\log{g}$) and metallicity (${\rm [Fe/H]}$) from individual and combined spectra; a full listing of all parameters provided by the analysis are summarized in Table 2 of \citet{Walker15mn}.  One difference in our present application of this approach is that the effective temperature ($T_{\rm eff}$) was pre-determined by the $V-I$ color and fixed as a constant during the fitting procedure (see \citealt{Song17} for details).   The present analysis also differs from \citet{Song17} in that we have adopted an extended wavelength range (5130--5192~\AA) to reflect a change in the observing parameters that exclude some nuisance spectral ghosts from the red ends of the target spectra.  For both NGC~419 and NGC~1846, we have fitted the spectra obtained from individual exposures as well as the spectra extracted from images consisting of the wavelength-calibrated sums of these individual exposures.

We corrected all the LOS velocities to the barycentric frame using the values calculated by the `radial\_velocity\_correction' function from Astropy. (We have compared the Astropy heliocentric corrections with those from independent algorithms, and we find agreement at the m/s level.)  As a check on our zero points, we applied the same Bayesian fitting procedure to twilight spectra obtained during the same observing runs (and often the same nights) as the target spectra.  These were used to identify any systematic offsets in $\log{g}$ and $\rm [Fe/H]$ (as for our target spectra, we fixed the solar effective temperature to its standard value of 5778~K). The twilight results are listed in \autoref{tab:twi}.  We have corrected all science $\log{g}$ and [Fe/H] values by the small offsets listed in the table. We have not corrected the heliocentric $v_{\rm los}$ offsets listed in \autoref{tab:twi} for two reasons. First, the offsets are comparable to the uncertainties in the systemic velocities of the clusters (see \autoref{tab:EM}). Second, the heliocentric offsets are nearly identical for the two runs in which the data were obtained for NGC~419 and NGC~1846. Consequently, the velocity offsets in \autoref{tab:twi} represent a negligible shift in $v_{\rm los}$ that we chose to ignore.

\subsubsection{Uncertainties in the LOS Velocities}
\label{sec:e_vlos}

\begin{figure*}
   \centering
    \includegraphics[width=0.45\textwidth]{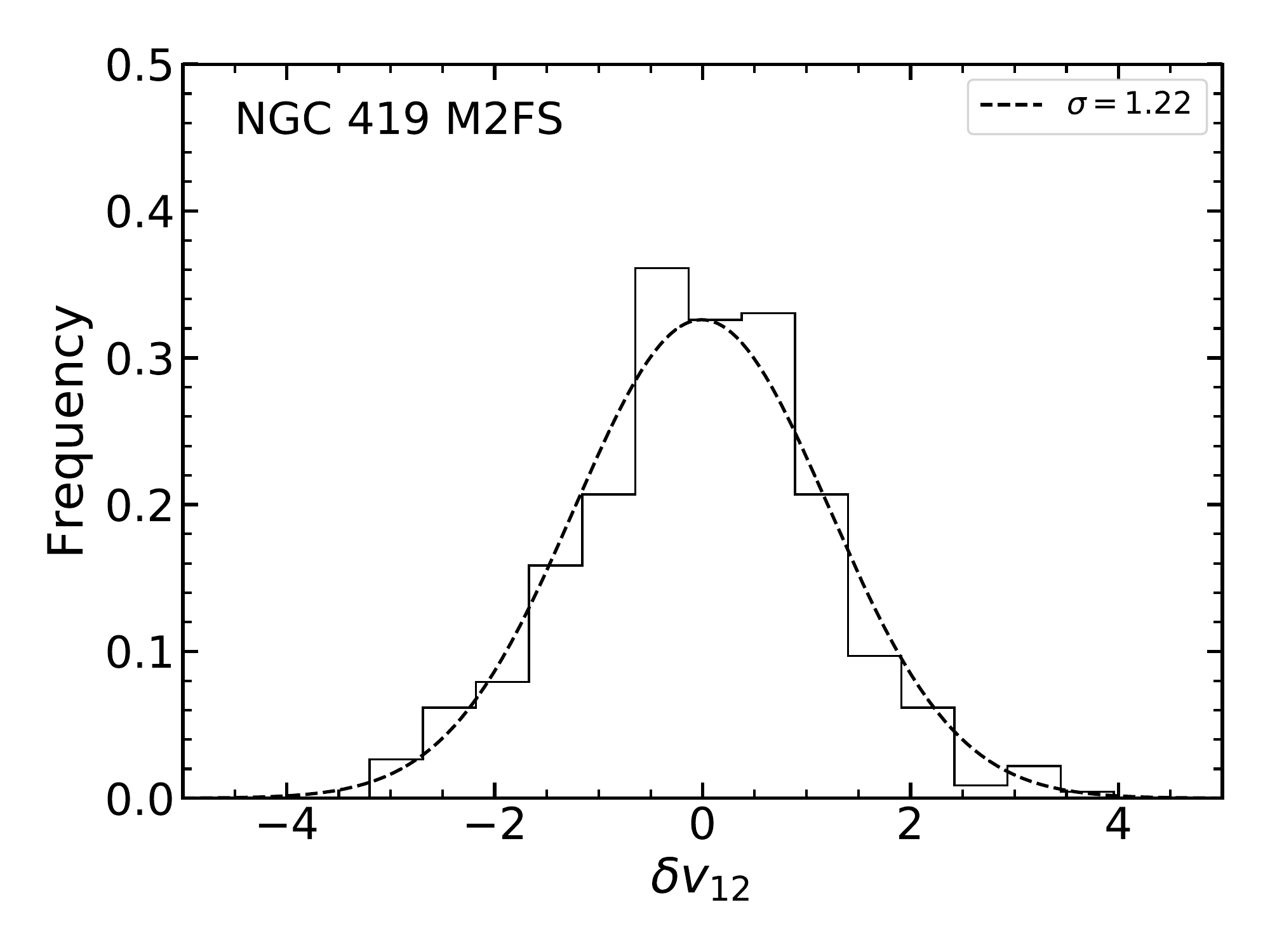}
    \includegraphics[width=0.45\textwidth]{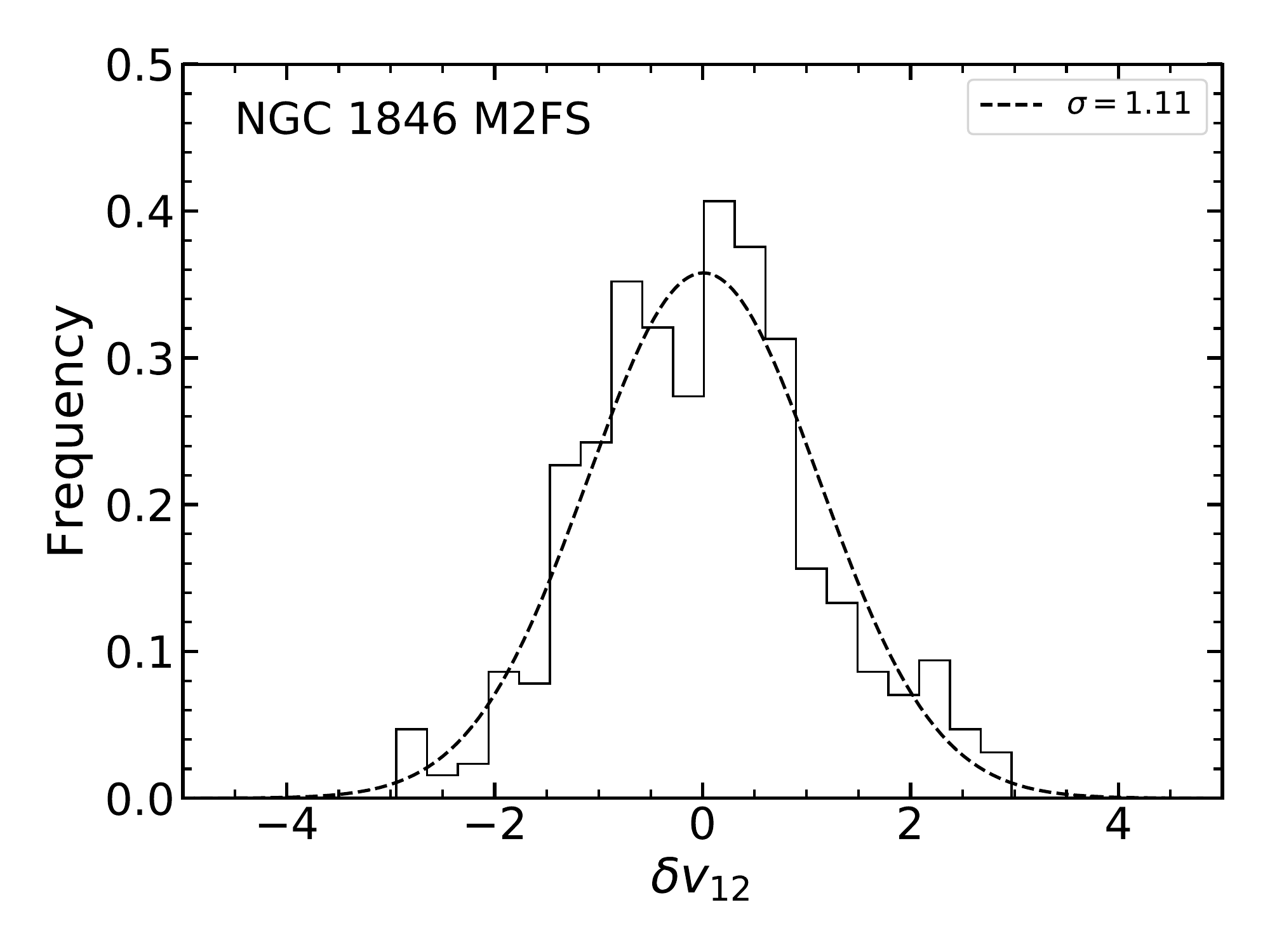}  
   \caption{Measured line-of-sight (LOS) velocity differences for targets observed in  individual M2FS exposures (see \autoref{sec:e_vlos} and \autoref{tab:obs}). The parameter $\delta v_{12}$---defined in \autoref{eq:delta}---represents the differences in measured velocities from individual exposures for a given star. The dashed lines are Gaussian fits to the histograms, and the standard deviation of each fitted Gaussian are given in the legend of each panel.  In this figure, we have combined results for all stars regardless of the mean S/N of their spectra.}
   \label{fig:cor_evlos}
\end{figure*}

\begin{figure*}
   \centering
    \includegraphics[width=0.45\textwidth]{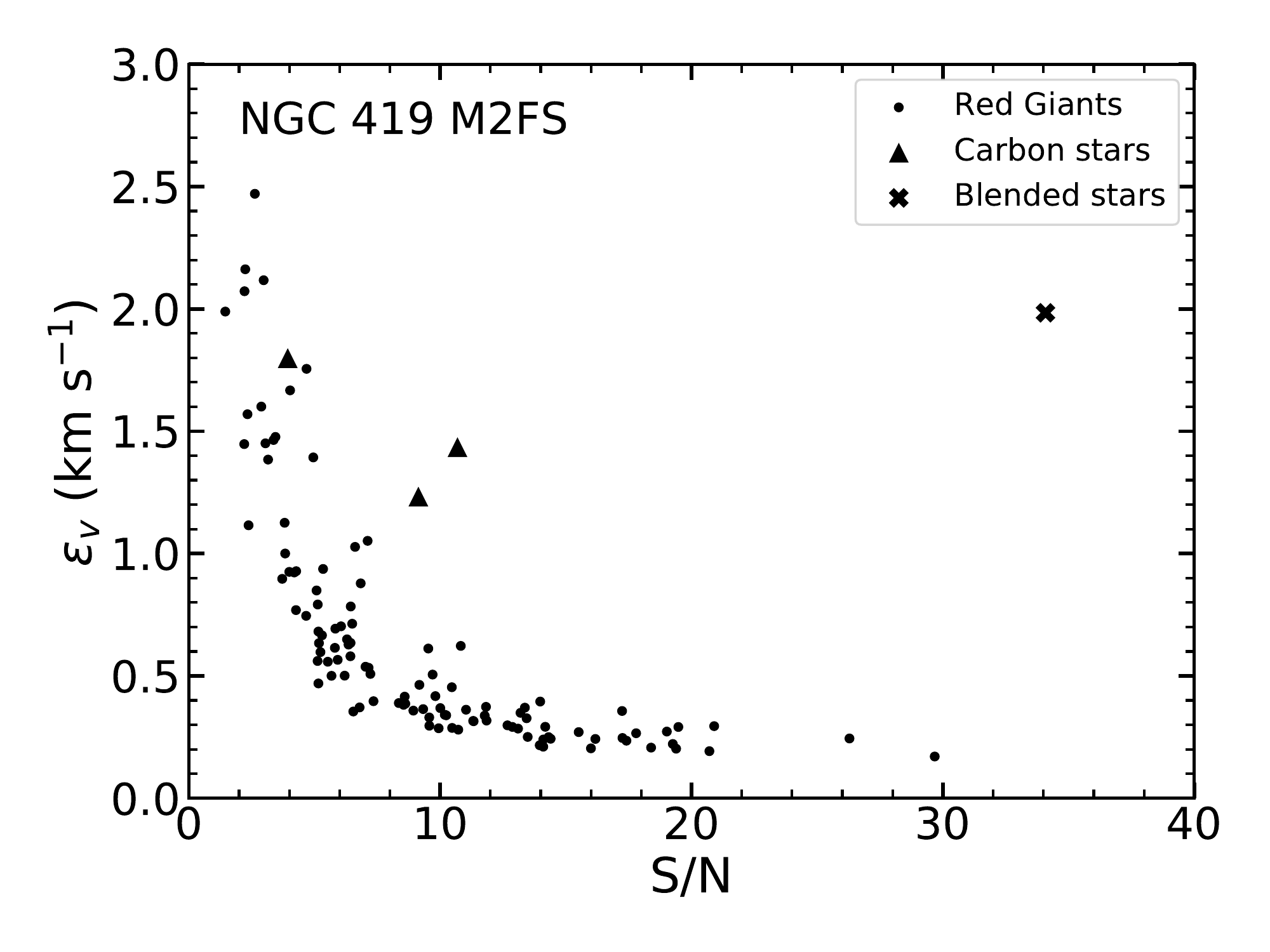}
    \includegraphics[width=0.45\textwidth]{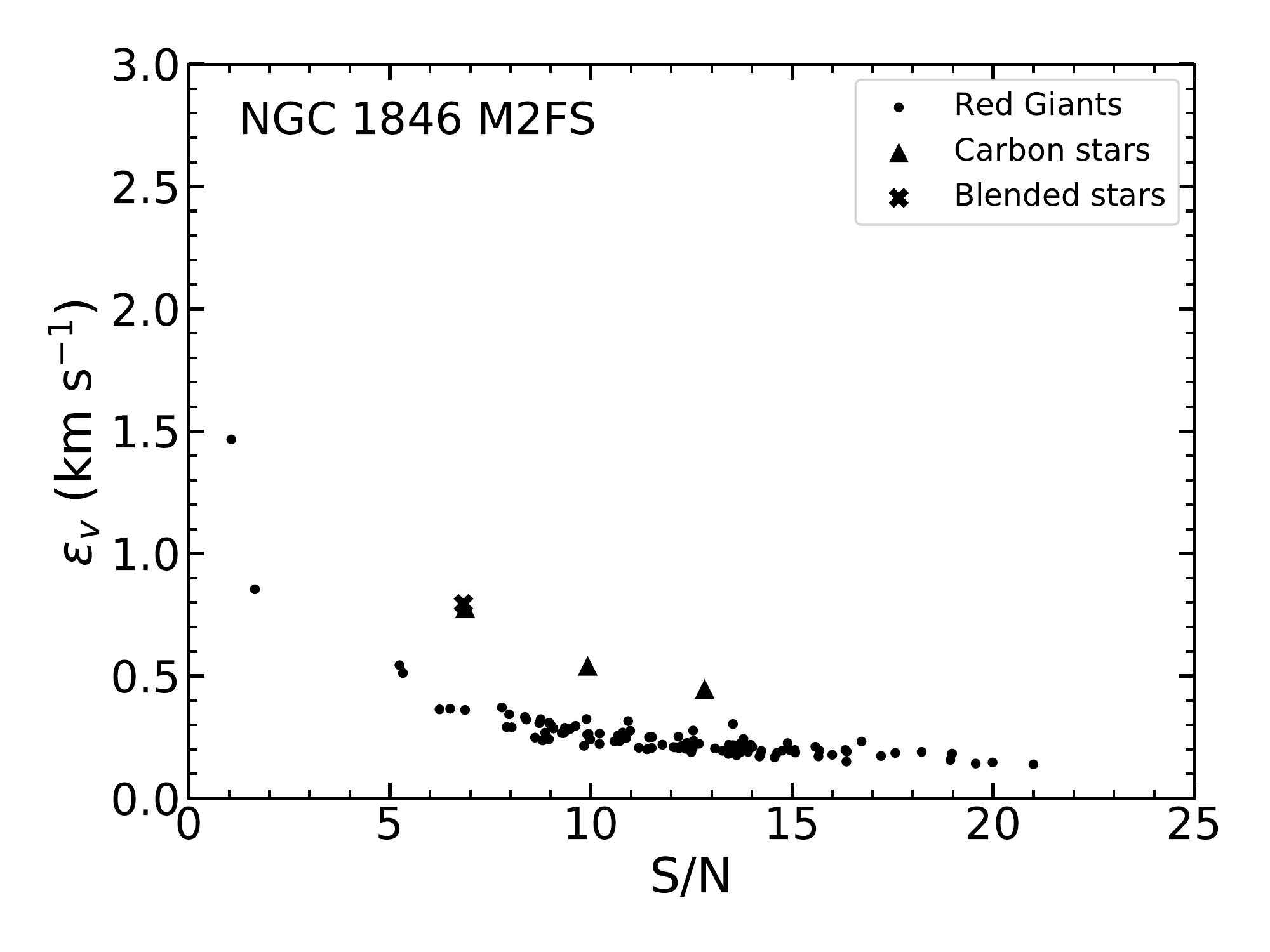}     
   \caption{Corrected LOS velocity uncertainties vs. median S/Ns per pixel for NGC~419 and NGC~1846. The dots correspond to normal red giants in the samples, while the triangles correspond to carbon stars and the crosses to the blended stars (see \autoref{sec:e_vlos} and \autoref{fig:spec_odd}).}
   \label{fig:e_vlos}
\end{figure*}

\begin{figure*}
   \centering
   \includegraphics[width=0.45\textwidth]{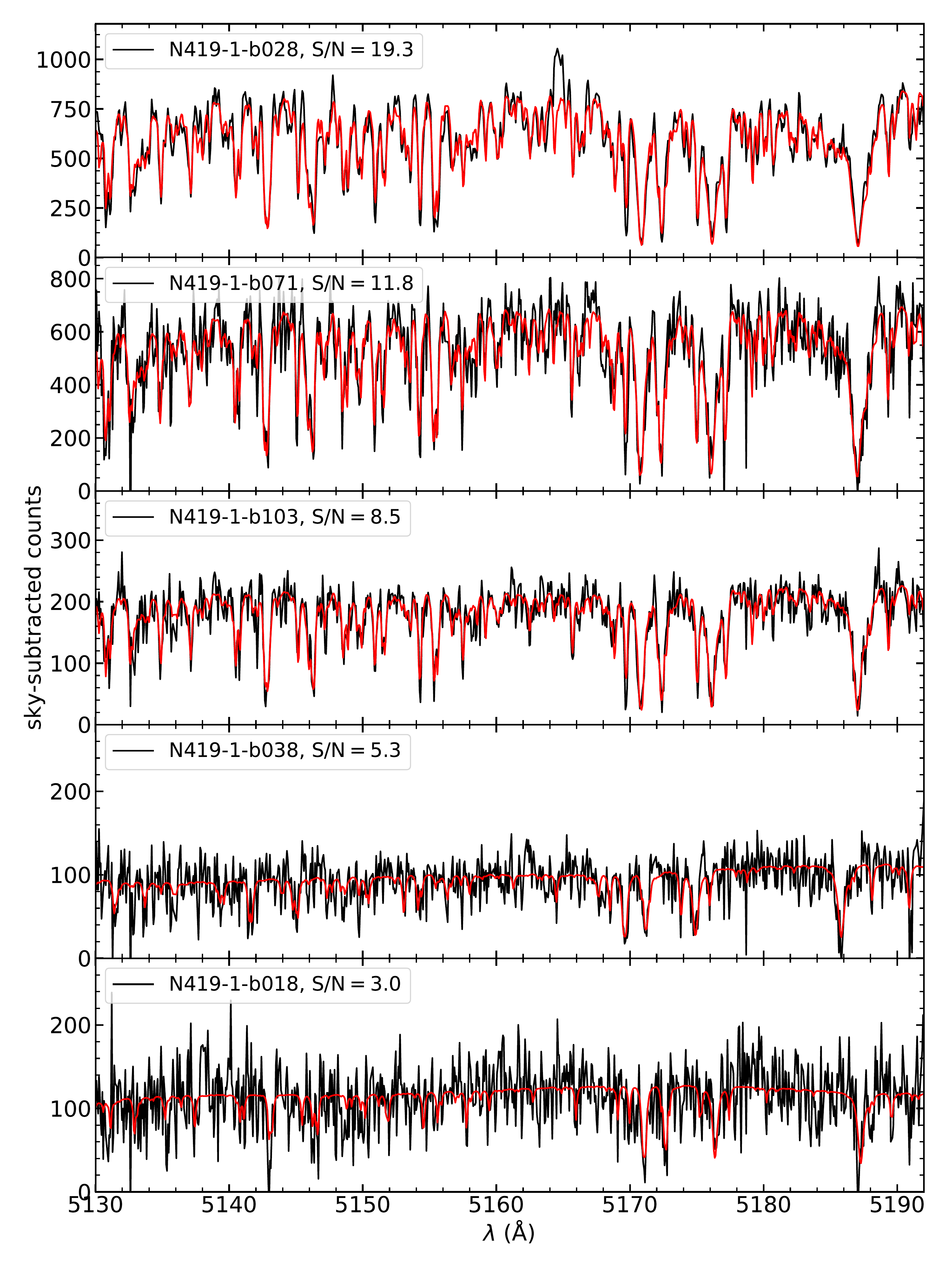}
   \includegraphics[width=0.45\textwidth]{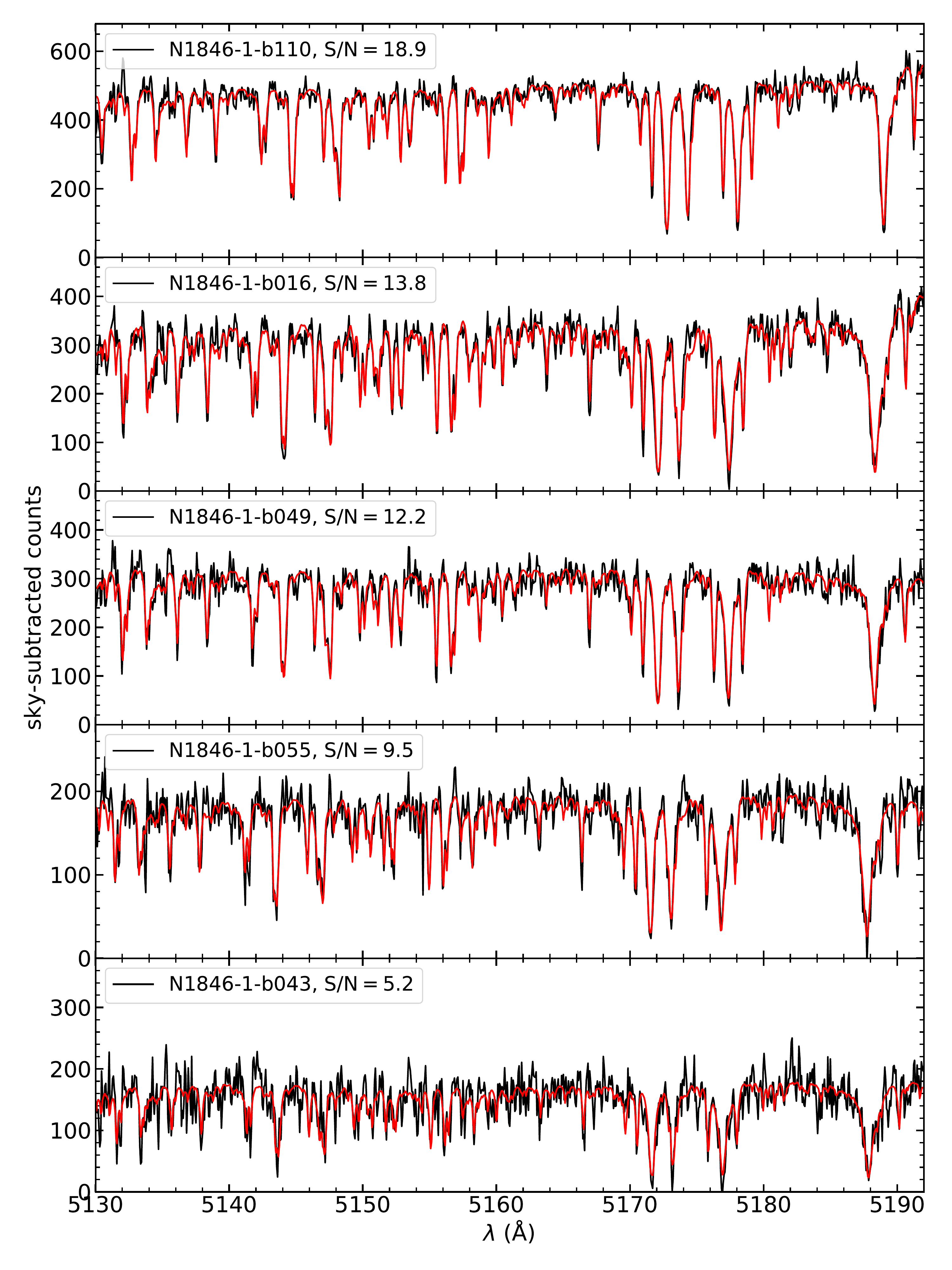}
   \caption{Representative M2FS spectra (black) for five stars observed with M2FS in NGC~419 (left panels) and five stars observed with M2FS in NGC~1846 (right panels).The spectra have been corrected for backgrounds as described in \autoref{sec:reduction}. The red lines are the best-fitting spectral models described in \autoref{sec:spec_fit}. The legend lists target ID and median S/N per pixel. The spectra shown here span the full range of S/N of M2FS spectra obtained in this study.}
   \label{fig:spec_sample}
\end{figure*}

\begin{figure*}
   \centering
   \includegraphics[width=0.45\textwidth]{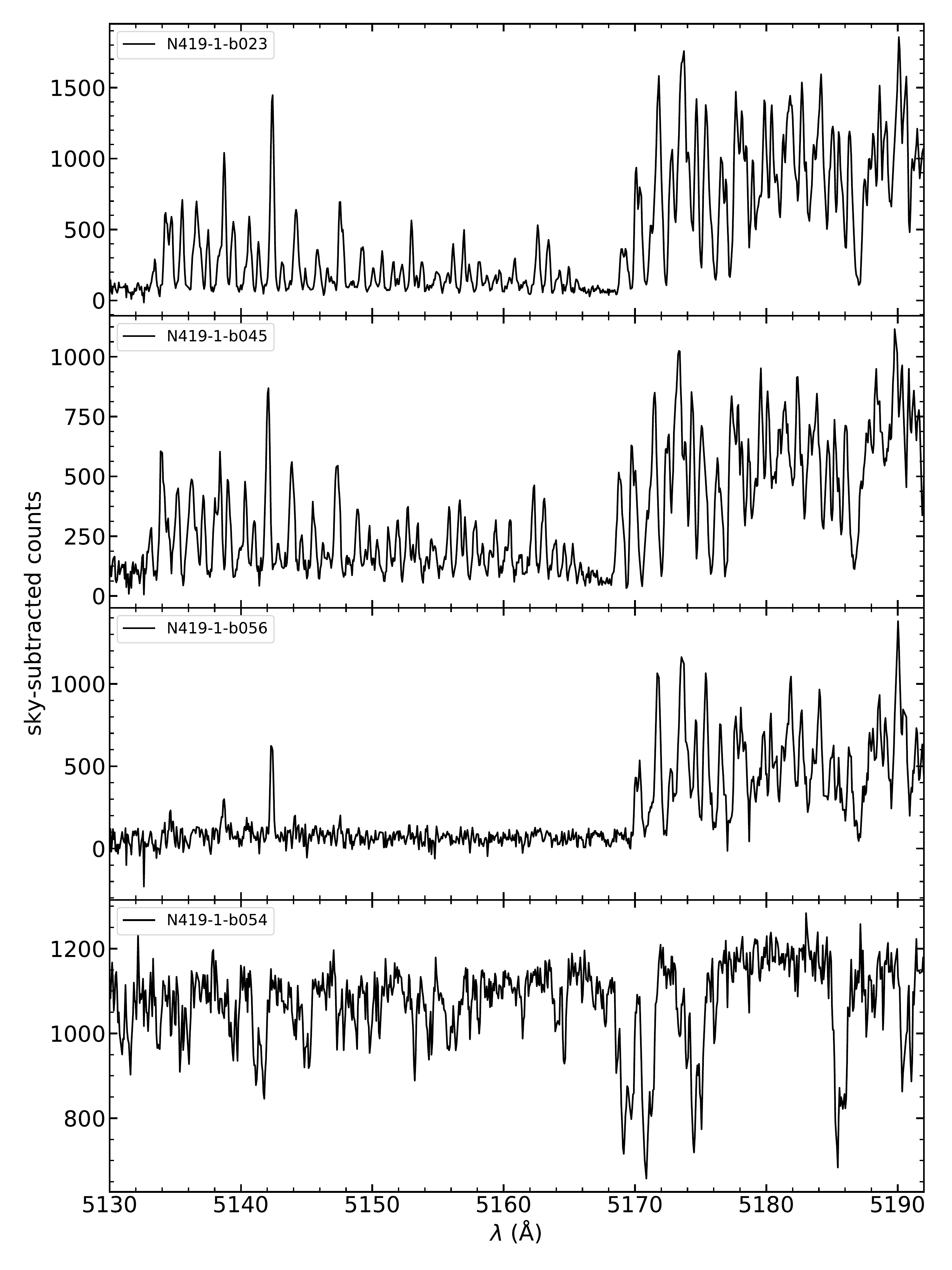}
   \includegraphics[width=0.45\textwidth]{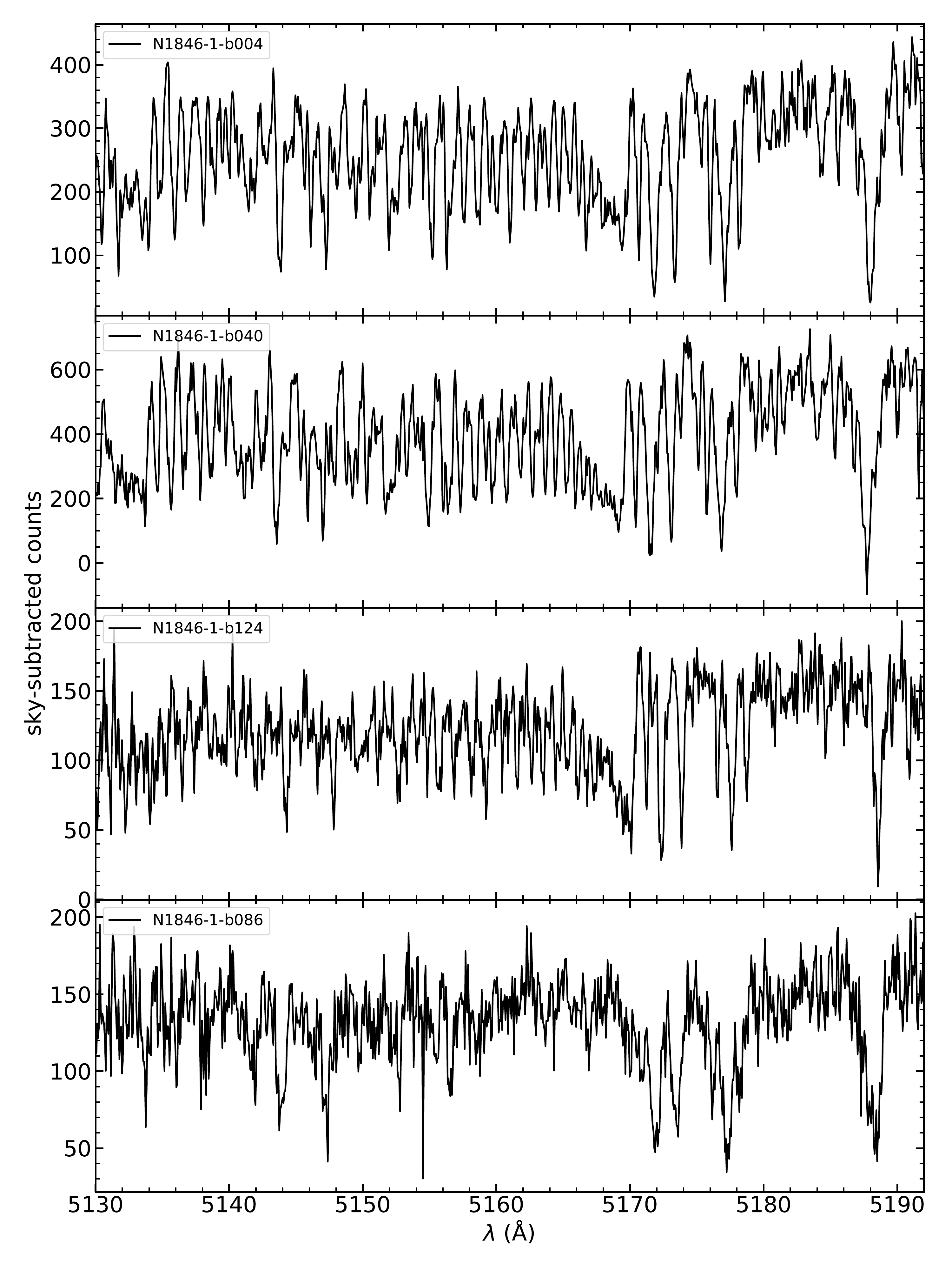}
   \caption{M2FS spectra of carbon stars and blended stars in NGC~419 (left panels) and NGC~1846 (right panels) noted in \autoref{fig:e_vlos}. The top three panels for each cluster show the spectra of the carbon stars.  The bottom panel for each cluster show the spectra of objects with clearly blended spectra from (at least) two unrelated stars (see \autoref{sec:e_vlos}). None of the targets corresponding to these spectra were used in our dynamical analyses of either cluster.  As noted in \autoref{sec:e_vlos}, none of the individual stars (four in total) within the blended spectra are likely members of either cluster.  }
   \label{fig:spec_odd}
\end{figure*}

\begin{table*}
\caption{M2FS Sample of NGC 419. (The full table is available online as supplementary material.)}
\label{tab:sample_n419}
\begin{threeparttable}
\begin{tabular}{cccccccc}
\hline
{ID} & {$\alpha_{2000}$} & {$\delta_{2000}$} & {$V$} & {$V-I$} & {$v_{\rm los}$} & {$P_{M_i}$} & {Sel. $^{\rm a}$}\\ 
{ } & {(h:m:s)} & {($^{\circ}$:$\arcmin$:$\arcsec$)} & {(mag)} & {(mag)} & {(km\,s$^{-1}$)} & { } & { } \\
\hline
N419-1-b001 & 01:08:44.19 & -73:00:40.0 & $16.887\pm0.047$ & $1.497\pm0.055$ & $188.41\pm0.20$ & $0.00$ & M \\
N419-1-b002 & 01:08:46.07 & -72:56:35.2 & $17.942\pm0.030$ & $1.024\pm0.056$ & $192.79\pm0.42$ & $0.00$ & M \\
N419-1-b003 & 01:08:45.46 & -72:55:15.4 & $17.982\pm0.030$ & $1.138\pm0.043$ & $153.30\pm0.39$ & $0.00$ & M \\
N419-1-b004 & 01:08:43.26 & -72:53:32.0 & $18.014\pm0.008$ & $0.855\pm0.010$ & $193.60\pm0.61$ & $0.06$ & H \\
N419-1-b005 & 01:08:42.25 & -72:53:12.3 & $18.769\pm0.010$ & $1.019\pm0.013$ & $144.09\pm1.05$ & $0.00$ & H \\
... \\
\hline
\end{tabular}
\begin{tablenotes}
 \item $^{\rm a}$ {Photometric source that the target were selected from: H stands for {\it HST} and M stands for MCPS.}
\end{tablenotes}
\end{threeparttable}
\end{table*}

\begin{table*}
\caption{Combined Sample of NGC 1846. (The full table is available online as supplementary material.)}
\label{tab:sample_n1846}
\begin{threeparttable}
\begin{tabular}{ccccccccc}
\hline
{ID} & {$\alpha_{2000}$} & {$\delta_{2000}$} & {$V$} & {$V-I$} & {$v_{\rm los}$} & {$P_{M_i}$} & {Sel. $^{\rm a}$} & {Spec. $^{\rm b}$} \\
{} & {(h:m:s)} & {($^{\circ}$:$\arcmin$:$\arcsec$)} & {(mag)} & {(mag)} & {(km\,s$^{-1}$)} &  {} & {} & {} \\
\hline
N1846-1-b001 & 05:07:51.21 & -67:33:03.8 & $17.251\pm0.307$ & $1.396\pm0.310$ & $287.97\pm0.19$ & $0.00$ & M & M2FS \\
... \\
N1846-1-r049 & 05:07:32.21 & -67:27:51.6 & $17.154\pm0.004$ & $1.401\pm0.006$ & $241.93\pm0.26$ & $1.00$ & H & M2FS \\
N1846-1-r079 & 05:07:30.12 & -67:27:27.7 & $17.115\pm0.004$ & $1.306\pm0.006$ & $238.29\pm0.17$ & $1.00$ & H & COMB \\
... \\
ACS-013-R & 05:07:33.59 & -67:26:41.2 & $16.99$ & $1.43$ & $238.10\pm0.46$ & $1.00$ & H & M13 \\
ACS-019-R & 05:07:14.49 & -67:28:16.4 & ... & ... & $278.73\pm0.47$ & $0.00$ & H & M13 \\
... \\
MCPS-007 & 05:05:34.36 & -67:26:34.2 & $16.810\pm0.063$ & $1.622\pm0.079$ & $253.77\pm0.85$ & $0.00$ & M & M13 \\
... \\
\hline
\end{tabular}
\begin{tablenotes}
 \item $^{\rm a}$ {Photometric source that the target were selected from: H stands for {\it HST} and M stands for MCPS.}
 \item $^{\rm b}$ {Spectroscopic source the LOS velocities were measured from: M2FS stands for this work, M13 stands for \citet{Mackey13}, and COMB stands for combined.}
\end{tablenotes}
\end{threeparttable}
\end{table*}

Given the anticipated small internal velocity dispersions of our target clusters, it is crucial to precisely estimate the uncertainties of our radial velocity measurements.   To do this, we have compared for every target the LOS velocities obtained from the individual exposures to measure two statistics that we have used to quantify the velocity uncertainties as a function of mean spectral S/N.   As noted in \autoref{tab:obs}, we obtained four individual exposures of science targets for both NGC~419 and NGC~1846.

The first of these statistics is the standard reduced $\chi$-square.  For the full datasets of a given cluster, we define this as
\begin{equation}
\chi^2_{\nu} = {1\over \nu} \sum_{i=1}^{N} \sum_{j=1}^4 { {(v_{i, j}-\bar{v_i})^2} \over {\varepsilon_{i,j}} },
  \label{eq:chi2}
\end{equation}
where $\varepsilon_{i,j}$ is the  $j$-th velocity error of the $i$-th target, $\bar{v_i}$ is the weighted average velocity of the individual velocities of the $i$-th target, and $\nu=N(n-1)$ is the total degree of freedom of this sample. For NGC~419, $\chi^2_{\nu}=1.54$, while for NGC~1846, $\chi^2_{\nu}=1.25$.

The second statistic we employed was taken from \citet{Kamann16} and is defined as
\begin{equation}
\delta v_{12} = { {v_2-v_1 }\over{\sqrt{\varepsilon^2_{v_1}+\varepsilon^2_{v_2}}} },
  \label{eq:delta}
\end{equation}
where $v_1$ and $v_2$ are a pair of repeat velocity measurements with the uncertainty of $\varepsilon_{v_1}$ and $\varepsilon_{v_2}$, respectively. The velocity uncertainties ($\varepsilon_{v}$) can be considered to be well-estimated when $\delta v_{12}$ is normally distributed with a standard deviation of one. Histograms of $\delta v_{12}$ are shown in the bottom panels of \autoref{fig:cor_evlos}, and the standard deviations ($\sigma_{\delta v_{12}}$) of $\delta v_{12}$ distributions are $1.22$ and $1.11$ for NGC~419 and NGC~1846, respectively.

Both statistics indicate that the individual velocity uncertainties obtained from our Bayesian fits were underestimated by approximately 23\% and 12\% for NGC~419 and NGC~1846, respectively. After applying correction factors of 1.23 and 1.12, respectively, we find that $\chi^2_{\nu}=1.02$ and $\sigma_{\delta v_{12}}=1.00$ for NGC~419, and $\chi^2_{\nu}=1.00$ and $\sigma_{\delta v_{12}}=1.00$ for NGC~1846. The median velocity uncertainties for the samples of NGC~419 and NGC~1846 are 0.38 and 0.22 \kms, respectively. The final barycentric LOS velocities of NGC~419 and NGC~1846 are listed in \autoref{tab:sample_n419} and \autoref{tab:sample_n1846}, respectively, along with their corrected uncertainties.  The table also lists the photometric data for every target taken from our measurements of {\it HST} data or the MCPS photometry of \citet{Zaritsky02, Zaritsky04} for NGC~419 and NGC~1846. respectively.

The analysis above assumes that all targets are non-variable and that their velocity uncertainties scale with S/N in the same systematic manner.  \autoref{fig:e_vlos} is a plot of the corrected velocity uncertainty as a function of the spectral S/N ratio.   It is evident that in both clusters a few stars deviate from the clear relationship between velocity uncertainty and S/N.  These outliers have been confirmed to be either carbon stars (see \autoref{fig:spec_sample} for representative spectra) or multiple stars too close together to be separated in any of the photometric/astrometric catalogs we used to identify targets, resulting in a composite spectrum.  The extracted spectra of these stars are shown in \autoref{fig:spec_odd} and they are noted explicitly in \autoref{fig:e_vlos}).  At a given S/N, such stars always have larger uncertainties than other targets due to the fact that (a) carbon star spectra are not incorporated in the SSPP stellar library \citep{Lee08a, Lee08b, Walker15mn}  used in our spectral fitting procedure, and (b) the fitting procedure assumes a single star and will be compromised when the spectrum is actually composite.   

We have chosen to ignore the carbon stars because of their larger uncertainties and the likelihood that they are velocity variable.  In the case of the blended stars, we have carried out two-star fits with reasonably satisfactory results.  However, in all cases, these composite cases appear to be non-members as both velocities are far from the mean velocities of their respective clusters and the velocity differences are too large for them to be plausible red-giant binaries. For these reasons, we have chosen not to include any of these stars (four in each cluster sample) in any of the subsequent analyses described in this paper.  Note that we retroactively removed these stars from the velocity uncertainty estimation described above; that analysis includes only what we consider to be spectra of `normal,' unblended red giants.

\subsubsection{Combined Sample for NGC~1846}
\label{sec:combine_sample}

\begin{figure}
   \centering
   \includegraphics[width=0.45\textwidth]{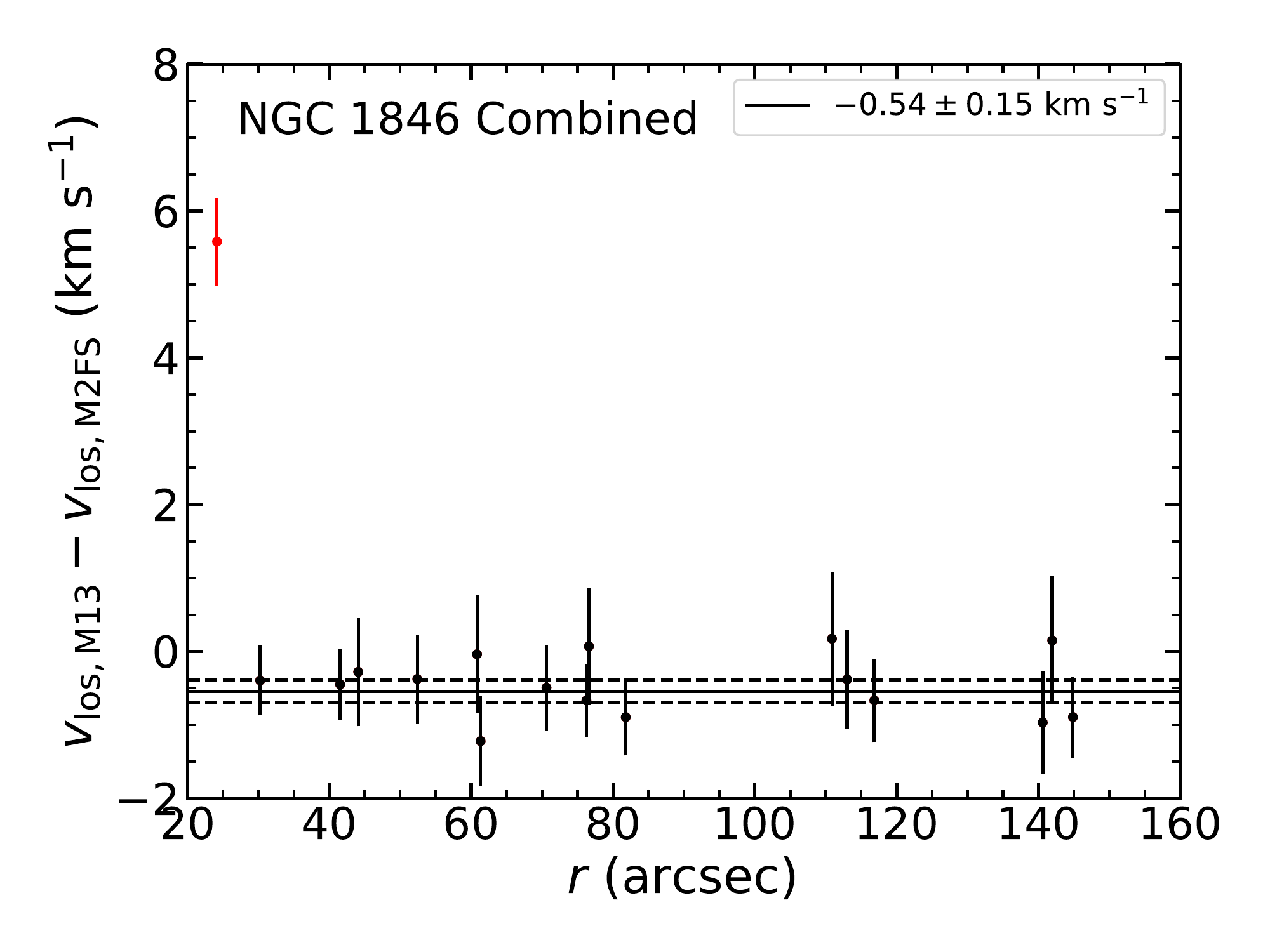}
   \caption{Differences of the LOS velocities measured for 17 stars in NGC~1846 in common to the present study and \citetalias{Mackey13}. The solid line and two dashed lines show the uncertainty-weighted mean and 1$\sigma$ uncertainty for the 16 stars, respectively. The 17th star---a probable spectroscopic binary---is marked in red and has been excluded in the calculation of the mean and 1$\sigma$ uncertainty velocity offset (see \autoref{sec:combine_sample}).}
   \label{fig:common_stars}
\end{figure}

Our  M2FS sample for NGC~1846 consists of 108 targets with useful spectra.   As it happens, 17 of the stars in this group are in common with the dataset from \citetalias{Mackey13}. As shown in \autoref{fig:common_stars}, apart from one exception among these common stars, the velocity differences from our analysis as \citetalias{Mackey13} was found to be small and stable; the exception was a star for which we measure a velocity difference $+5.58$ \kms\ between the two datasets. Specifically,  \citetalias{Mackey13} measured $243.39\pm0.57$ \kms\ and we measured $237.81\pm0.17$ \kms\ with M2FS for two independent measurements of this star taken 9.23 years apart. The latter star is almost certainly a true binary; its spectrum is not obviously composite and its implied velocity amplitude of a few {\it \kms} over a period of about 5 years is consistent with a main-sequence companion slightly less massive than the red giant that dominates the spectrum.  For the other 16 stars common to the two samples, we find a small but significant velocity difference of $-0.54\pm0.15$ \kms\ in the sense \citetalias{Mackey13} {\it minus} M2FS.  The scatter in this mean offset is small and entirely consistent with the combined median uncertainties of the \citetalias{Mackey13} sample (0.57 \kms) and the M2FS sample (0.30 \kms).  Consequently, we have defined a `Combined Sample' consisting of the \citetalias{Mackey13} sample corrected by the velocity difference noted above (\autoref{fig:common_stars}).  For the stars in common to the two samples, we have calculated their error-weighted mean velocities and velocity uncertainties; these stars are noted and their weighted mean values listed in \autoref{tab:sample_n1846}; all other stars from \citetalias{Mackey13} but not in our M2FS sample are also noted in this table.   The final Combined Sample for NGC~1846 contains 196 stellar targets, including the binary star.

\subsection{Systemic Velocity and Velocity Dispersion}
\subsubsection{The EM alogrithm}
\label{sec:EM}

\begin{figure*}
   \centering
    \includegraphics[width=0.45\textwidth]{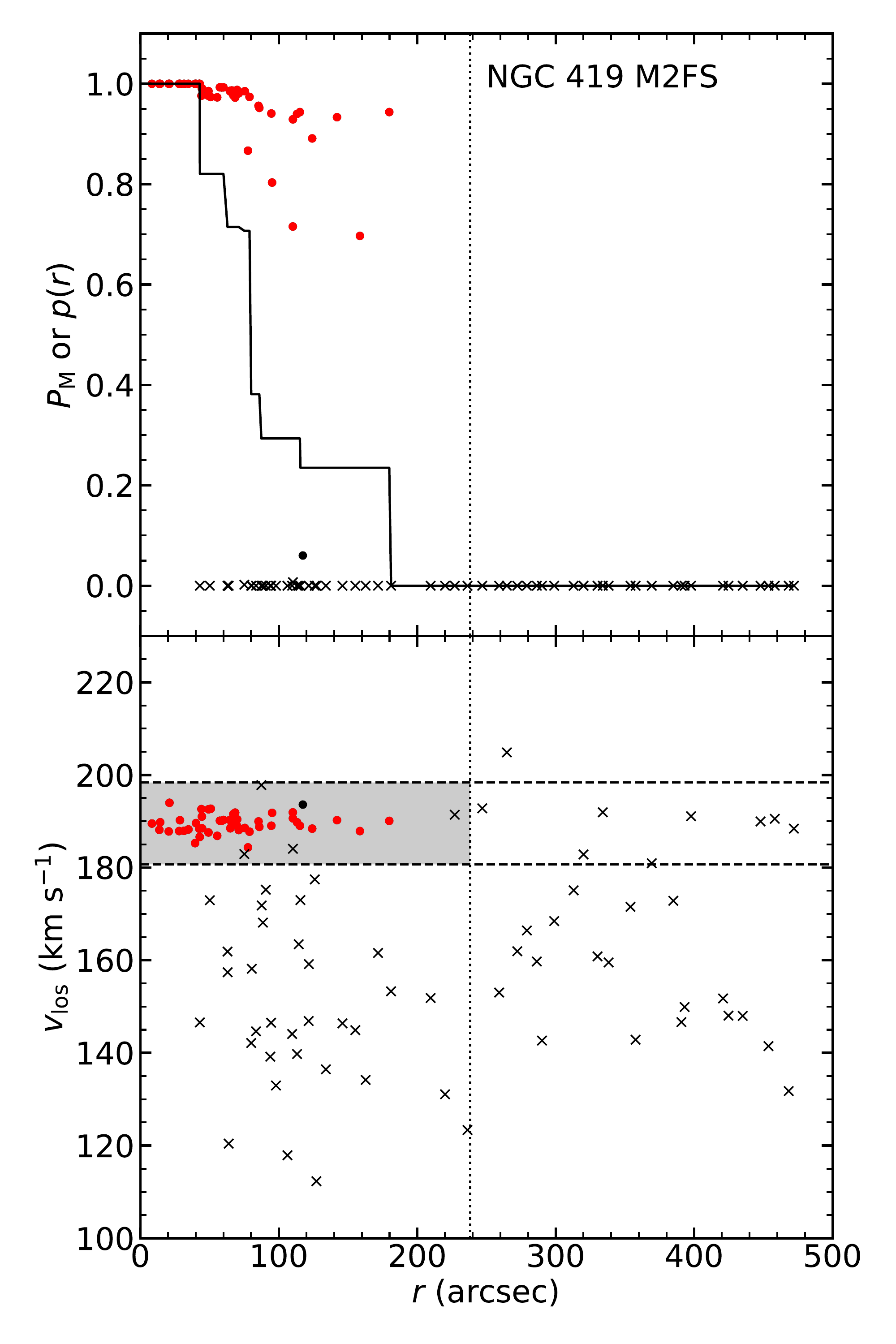}
    \includegraphics[width=0.45\textwidth]{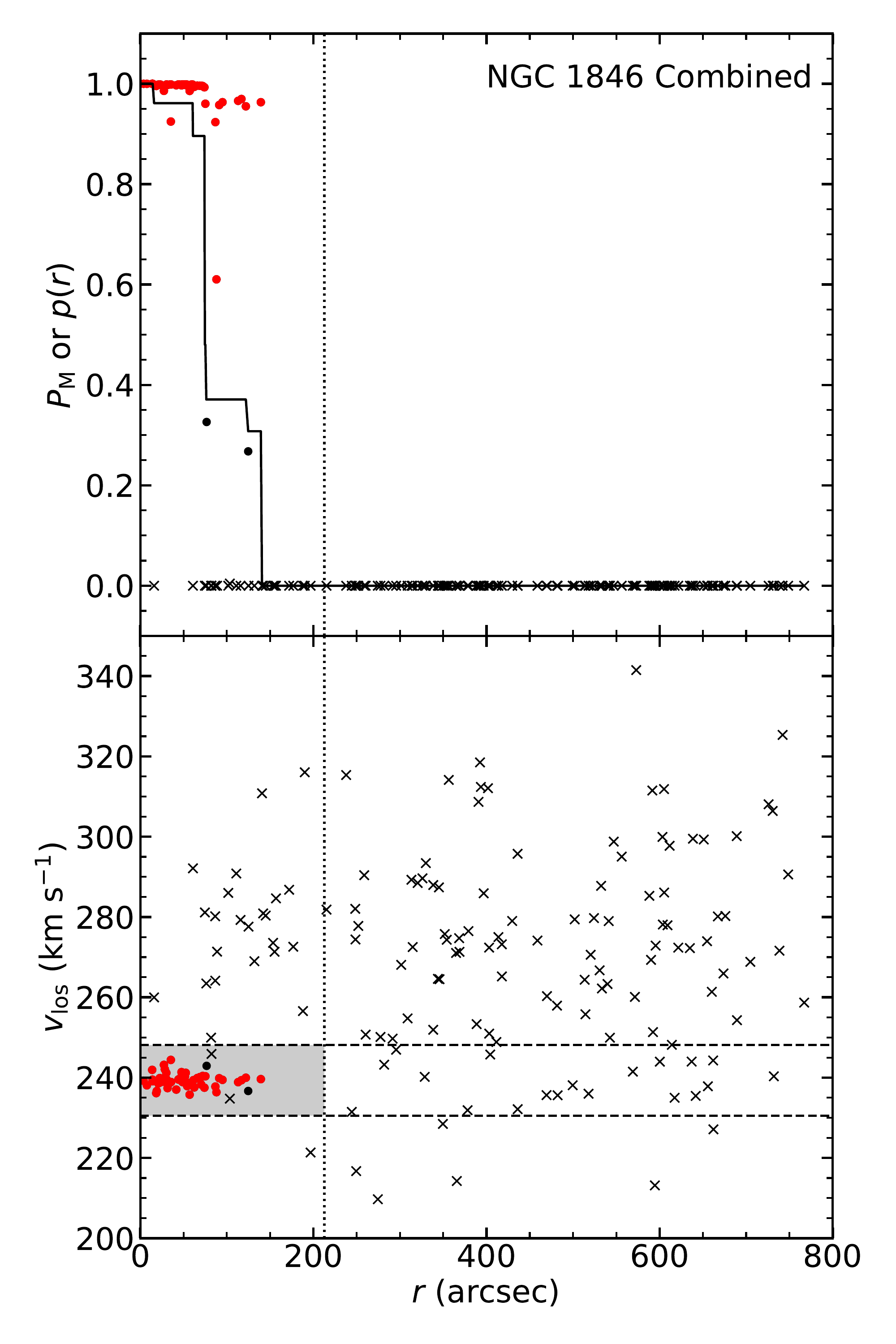}    
   \caption{Membership probabilities (top) and LOS velocities (bottom) for stars as a function of the projected distance from their respective cluster centers. The adopted cluster centers are listed in \autoref{tab:obs}. The top panels show the unconditional probability function $p(r)$ (solid line) from the EM algorithm (see \autoref{sec:EM}).  The shaded regions in the bottom panels denote the `box' regions we defined in \autoref{sec:box}. Red dots correspond to stars with $P_{\rm M} \geq 0.5$, black dots correspond to stars with $0.05 \le P_{\rm M}<0.5$, and crosses correspond to stars with $P_{\rm M}<0.05$. The vertical lines denote the locations of the tidal radii.}
   \label{fig:mem}
\end{figure*}

\begin{table*}
 \caption{Velocity and Dispersion Results from the EM method.}
 \label{tab:EM}
 \begin{threeparttable}
 \begin{tabular}{ccccccccccc}
\hline
    {Cluster}  &  
    {Dataset}  &  
    {Binary?$^{\rm a}$} &  
    {RC$^{\rm b}$} &  
    { $V_{\rm sys}$}  &  
    {$\sigma_{\rm p,\,0}$}  &  
    {$V_{\rm field}$}  &  
    {$\sigma_{\rm field}$}  &  
    {$N_{\rm total}$}  &  
    {$N_{\rm tidal}$}
    \\
    {} &  
    {} &  
    {} &  
    {} &  
    {$\rm (km\,s^{-1})$}  &  
    {$\rm (km\,s^{-1})$}  &  
    {$\rm (km\,s^{-1})$}  &  
    {$\rm (km\,s^{-1})$}  &  
    {}  &  
    {} \\
\hline 
NGC 419  & M2FS     & ... & N & $189.5^{+0.3}_{-0.3}$ & $2.42^{+0.41}_{-0.27}$ & $160.3^{+2.8}_{-2.7}$ & $21.7^{+1.4}_{-1.6}$ & 111 & 83 \\
         & M2FS     & ... & Y & $189.5^{+0.3}_{-0.3}$ & $2.20^{+0.33}_{-0.19}$ & $160.6^{+2.8}_{-2.7}$ & $21.7^{+1.1}_{-1.8}$ & 111 & 83 \\
\\
NGC 1846 & Combined & N & N & $239.3^{+0.2}_{-0.2}$ & $2.05^{+0.32}_{-0.30}$ & $269.5^{+2.2}_{-2.0}$ & $25.2^{+1.4}_{-1.5}$ & 195 & 80 \\
         & Combined & N & Y & $239.3^{+0.3}_{-0.2}$ & $1.94^{+0.29}_{-0.24}$ & $269.5^{+2.2}_{-2.0}$ & $25.2^{+1.2}_{-1.5}$ & 195 & 80 \\
         & Combined & Y & N & $239.3^{+0.2}_{-0.2}$ & $2.02^{+0.33}_{-0.31}$ & $269.5^{+2.2}_{-2.0}$ & $25.2^{+1.4}_{-1.4}$ & 196 & 81 \\
\\
         & M2FS     & Y & N & $239.3^{+0.3}_{-0.3}$ & $1.85^{+0.28}_{-0.26}$ & $268.4^{+3.1}_{-2.8}$ & $24.6^{+1.8}_{-2.4}$ & 108 & 69 \\
         & M2FS     & Y & Y & $239.3^{+0.2}_{-0.2}$ & $1.79^{+0.29}_{-0.35}$ & $268.3^{+3.3}_{-3.1}$ & $24.6^{+1.9}_{-2.4}$ & 108 & 69 \\
\\
         & Ma13      & N & N & $238.6^{+0.5}_{-0.5}$ & $2.27^{+0.64}_{-0.83}$ & $270.4^{+2.9}_{-2.9}$ & $25.5^{+1.5}_{-2.0}$ & 105 & 29 \\
         & Ma13      & N & Y & $239.0^{+0.5}_{-0.5}$ & $2.13^{+0.42}_{-0.31}$ & $270.4^{+3.0}_{-2.8}$ & $25.5^{+1.4}_{-1.9}$ & 105 & 29 \\
         & Ma13      & Y & N & $238.9^{+0.5}_{-0.6}$ & $2.59^{+0.47}_{-0.68}$ & $270.5^{+2.7}_{-2.9}$ & $25.5^{+1.4}_{-2.0}$ & 106 & 30 \\
         & Ma13      & Y & Y & $239.2^{+0.4}_{-0.5}$ & $2.15^{+0.33}_{-0.29}$ & $270.5^{+2.9}_{-2.6}$ & $25.5^{+1.6}_{-1.9}$ & 106 & 30 \\
\hline
\end{tabular}
\begin{tablenotes}
\item $^{\rm a}$ {This flag denotes whether the confirmed binary star in NGC~1846 is excluded (N) or included (Y). This star is `N1846-1-r079' in \autoref{tab:sample_n1846}, and the individual velocities are summarized in \autoref{sec:combine_sample} and \autoref{fig:common_stars}.}
\item $^{\rm b}${This flag indicates whether the rotational correction described in \autoref{sec:rot} was applied.}
\end{tablenotes}
\end{threeparttable}
\end{table*}

We used an expectation-maximization (EM) algorithm \citep{Walker09, Kimmig15} to simultaneously constrain the systemic velocity and the model-dependent central velocity dispersion of each cluster. 
This method, which can efficiently distinguish cluster members from field stars, iteratively determines the cluster membership probability for each star considering whether its velocity and location is consistent with a selected dynamical model. 
As this process iterates, membership probabilities and the parameters of the selected model are updated until a convergence criterion is achieved. 

In our implementation of this approach, we constrained the systemic velocity and the projected central velocity dispersion of the cluster from its member population, simultaneously with the mean velocity and the velocity dispersion of the field stellar population (that is, nonmembers). 
In the present case, the EM analysis of the kinematic and positional data for targets of a given cluster involves the estimation of four parameters 
\begin{equation}
\zeta\equiv\{\langle V \rangle_{\rm mem},\sigma_{V_0,\rm mem}^2,\langle V \rangle_{\rm non},\sigma_{V_0,\rm non}^2\}.
\label{eq:param}
\end{equation}
These refer to the mean velocities and velocity dispersions of the cluster (member, `mem') and field (non-member, `non') populations assumed to comprise our kinematic samples.  The expected log-likelihood used in the EM approach and given our data can be written as
\begin{eqnarray}
  E(\ln \mathcal{L}(\zeta)|S)=\displaystyle \sum_{i=1}^{N}P_{M_i} \ln \biggl [p_{\rm mem}(V_i)p(r_i)\biggr ]\nonumber\\
  +\displaystyle \sum_{i=1}^N(1-P_{M_i})\ln \biggl [p_{\rm non}(V_i)[1-p(r_i)]\biggr], \hspace{0.2in}
  \label{eq:mstep}
\end{eqnarray}
where $S\equiv\{V_i,\,\varepsilon_{V_i},\,r_i\}_{i=1}^N$ represents the full dataset,  $P_{M_i}$ is a normalized cluster-membership probability of the $i$-th star, and $p_{\rm mem}(V_i)$ and $p_{\rm non}(V_i)$ are the cluster membership and non-membership probabilities of the same star. The term $p(r_i)$, referred as the `unconditional probability function' by \citet{Walker09}, represents a non-increasing component of the membership probability that considers the observational selection, the assumed dynamical model, and the uncertainty of the extent of a cluster.

One iteration of the EM algorithm starts with the expectation step (E step) and ends with the maximization step (M step). 
In the E step, $P_{M_i}$ is estimated from a combination of $p_{\rm mem}(V_i)$, $p_{\rm non}(V_i)$ and $p(r_i)$ through
\begin{eqnarray}
  P_{M_i}\equiv P(M_i=1|V_i,r_i)\hspace{1.15in}\nonumber\\
  =\frac{p_{\rm mem}(V_i)p(r_i)}{p_{\rm mem}(V_i)p(r_i)+p_{\rm non}(V_i)[1-p(r_i)]}, 
  \label{eq:estep}
\end{eqnarray}
 where 
\begin{equation}
  p_{\rm mem}(V_i) = \frac{\exp \biggl [-\frac{1}{2} \biggl (\frac{[V_i-\langle V \rangle_{\rm mem}]^2}{\sigma_{V_0,\rm mem}^2+\varepsilon_{V_i}^2}\biggr ) \biggr ]}{\sqrt{2\pi (\sigma_{V_0,\rm mem}^2+\varepsilon_{V_i}^2)}},
  \label{eq:pmem}
\end{equation}
and
\begin{equation}
  p_{\rm non}(V_i) = \frac{\exp \biggl [-\frac{1}{2} \biggl (\frac{[V_i-\langle V \rangle_{\rm non}]^2}{\sigma_{V_0,\rm non}^2+\varepsilon_{V_i}^2}\biggr ) \biggr ]}{\sqrt{2\pi (\sigma_{V_0,\rm non}^2+\varepsilon_{V_i}^2)}}.
  \label{eq:pnon}
\end{equation}
The term ${p}(r_i)$ in our study has the format of
\begin{equation}
{p}(r_i)=p_{\rm dyn}(r_i)\cdot\displaystyle \min_{1\leq u \leq i}\biggl [\displaystyle\max_{i\leq v \leq N}\frac{\Sigma_{j=u}^v{P}_{M_j}}{v-u+1}\biggr ],
  \label{eq:monotonic}
\end{equation}
where the first part makes sure that a member star must be located within its tidal radius, i.e.
\begin{equation}
p_{\rm dyn}(r_i) = \left\{ \begin{array}{rl}
 1 &\mbox{ if $r_i\leq r_t$} \\
 0 &\mbox{ if $r_i>r_t$}
       \end{array} \right. ,
\end{equation}
while the second part reflects the non-increasing feature of ${p}(r_i)$ that follows the discussion by \citet{Walker09}. 

In the M step, both $p(r_i)$ and $\zeta$ are updated so that they are on track to converge to the best-fit results. 
To update $p(r_i)$, we simply recalculate \autoref{eq:monotonic}. For the member parts in $\zeta$, we have the following equation pair 
\begin{equation}
  \langle {V} \rangle_{\rm mem}=\frac{\displaystyle\sum_{i=1}^{N}\frac{P_{M_i}V_i}{1+\varepsilon_{V_i}^2/{\sigma}_{V_0,\rm mem}^{2}}}{\displaystyle\sum_{i=1}^N\frac{P_{M_i}}{1+\varepsilon_{V_i}^2/{\sigma}_{V_0,\rm mem}^{2}}}, 
  \label{eq:EM_mean}
\end{equation}
and
\begin{equation}
  {\sigma}_{V_0,\rm mem}^{2}=\frac{\displaystyle\sum_{i=1}^{N}\frac{P_{M_i}[V_i-\langle {V}\rangle_{\rm mem}]^2}{(1+\varepsilon_{V_i}^2/{\sigma}_{V_0,\rm mem}^{2})^2}}{\displaystyle\sum_{i=1}^N\frac{P_{M_i}}{1+\varepsilon_{V_i}^2/{\sigma}_{V_0,\rm mem}^{2}}},
  \label{eq:EM_disp}
\end{equation}
while $\langle V \rangle_{\rm non}$ and $\sigma_{V_0,\rm non}^2$ are in the same forms as in the pair above but with $P_{M_i}$ replaced by $(1-P_{M_i})$.

When considering the dynamical model of a cluster, $\sigma_{V_0,\rm mem}$ should be replaced by $\sigma_{\rm dyn}(r_i)\sigma_{V_0,\rm mem}$, where the additional factor $\sigma_{\rm dyn}(r_i)$ represents the projected velocity dispersion profile of a cluster with a central value of 1 \kms. Thus \autoref{eq:EM_disp} becomes
\begin{equation}
  {\sigma}_{V_0,\rm mem}^{2}=\frac{\displaystyle\sum_{i=1}^{N}\frac{P_{M_i}[V_i-\langle {V}\rangle_{\rm mem}]^2}{\{1+\varepsilon_{V_i}^2/[\sigma_{\rm dyn}(r_i)\sigma_{V_0,\rm mem}]^2\}^2}}{\displaystyle\sum_{i=1}^N\frac{P_{M_i}\sigma_{\rm dyn}^2(r_i)}{1+\varepsilon_{V_i}^2/[\sigma_{\rm dyn}(r_i)\sigma_{V_0,\rm mem}]^{2}}}.
  \label{eq:EM_disp_dyn}
\end{equation}

In the present analysis we have adopted single-mass King models \citep[][K66]{King66} to generate $\sigma_{\rm dyn}(r_i)$. The structural parameters of a K66 model were transformed from those of the best-fit K62 profile, under the assumption that both agreed on three basic parameters of a surface brightness profile: the central value, the core radius where the value is half of the central value, and the total luminosity. The transformed K66 parameters derived from the K62 parameters we adopted in \autoref{sec:reduction} are also listed in \autoref{tab:basic}. In actual practice, we used the code \texttt{LIMEPY} \citep{limepy} to calculate the appropriate K66 models based on the transformed K62 parameters listed for both NGC~419 and NGC~1846 in Table 1.  

At the start of the EM analysis for any dataset, we initialized ${p}(r_i)=0.5$ and $P_{M_i}=0.5$ for all stars and estimated the initial membership parameters in $\zeta$ from the stars within its K66 tidal radius and assumed non-membership for the rest.  We iterated until a convergence criterion of $\Delta \zeta/\zeta<1\times10^{-5}$ from one iteration to the next.  Typically this condition was satisfied within 10-20 iterations. The errors on $\zeta$ are estimated via bootstrapping, in which 1000 realizations are randomly sampled from the original dataset.

We illustrate all key aspects of the EM analysis in \autoref{fig:mem}. The top panels compare the membership probabilities $P_{\rm M}$ with the $p(r_i)$ profiles, while the bottom panels plot radial velocity versus distance from the center of each cluster. 
In both panels, all targets are marked differently by their $P_{M_i}$: red dots correspond to those with $P_{\rm M} \geq 0.5$, black dots correspond to those with $0.05 \le P_{\rm M}<0.5$, and black crosses correspond to those with $P_{\rm M}<0.05$.
\autoref{tab:EM} lists the best-fit results in the form of $\zeta=\{V_{\rm sys}, \sigma_{\rm p, 0}, V_{\rm field}, \sigma_{\rm field}\}$, where now $V_{\rm sys}$ is the systemic cluster velocity, $\sigma_{\rm p, 0}$ is the projected cluster central velocity dispersion, $V_{\rm field}$ is the mean velocity of the field population, and $\sigma_{\rm field}$ is the projected velocity dispersion of the field population. The table also lists  the total sample size and the number of targets located within the K66 tidal radius.  Only the M2FS dataset was considered for NGC~419.  For NGC~1846 the situation is more complicated, and we list results for the \citetalias{Mackey13}, M2FS and Combined Samples.  In the \citetalias{Mackey13} and Combined cases, we further consider samples that include and exclude the binary star described in \autoref{sec:combine_sample}.  Note that for the M2FS sample, we obtained precisely the same results with the binary star included or excluded given that in that dataset this star's velocity is very close to the mean cluster velocity derived by the EM algorithm.

\subsubsection{Assigning Cluster Membership}
\label{sec:box}

\begin{figure*}
   \centering  
    \includegraphics[width=0.45\textwidth]{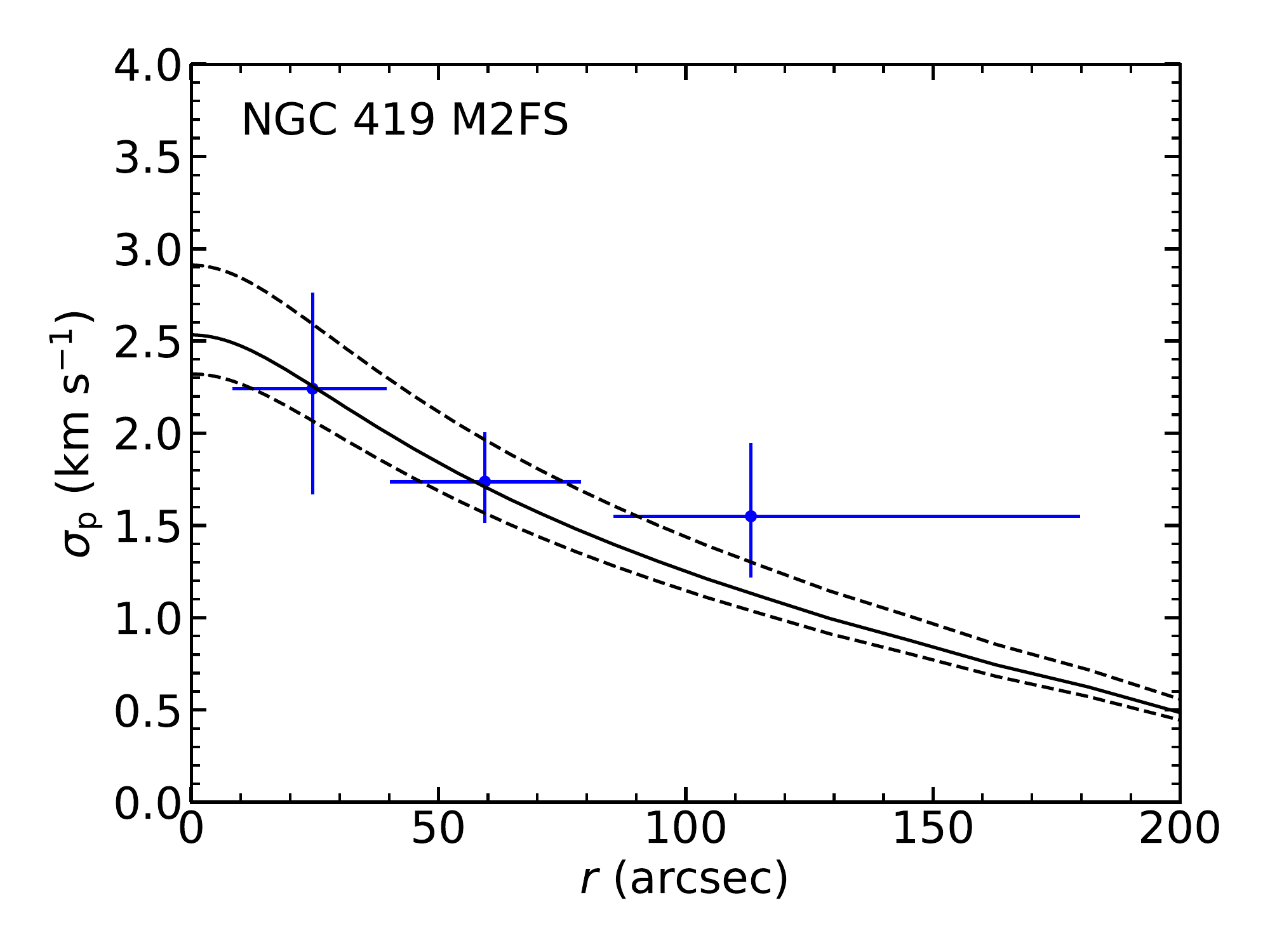}
    \includegraphics[width=0.45\textwidth]{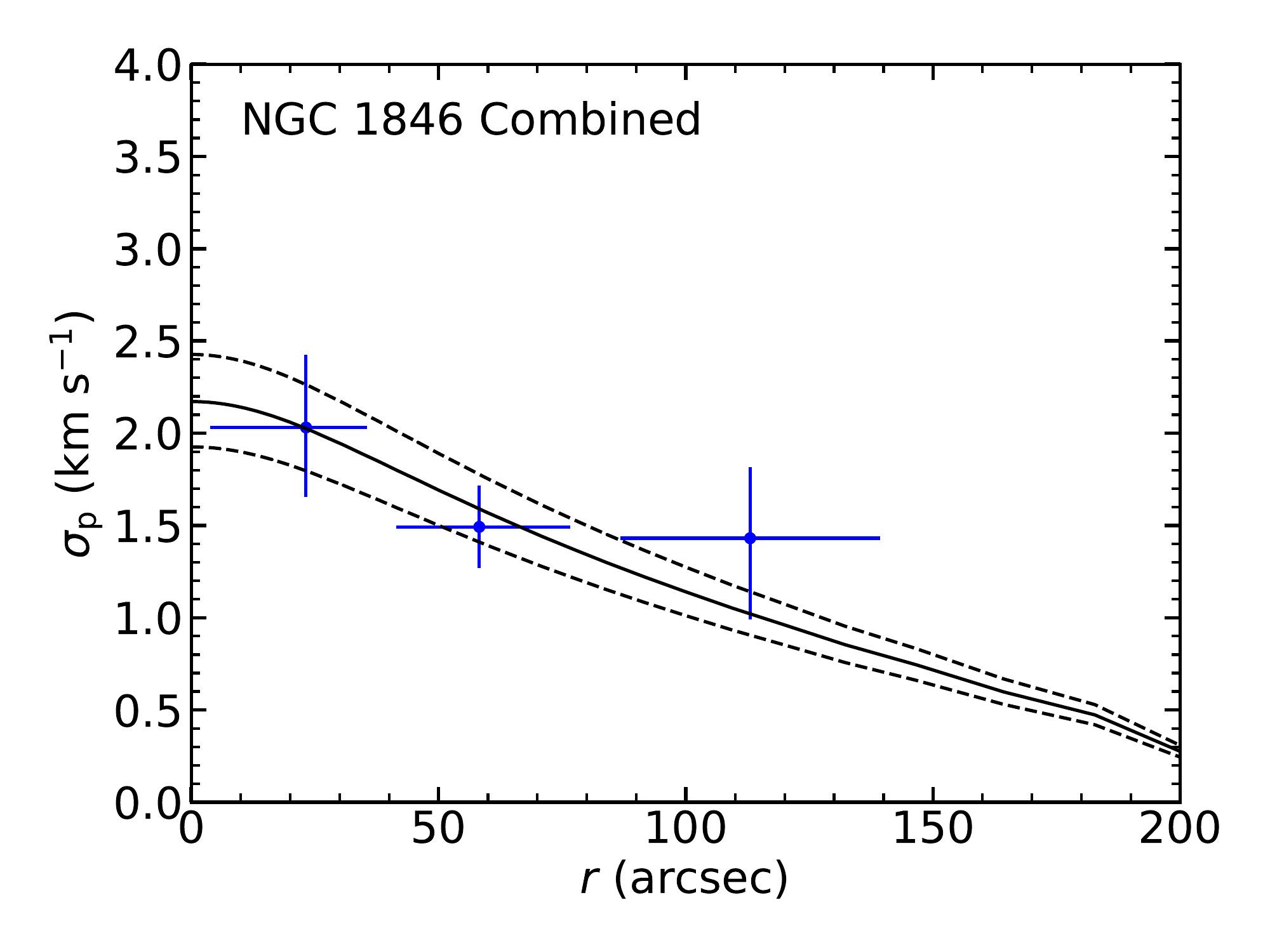}
    \includegraphics[width=0.45\textwidth]{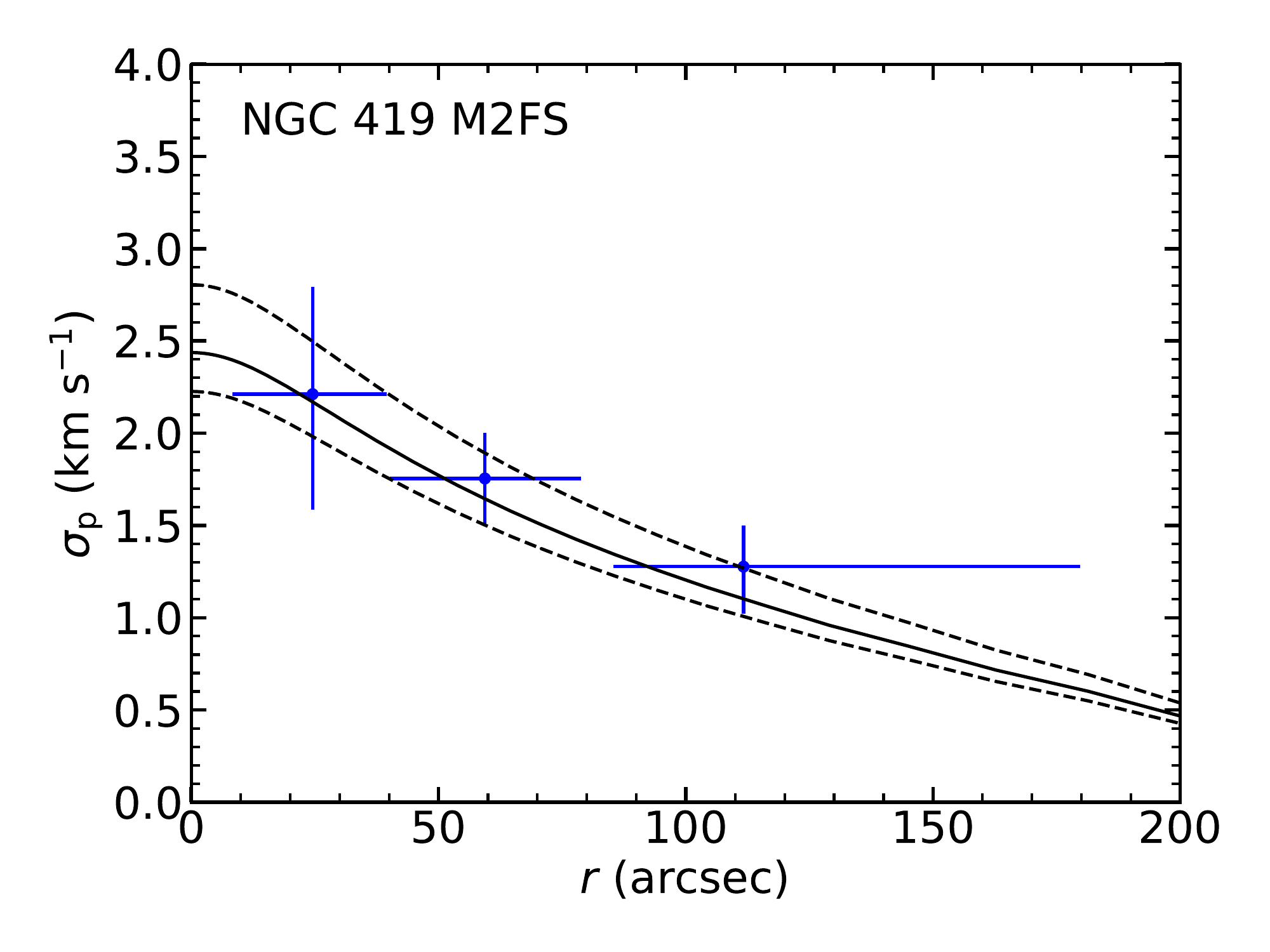}
    \includegraphics[width=0.45\textwidth]{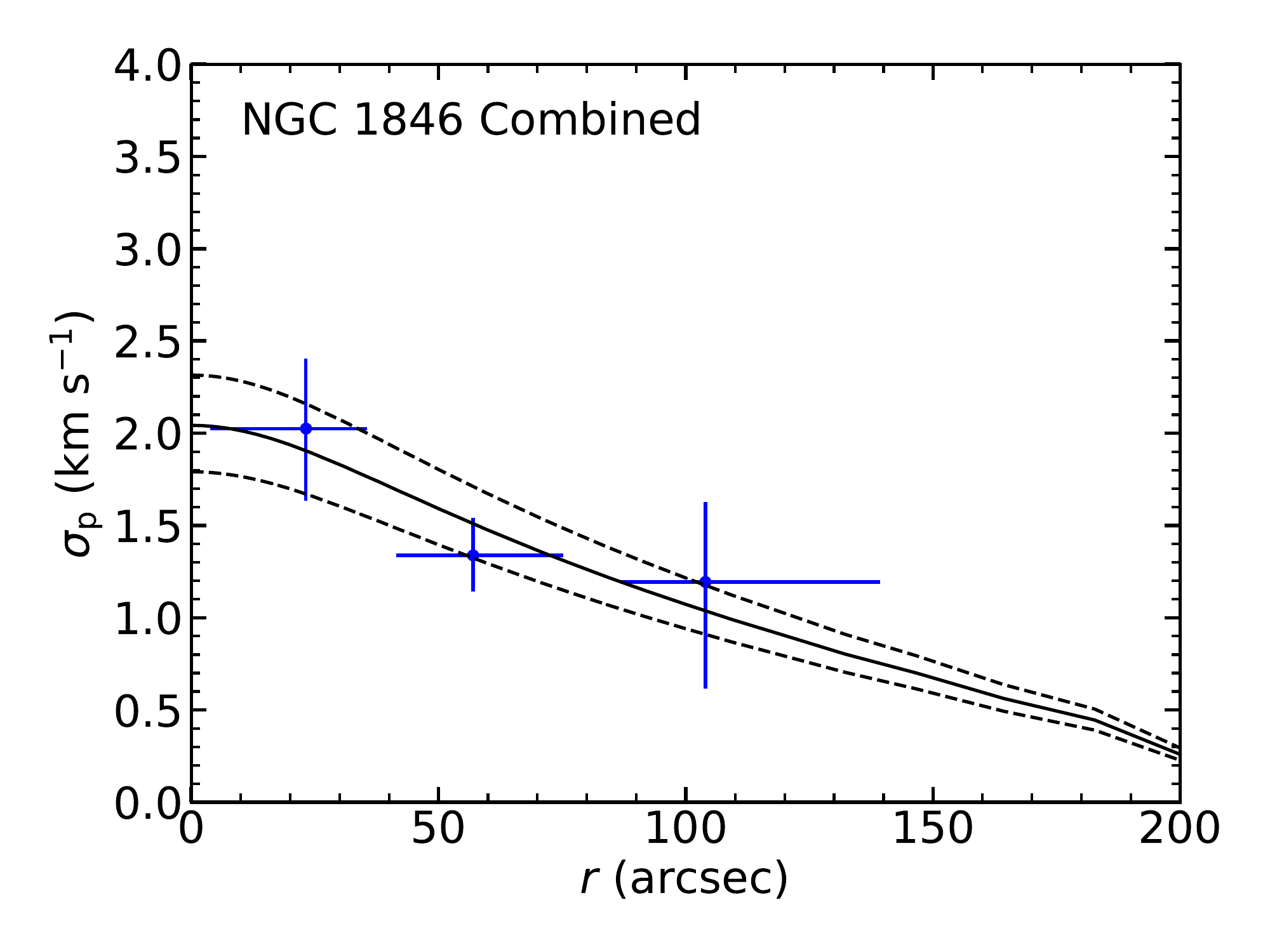}
   \caption{Binned velocity dispersion profiles (blue dots) with best-fit K66 models (solid lines) scaled by the PM05 values (top panels) and the PM50 values (bottom panels). The adopted cluster centers are listed in \autoref{tab:obs}. The two dashed lines in each panel denote the 1$\sigma$ uncertainties on the central velocity dispersion. The vertical `error bars' are the 1$\sigma$ uncertainties of the $\sigma_{\rm p}$ values in each bin. The horizontal `error bars' represent the radial range of stars in each bin.}
   \label{fig:vd_profile}
\end{figure*}

\begin{figure*}
   \centering    
   \includegraphics[width=0.45\textwidth]{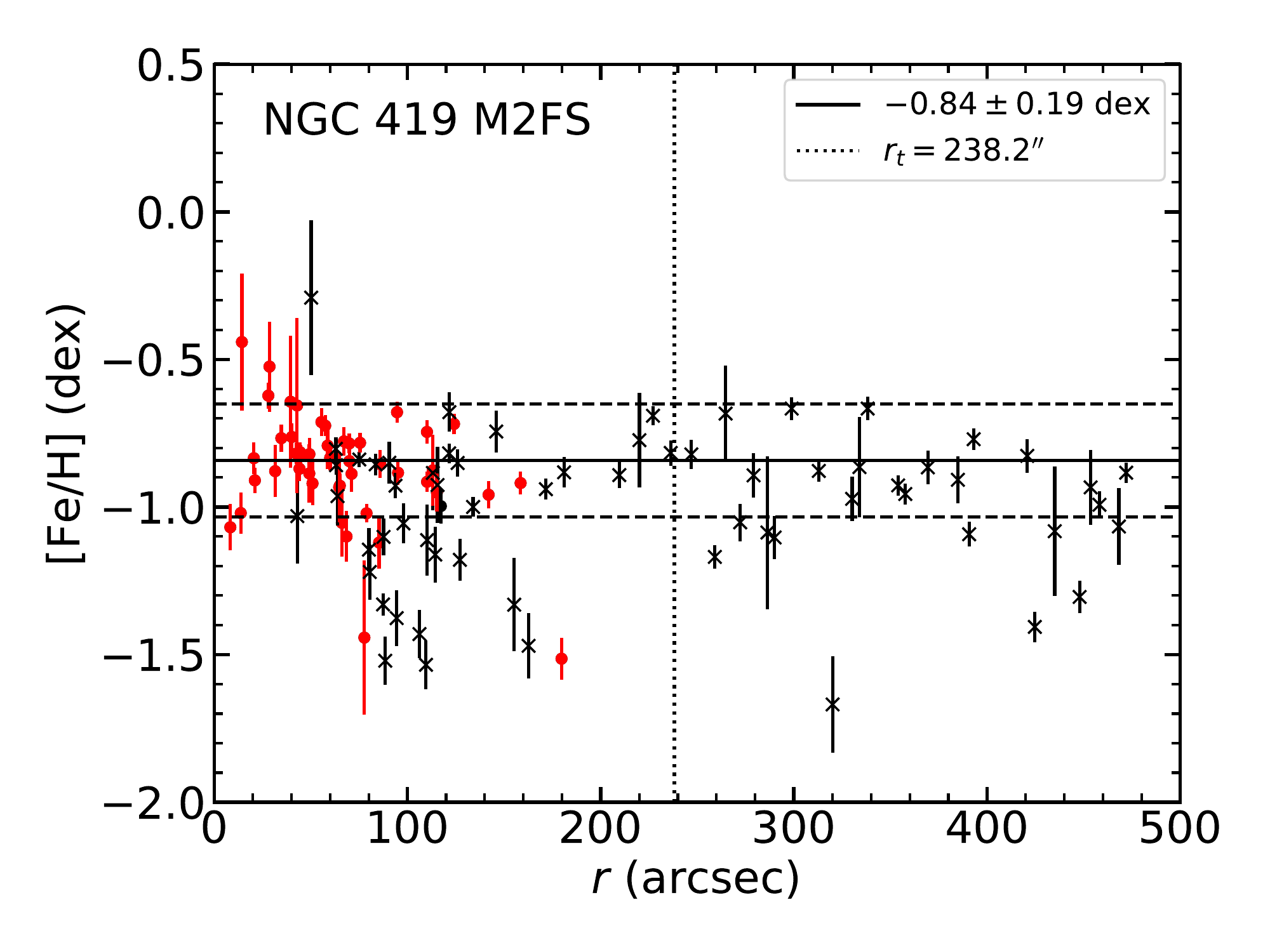}
   \includegraphics[width=0.45\textwidth]{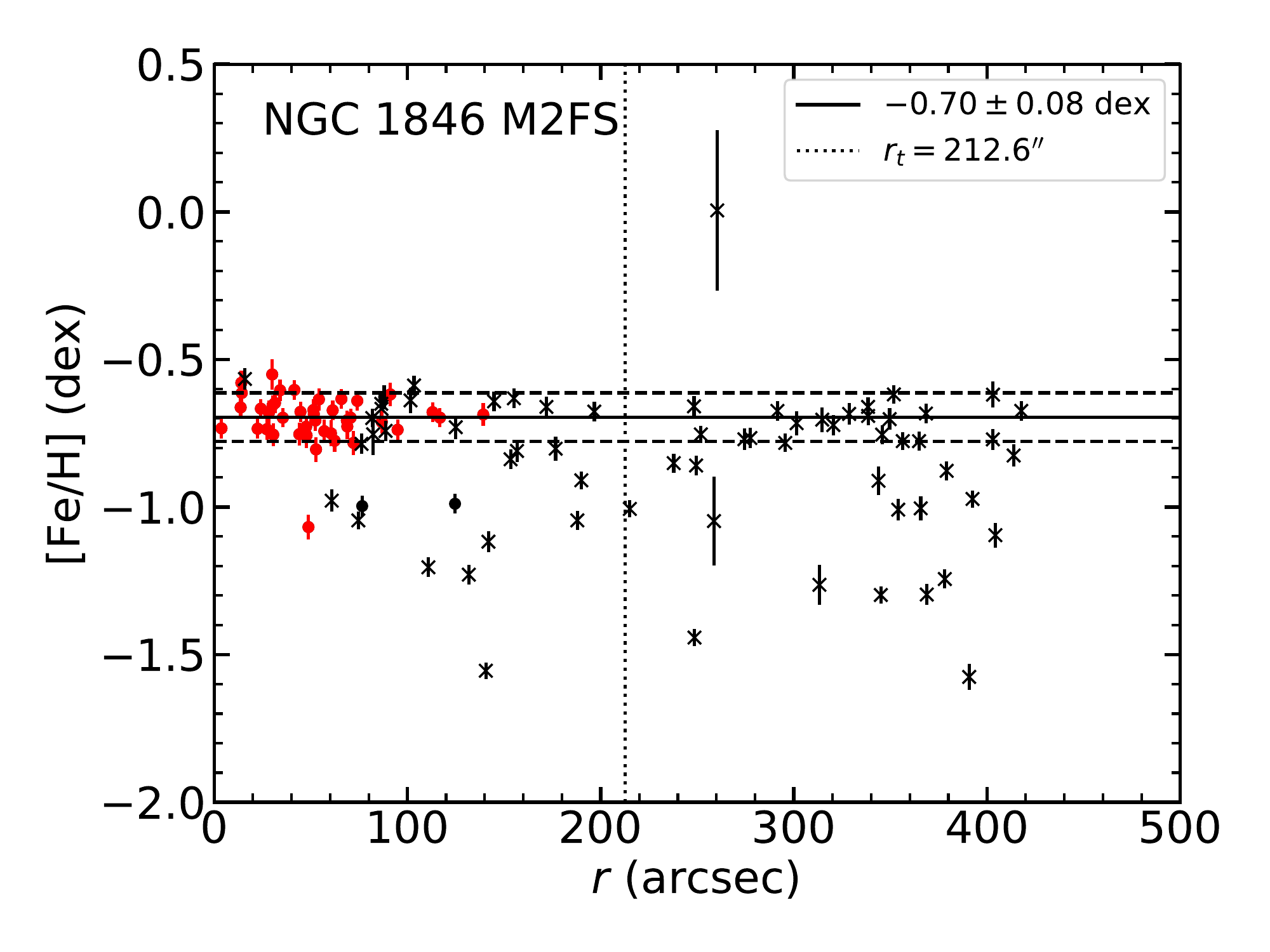}      
   \caption{Metallicity of stars in NGC~419 (left panel) and NGC~1846 (right panel) as a function of the projected distances from the respective cluster centers. The adopted cluster centers are listed in \autoref{tab:obs}. As in \autoref{fig:mem}, red dots correspond to stars with $P_{\rm M} \geq 0.5$, black dots correspond to stars with $0.05 \le P_{\rm M}<0.5$, and black crosses correspond to stars with $P_{\rm M}<0.05$. In each panel, the vertical dotted line denotes the cluster tidal radius. The solid horizontal line and two dashed horizontal lines denote the mean and standard deviation of the metallicities of all stars with $P_{\rm M} \geq 0.5$.  Note that at most two stars in the PM50 samples (one in each cluster) could be argued to be non-members based on their metallicity (see \autoref{sec:box} for further discussion).}
   \label{fig:FeH}
\end{figure*}

\begin{table*}
\caption{Velocity and Dispersion Results from the PM50, PM05 and Box methods$^{\rm a}$.}
\label{tab:box}
\begin{threeparttable}
\begin{tabular}{ccccccccc}
\hline
    {Cluster} & 
    {Dataset} & 
    {Method} & 
    {Binary?$^{\rm b}$} &  
    {RC$^{\rm c}$} &  
    {$N_{\rm mem}$} &  
    {$V_{\rm sys}$} &  
    {$\sigma_{\rm p,\,0}$} 
    \\ 
    {} &  
    {} & 
    {} &  
    {} & 
    {} &  
    {} &  
    {$\rm (km\,s^{-1})$} &  
    {$\rm (km\,s^{-1})$} \\ 
\hline
NGC 419  & M2FS     & Box             & ... & N & 51 & $189.6^{+0.3}_{-0.4}$ & $3.57^{+0.59}_{-0.54}$ \\
         & M2FS     & PM05            & ... & N & 47 & $189.6^{+0.3}_{-0.3}$ & $2.53^{+0.39}_{-0.20}$ \\
         & \bf M2FS     & \bf PM50            & \bf ... & \bf N & \bf 46 & \bf $\bf 189.5^{+0.3}_{-0.3}$ & \bf $\bf 2.44^{+0.37}_{-0.21}$ \\
\\
         & M2FS     & Box             & ... & Y & 51 & $189.5^{+0.3}_{-0.3}$ & $3.44^{+0.66}_{-0.56}$ \\
         & M2FS     & PM05            & ... & Y & 46 & $189.5^{+0.3}_{-0.3}$ & $2.24^{+0.33}_{-0.14}$ \\
         & M2FS     & PM50            & ... & Y & 45 & $189.5^{+0.3}_{-0.3}$ & $2.22^{+0.34}_{-0.14}$ \\
\\                                                
NGC 1846 & Combined & Box             & N & N & 56 & $239.3^{+0.3}_{-0.3}$ & $2.52^{+0.33}_{-0.32}$ \\	
		 & Combined & PM05            & N & N & 54 & $239.3^{+0.2}_{-0.2}$ & $2.17^{+0.25}_{-0.24}$ \\
         & \bf Combined & \bf PM50            & \bf N & \bf N & \bf 52 & \bf $\bf 239.3^{+0.2}_{-0.2}$ & \bf $\bf 2.04^{+0.28}_{-0.24}$ \\ 
\\
         & Combined & Box             & N & Y & 56 & $239.4^{+0.3}_{-0.2}$ & $2.43^{+0.32}_{-0.32}$ \\ 
         & Combined & PM05            & N & Y & 54 & $239.3^{+0.2}_{-0.2}$ & $2.07^{+0.26}_{-0.23}$ \\ 
         & Combined & PM50            & N & Y & 52 & $239.3^{+0.2}_{-0.2}$ & $1.93^{+0.25}_{-0.23}$ \\ 
\\                                            
		 & Combined & Box             & Y & N & 57 & $239.3^{+0.3}_{-0.3}$ & $2.49^{+0.32}_{-0.32}$ \\
		 & Combined & PM05            & Y & N & 55 & $239.3^{+0.2}_{-0.2}$ & $2.15^{+0.27}_{-0.23}$ \\
		 & Combined & PM50            & Y & N & 53 & $239.3^{+0.2}_{-0.2}$ & $2.02^{+0.28}_{-0.25}$ \\ 
\\                                            
   		 & M2FS     & Box             & Y & N & 46  & $239.3^{+0.3}_{-0.3}$ & $2.48^{+0.36}_{-0.39}$ \\
   		 & M2FS     & PM05            & Y & N & 44  & $239.3^{+0.2}_{-0.2}$ & $2.04^{+0.23}_{-0.25}$ \\
		 & M2FS     & PM50            & Y & N & 41  & $239.4^{+0.2}_{-0.2}$ & $1.80^{+0.23}_{-0.24}$ \\
\\
		 & M2FS     & Box             & Y & Y & 46  & $239.3^{+0.3}_{-0.3}$ & $2.45^{+0.35}_{-0.39}$ \\
		 & M2FS     & PM05            & Y & Y & 44  & $239.3^{+0.2}_{-0.2}$ & $2.00^{+0.23}_{-0.26}$ \\
		 & M2FS     & PM50            & Y & Y & 41  & $239.4^{+0.2}_{-0.2}$ & $1.75^{+0.22}_{-0.25}$ \\
\\	                                              
         & Ma13      & Box, PM05       & N & N & 21  & $238.7^{+0.4}_{-0.5}$ & $2.39^{+0.53}_{-0.64}$ \\
         & Ma13      & PM50            & N & N & 20  & $238.5^{+0.4}_{-0.4}$ & $2.17^{+0.59}_{-0.68}$ \\
         & Ma13      & Box, PM05, PM50 & N & Y & 21  & $239.1^{+0.4}_{-0.4}$ & $2.21^{+0.40}_{-0.32}$ \\
         & Ma13      & Box, PM05, PM50 & Y & N & 22  & $239.0^{+0.5}_{-0.5}$ & $2.64^{+0.47}_{-0.51}$ \\
         & Ma13      & Box, PM05, PM50 & Y & Y & 22  & $239.2^{+0.4}_{-0.5}$ & $2.16^{+0.33}_{-0.25}$ \\
\hline
\end{tabular}
\begin{tablenotes}
\item $^{\rm a}$ {The rows in bold highlight the best-fit results we adopted in the following sections and tables.}
\item $^{\rm a}$ {This flag denotes whether the confirmed binary star in NGC~1846 is excluded (N) or included (Y). This star is `N1846-1-r079' in \autoref{tab:sample_n1846}, and the individual velocities are summarized in \autoref{sec:combine_sample} and \autoref{fig:common_stars}.}
\item $^{\rm b}${This flag indicates whether the rotational correction described in \autoref{sec:rot} was applied.}
\end{tablenotes}
\end{threeparttable}
\end{table*}

As a check on the EM algorithm, we have calculated the systemic velocities and projected central cluster dispersions for NGC~419 and NGC~1846 using three methods we refer to as the `Box', `PM05' and `PM50' estimates.

The Box value is simply calculated by using all targets in a rectangular area (the `Box') in the $v_{\rm los}$-$r$ plane (shaded area in \autoref{fig:mem}).  The boundaries of this region are a minimum and maximum systemic velocity ($V_{\rm sys,min}$ and $V_{\rm sys,max}$) and the adopted K66 tidal radius (see \autoref{tab:basic}).   The velocity range was determined by assuming (conservatively) that each cluster has a mass of $10^6\ \rm M_{\sun}$.  Using the K66-model realization in LIMEPY \citep{limepy} and the respective cluster structural parameters (\autoref{tab:basic}), we estimated the central velocity dispersion, $\sigma_{10^6}$.   We then set $V_{\rm sys,max}$ and $V_{\rm sys,min}$ as $V_{\rm sys}-n\sigma_{10^6}$ and $V_{\rm sys}+n\sigma_{10^6}$, respectively, where $V_{\rm sys}$ is the straight mean of velocities in an initial estimate of the box boundaries.  For both clusters, the term $n\sigma_{10^6}$ came out to be 8.8 \kms\ (adopting $n = 1$).  The final Box samples converged quickly using this approach with any reasonable first guess for $V_{\rm sys}$.  

The PM05 and PM50 samples represent the stars with normalized cluster membership probabilities as determined by the EM algorithm (\autoref{sec:EM}) of greater than 5\% (i.e., $P_{M_i}\ge0.05$) and 50\% (i.e., $P_{M_i}\ge0.5$), respectively. What differs here from the EM analysis---where membership probabilities are used to weight individual stars---is that for the PM05 and PM50 samples stars that satisfy these criteria are considered to be certain members and all other are certain non-members which are dropped from the analysis.
In practice, for the PM05 and PM50 samples we iterated \autoref{eq:EM_mean} and \autoref{eq:EM_disp_dyn} until the velocity dispersions converged to better than 0.0025 \kms\ between successive iterations. 

The Box, PM05 and PM50 samples are identified in the lower panels of \autoref{fig:mem}.
We list these results for the Box, PM05 and PM50 samples in \autoref{tab:box}. 
It should be clear that by  regarding all targets in the Box sample as cluster members, the corresponding Box dispersion measurement represents the
maximum $\sigma_{\rm p,0}$ we derive from our sample.   
On the other hand, we calculated the PM50 and PM05 values for avoiding the bias of the EM algorithm, which is caused by the unequal-weight term $P_{M_i}$ and tends to underestimate $\sigma_{\rm p,0}$ for NGC~419 and NGC~1846.
Comparing \autoref{tab:EM} and \autoref{tab:box}, we found that the PM50 values are roughly equal to the results of the EM algorithm, while the PM05 values are slightly larger. 
For all subsequent analysis in this paper, we will work exclusively with the PM50 values to explore the dynamical priperties of the clusters.  For reference, normalized membership probabilities, $P_{M_i}$, are listed in \autoref{tab:sample_n419} and \autoref{tab:sample_n1846} for NGC~419 and NGC~1846, respectively.

In \autoref{fig:vd_profile}, we show the three-bin velocity dispersion profiles comparing with the best-fit K66 model constrained from both PM05 (top panels) and PM50 (bottom panels) values. The velocity dispersion profiles were constructed by dividing all targets with $P_{M_i}\geq0.05$ (top panels) and $P_{M_i}\geq0.5$ (bottom panels) into three bins, respectively. The radius range of each bin is indicated by the horizontal bars of $r$, and the central dots in radius are chosen as the median $r$ value of the stars in each bin. It is worth stressing that we did not fit any of the kinematic data for NGC~419 or NGC~1846 to their respective K66 velocity dispersion profiles;  \autoref{fig:vd_profile} simply indicates that the PM05 and PM50 samples are consistent with the underlying models and, additionally, that the PM50 sample agrees best with those models.  

There is one other criterion---metallicity---that we can in principle apply to assess cluster membership.  \autoref{fig:FeH} shows a plot of the metallicities for all stars in both NGC~419 and NGC~1846 as a function of distance from the centers of the clusters.  The metallicities were determined from the spectral fitting of individual spectra as described in \autoref{sec:spec_fit}.  We stress that the absolute metallicity values may suffer from systematic offsets that depend on the specific stellar model atmospheres and adopted effective temperatures that we are using here (see \autoref{sec:spec_fit}).  However, the {\it relative} metallicities should be reliable as all stars are analyzed in the same manner.   The color/symbol coding is the same as for \autoref{fig:mem} so that one can easily distinguish the Box, PM05 and PM50 samples.   The key features to note are that (a) the metallicity distribution of the field population (located beyond the tidal radius as denoted by the vertical dotted line in \autoref{fig:FeH}) is skewed to low metallicity as expected from a composite disk+halo population in both Magellanic Clouds (e.g., \citealp{Cole05, Carrera11, Song17} for the LMC, and \citealp{Dobbie14} for the SMC), (b) the metallicity distributions of the clusters are much more compact, and plausibly more nearly Gaussian in form, and (c) the S/N of the NGC~419 data is typically a bit lower and hence the measured metallicity distribution of that cluster is clearly broader.  

In both clusters, one star is plausibly many sigma below the mean cluster metallicity; on this basis, these stars appear to be likely non-members of their clusters despite their both being in the PM50 samples based on their kinematics.   Because of this, the removal of either star has negligible effect on the systemic velocity or projected velocity dispersion in either NGC~419 (0.0 and 0.0 \kms, respectively) or NGC~1846 (0.0 and 0.02 \kms).  We will return to a comprehensive discussion of the absolute metallicity estimates from our M2FS spectra when we complete the dynamical analyses of all 20+ clusters in our full sample (see \autoref{sec:cluster_select}).

Comparing the results in \autoref{tab:box} reveals that the PM50 sample returns essentially the same kinematic results as the EM algorithm (\autoref{sec:EM}).   This implies that the EM algorithm is applying weighting that closely mimics what one would do by assigning full membership to stars with membership probabilities $>50\%$.  The PM50 approach also helps to label stars definitively as members or non-members which may be useful for certain types of studies or follow-up observations.   Based on this exercise using our NGC~419 and NGC~1846 datasets, we plan in our future papers on MC cluster M/L ratios to base our key dynamical results on the EM estimates of the kinematic properties of the clusters in our full sample but also to report the PM50 samples in order to identify explicitly the stars we consider to most likely be members of their respective clusters.

\subsubsection{Recovery of the Central Velocity Dispersion }
\label{sec:recover}

\begin{figure*}
   \centering
   \includegraphics[width=0.45\textwidth]{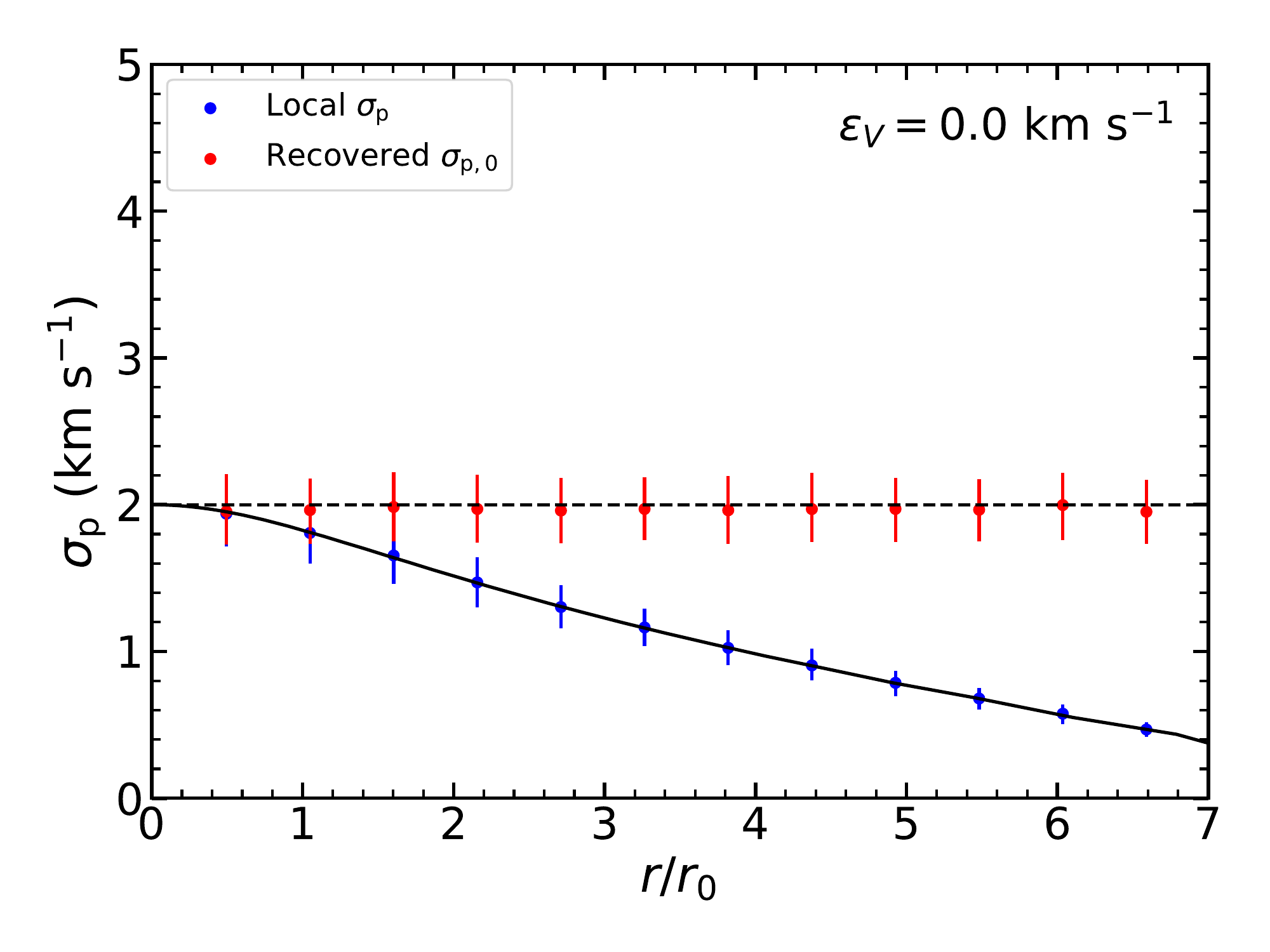}
   \includegraphics[width=0.45\textwidth]{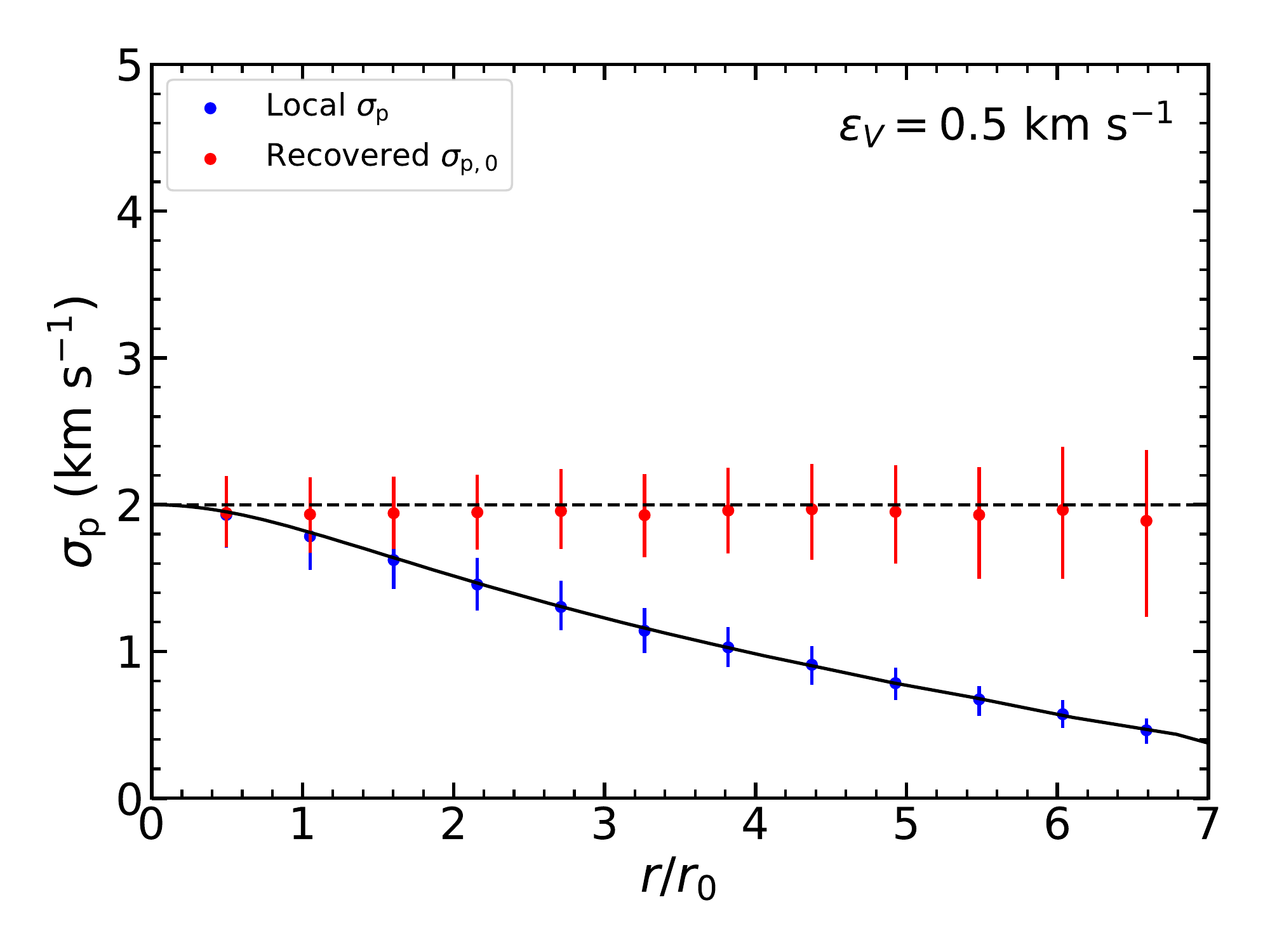}
   \includegraphics[width=0.45\textwidth]{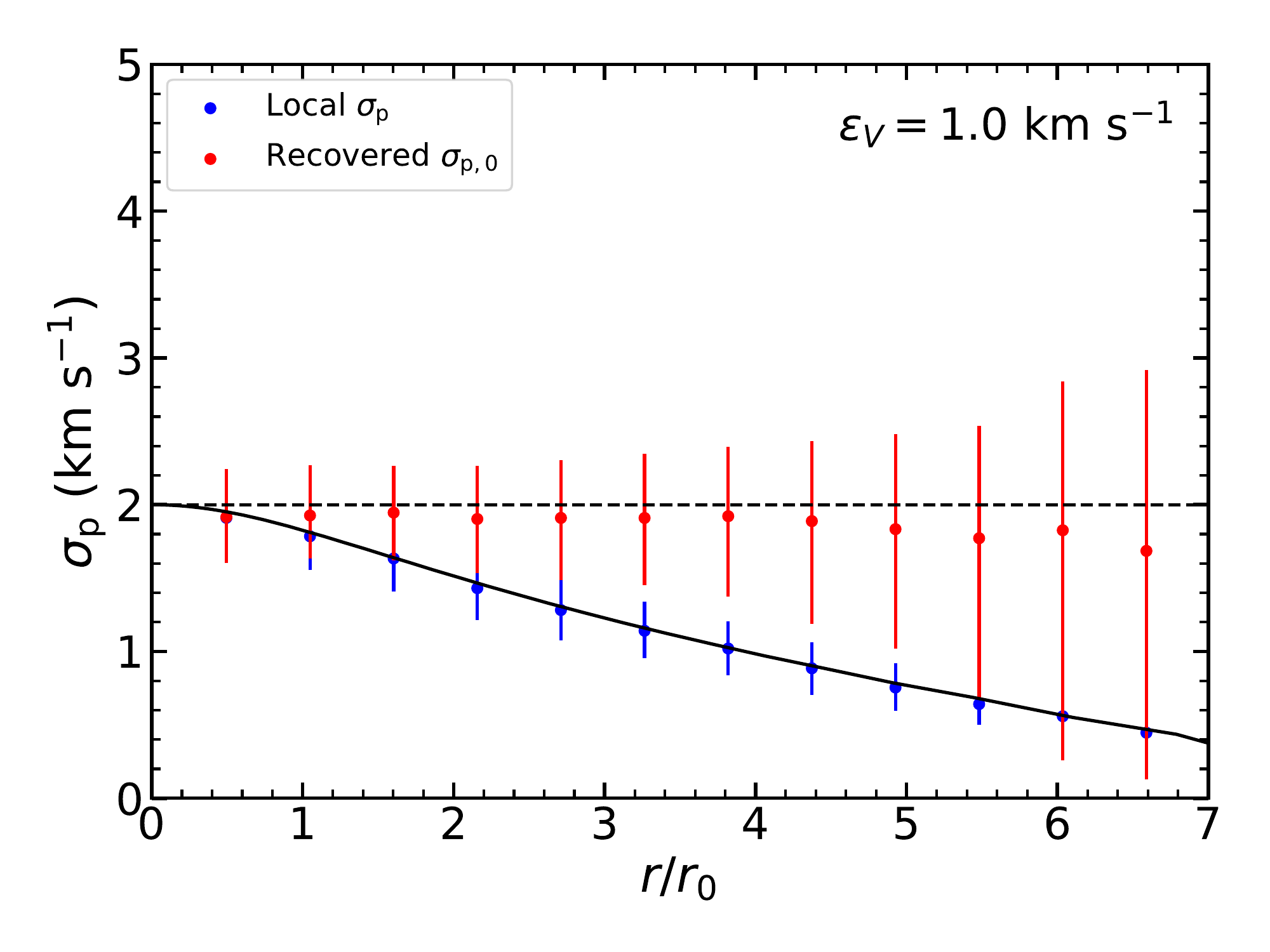}
   \includegraphics[width=0.45\textwidth]{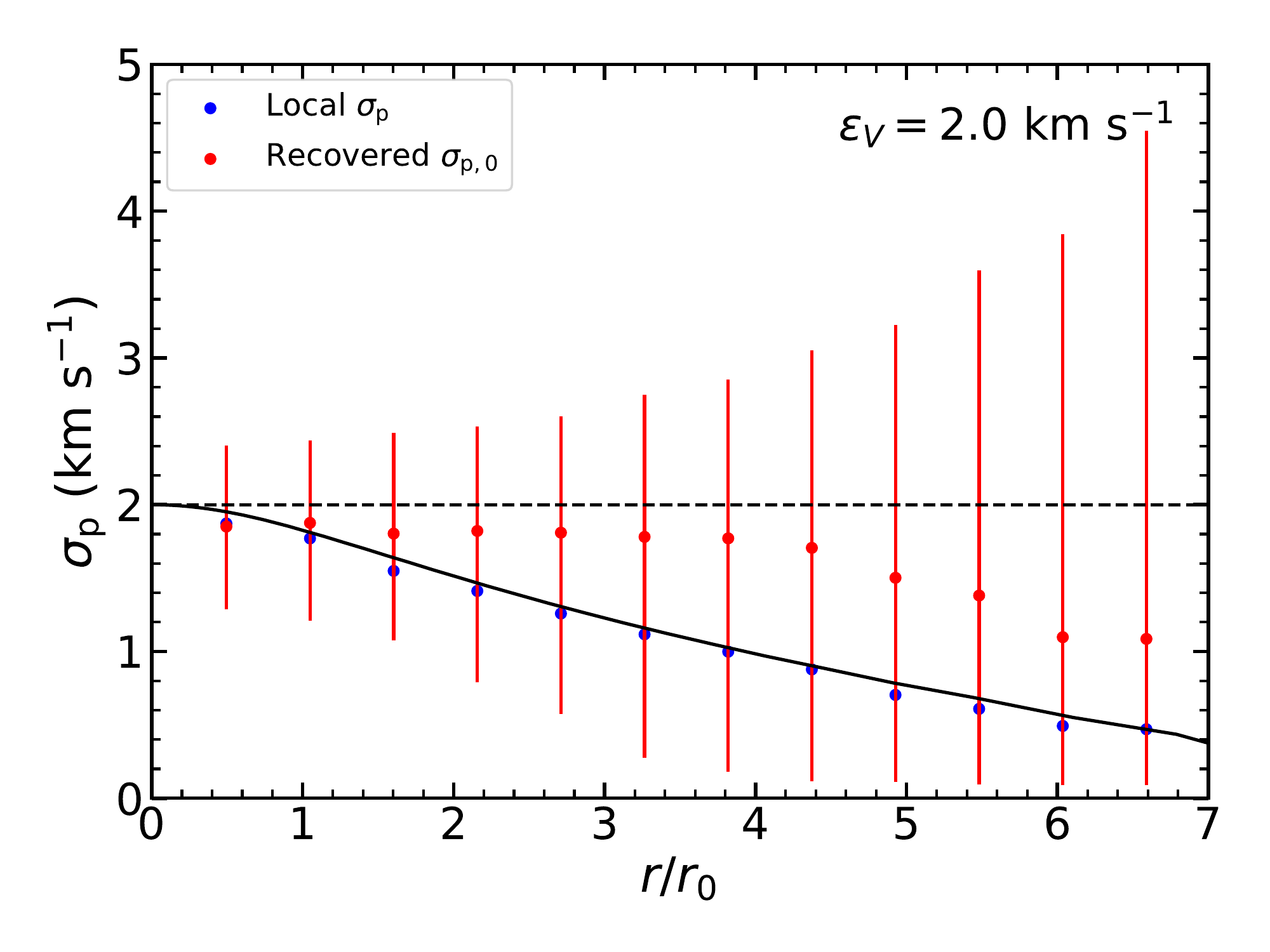}
   \caption{Tests of reliability of the EM algorithm to recover $\sigma_{\rm p,\,0}$ for different radii.  Within a given panel, the velocity uncertainties, $\varepsilon_V$, were taken to be constant, varying from 0.0 to 2.0 \kms\ as noted. The solid lines show the K66 profiles for NGC~1846. The dashed line shows the assumed 2 \kms\  central velocity dispersion. Blue dots correspond to the velocity dispersion measured from 30 members at various radii. Red dots show the recovered $\sigma_{\rm p,\,0}$ for each sample from a single bin at the same radius. The error bars show the 1$\sigma$ uncertainties (that is, the 67\% confidence ranges) of $\sigma_{\rm p,\,0}$ based on bootstrapping 30 members from a given bin with their associated $\varepsilon_V$.  For $\varepsilon_V \leq 0.5$ \kms, the EM algorithm returns the correct central dispersion without bias and to reasonable precision.  By $\varepsilon_V \sim 1$ \kms, the method remains reasonably unbiased, but the implied error on $\sigma_{\rm p,\,0}$ becomes comparable to the inferred dispersion when only 30 tracers are available per bin.}
   \label{fig:test_v0}
\end{figure*}

\begin{figure*}
   \centering
   \includegraphics[width=0.45\textwidth]{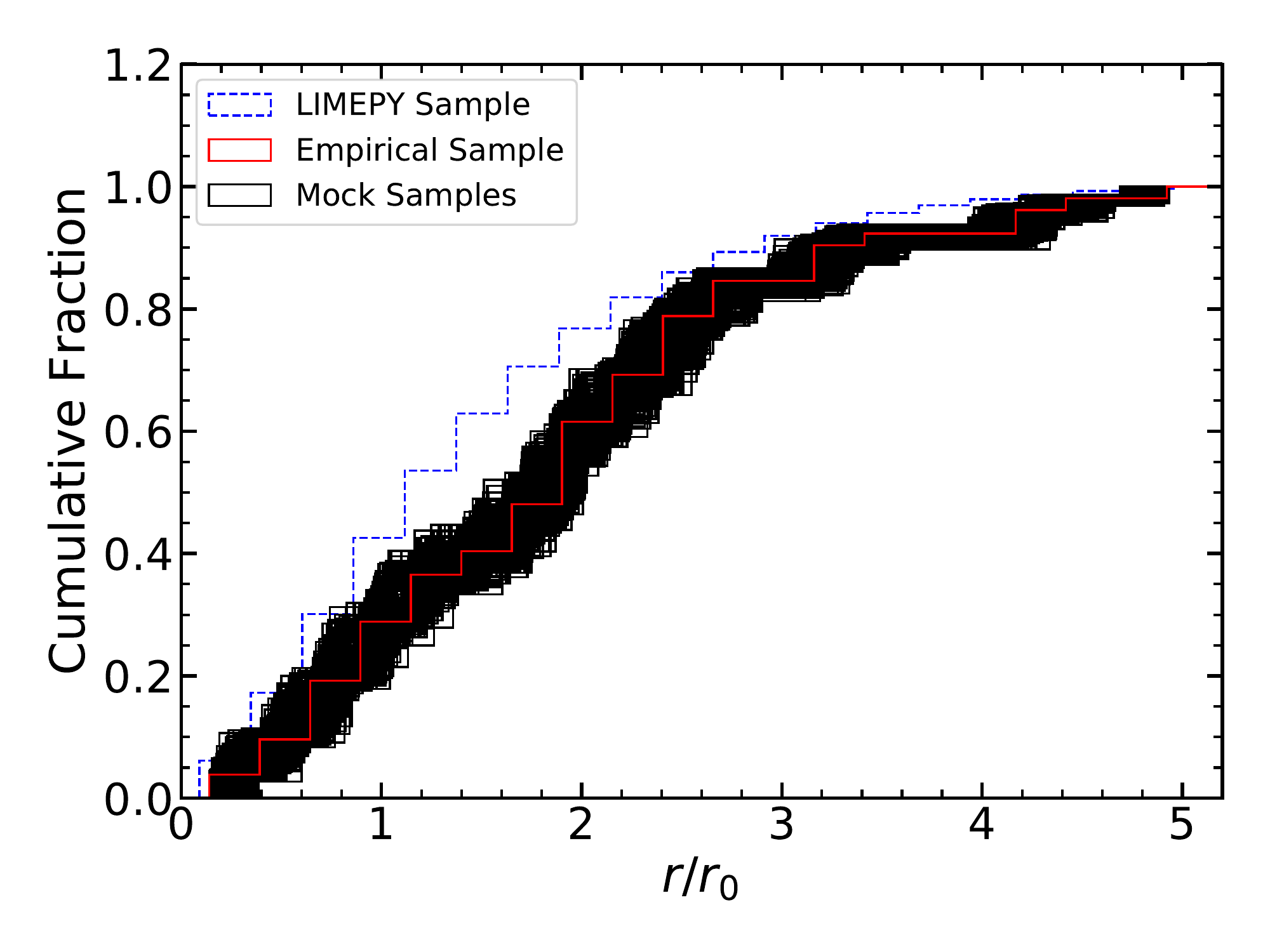}
   \includegraphics[width=0.45\textwidth]{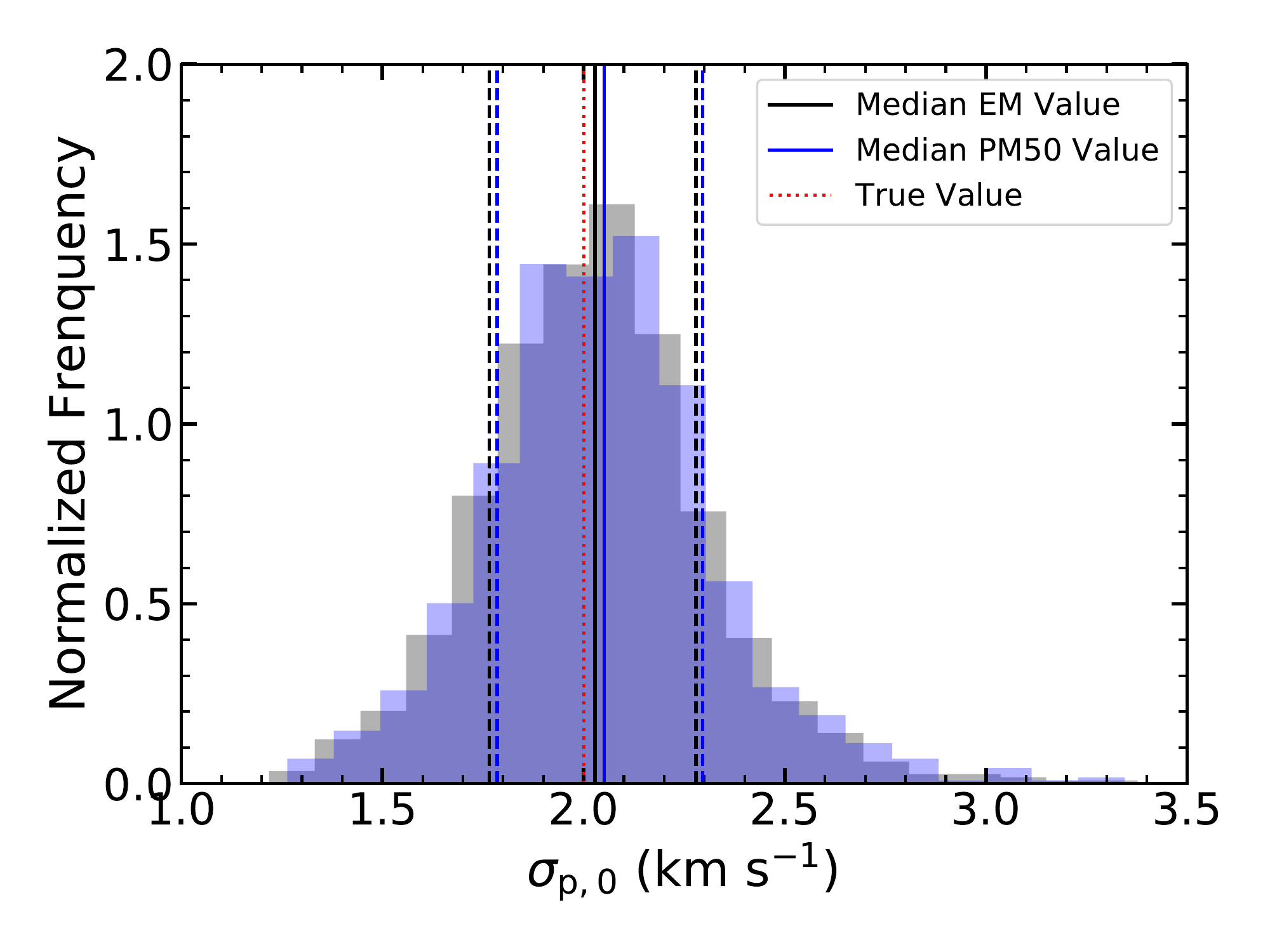}
   \caption{Tests of the EM algorithm for recovering $\sigma_{\rm p,0}$ with 1000 mock samples from the empirical selection profile of the M2FS sample for NGC~1846. In the left panel, the blue line corresponds to the normalized cumulative number of stars as a function of distance from the cluster center.  The red line corresponds to the cumulative empirical selection profile of NGC~1846 based on the targets observed with M2FS. The black lines correspond to the mock samples we simulated in the tests. The right panel shows a histogram of best-fit $\sigma_{\rm p,0}$ values from our mock samples using the EM algorithm (gray; individual targets are given their weights by the algorithm itself, see \autoref{sec:EM}) and the PM50 value (blue; all targets with $P_{\rm M} > 0.5$ are assumed to be certain members, i.e. full weight,  while all others are considered to be certain non-members, i.e. zero weight; see \autoref{sec:box}). The vertical lines in the right panel denote the median values (solid lines) and $1\sigma$ (68\% confidence) limits for the EM (black) and PM50 (blue) cases. The true input value of the simulation (vertical red dotted line) is $\sigma_{\rm p,0}=2$ \kms.  \autoref{sec:recover} lists the results plotted in this figure.}
   \label{fig:test_v1}
\end{figure*}

We now turn to some tests to determine how well we recover the projected central velocity dispersion ($\sigma_{\rm p,0}$) via \autoref{eq:EM_disp_dyn} and the reliability of the EM algorithm introduced in \autoref{sec:EM} and under the assumption of a K66 dynamical model. Our procedure is based on simulations of the observed dataset generated by the dynamical-model sampling routine \texttt{LIMEPY.SAMPLE}. To mimic NGC~1846, we constructed a mock K66-based cluster with the same structural parameters as listed in \autoref{tab:basic}, and adopted $V_{\rm sys} = 239$ \kms\ and $\sigma_{\rm p,0} = 2$ \kms\ which correspond to rounded values for these parameters from \autoref{tab:box}.  We did not carry out simulations specifically for NGC~419 since none of the issues we address below are specifically related either of the clusters in our sample and the statistical properties of NGC~419 and NGC~1846 are similar.

We first tested whether the recovery of $\sigma_{\rm p,0}$ exhibits bias due to the distribution of the  projected distances of tracers from the cluster center. As an extreme ideal case, we set up iterations in which we selected 30 cluster members at a given radius from the center and assumed that their line-of-sight velocities can be perfectly measured (i.e. $\varepsilon_{V}=0$). Both the sample velocity dispersions and their corresponding recovered central values were calculated by \autoref{eq:EM_mean} and \autoref{eq:EM_disp_dyn} iteratively until the value converged to better than 0.0025 \kms\ between successive iterations. In the top left panel of \autoref{fig:test_v0}, we show the median values and the 1$\sigma$ errors estimated from 1000 different samplings at each of several radii.   There is no significant bias along the $r$-axis apart from a slight tendency to underestimate the central dispersion from samples obtained exclusively at radii approaching the tidal radius of the system. 

To explore this further, we ran similar tests but now assuming non-zero velocity uncertainties ($\varepsilon_{V}$). To show this, we selected as before 30 tracers at a given radius for 1000 times, but for each tracer we replaced its true velocity ($V_i$) with a random value from a normal distribution of fixed dispersion ($\varepsilon_{V_i}$).  We carried out these tests for for $\varepsilon_{V_i}$ from 0.5 to 2.0 \kms\ and the results are summarized in \autoref{fig:test_v0}.  As $\varepsilon_{V_i}$ is increased beyond 0.5 \kms, it is evident that estimates based only on stars in the outer parts of the cluster of the central velocity dispersions become extremely unreliable.  For NGC~1846, the median velocity uncertainty is 0.33 \kms for the Combined Sample, close to the case of $\varepsilon_{V_i}=0.5$ \kms\ in the top right panel of \autoref{fig:test_v0}. The sample for NGC~419 exhibits a mean $\varepsilon_{V_i} \leq 1.0$ \kms\ (see \autoref{fig:e_vlos}), so the behavior in that case resembles most closely the results in the bottom left panel of \autoref{fig:test_v0}.   


Of course, in practice we sample stars over a range of radii for both clusters (these correspond to the shaded areas in \autoref{fig:mem}).  The previous test implies that our mean bias due to our EM analysis is $< 0.2$ \kms.  However, we can do better by adopting an `empirical selection profile' (ESP)  that tracks how many members---here defined as stars with with $P_{M_i}\geq0.50$ (see \autoref{sec:box})---were observed at given radius from the Combined Sample of NGC~1846, and then using the ESP when selecting members from the mock K66 cluster (left panel of the \autoref{fig:test_v1}). After 1000 samplings, the median value with their corresponding 1$\sigma$ confidence interval is  $1.97^{+0.19}_{-0.20}$ \kms. Since the model dispersion was taken to be 2.0 \kms, we find no significant bias caused by our sample selection strategy. 

Incorrect or biased identification of field-star contamination can also bias derived kinematic parameters for clusters like NGC~419 and NGC~1846.  To test if this is strongly affecting our analyses, we have generated mock kinematic samples based on the Combined Sample for NGC~1846.  

These mock samples consisted of clusters members and unassociated field stars, with the total number of cluster members in each set to a Poisson random deviate of 52, the actual number of members in our Combined PM50 sample for NGC~1846 (see \autoref{sec:box}).  Kinematics and positions for these members were then drawn at random using LIMEPY assuming (a) the structural parameters from a K66 model adopted for NGC~1846 (see \autoref{tab:basic}), (b) a systemic velocity and central projected velocity dispersion of 239 \kms\ and 2.0 \kms, respectively, for the mock cluster, and (c) the same spatial sampling profile of mock sample members as for the Combined Sample for NGC~1846 (see left panel of \autoref{fig:test_v1}).  All remaining stars---which brought the total in the mock sample to 195---were drawn from a `background' distribution with a systemic velocity and projected velocity dispersion of 269.5 \kms\ and 25.2 \kms, respectively (see \autoref{tab:EM}).  These field stars were distributed uniformly over the field from which the Combined Sample was drawn.  The velocity errors of stars in the mock sample were assigned the uncertainties of stars in the Combined Sample, and an appropriate Gaussian deviate was added to each mock velocity.  

From our analyses of 1000 such samples, we find that both the EM algorithm (see \autoref{sec:EM}) and the PM50 sample (see \autoref{sec:box}) return nearly the systemic velocity and central dispersion for the simulated cluster to good precision, of order 0.3 \kms. The results, summarized in \autoref{fig:test_v1}, from the EM algorithm are $V_{\rm sys}=239.0^{+0.3}_{-0.2}$ \kms\ and $\sigma_{\rm p,0}=2.03^{+0.26}_{-0.25}$ \kms. From the PM50 analysis, the results are $V_{\rm sys}=239.0^{+0.3}_{-0.3}$ \kms\ and $\sigma_{\rm p,0}=2.05^{+0.27}_{-0.25}$ \kms. The errors are the 1-$\sigma$ (68.2\%) confidence ranges of the various parameters determined from the simulations (\autoref{fig:test_v1}).

The simulation results suggest a possible bias such that the derived dispersion is about $0.04\pm0.02$ \kms\ higher than the true cluster dispersion.   Field stars that happen to be close to the cluster will reduce the inferred dispersion only if they have velocities within 1$\sigma$ of the cluster mean.  However, such stars can increase the dispersion over a wider velocity range, about $\pm 3 \sigma$.  Since the field stars have essentially a flat distribution over the velocity range inhabited by cluster members, they will more often---by about a factor of two---increase the dispersion when mistaken as cluster members.  This bias is small compared to the errors inherent in our kinematic results (see \autoref{sec:EM} and \autoref{tab:EM}), so we will ignore this bias in this paper.   However, the more ambiguous cluster/field separation is---say in a low-density cluster or high-density field---the more likely this bias may lead to statistically significant overestimates in velocity dispersions estimates of star clusters.

\section{Cluster Masses and Mass-to-light Ratios}
\label{sec:mass}
\subsection{Methodology}
\label{sec:MLv}

\begin{table*}
\caption{Mass, Luminosity and $M/L_V$.}
\label{tab:MLv}
\begin{threeparttable}
\begin{tabular}{cccccccc}
\hline
    {Cluster} & 
    {Dataset} & 
    {${M_{\rm tot}}$} & 
    {${L_{V,\,\rm tot}}$} & 
    {$M/L_{\rm V}$} & 
    {$\log{M_{\rm tot}}$} & 
    {$\log{L_{V,\,\rm tot}}$} & 
    {$\log{M/L_{\rm V}}$} 
    \\
    {} & 
    {} & 
    {$(\times 10^5\ {\rm M}_{\sun})$} & 
    {$(\times 10^5\ {\rm L}_{\sun})$} & 
    {$({\rm M}_{\sun}\ {\rm L}_{\sun}^{-1})$} & 
    {$({\rm M}_{\sun})$} & 
    {$({\rm L}_{\sun})$} & 
    {$({\rm M}_{\sun}\ {\rm L}_{\sun}^{-1})$} \\
\hline
    NGC 419  &  M2FS & $0.76^{+0.25}_{-0.13}$ & $3.46^{+0.71}_{-0.71}$ & $0.22^{+0.08}_{-0.05}$ & $4.88^{+0.14}_{-0.08}$ & $5.54^{+0.09}_{-0.09}$ & $-0.66^{+0.16}_{-0.10}$\\
    NGC 1846 & M2FS  & $0.42^{+0.11}_{-0.12}$ & $1.67^{+0.47}_{-0.47}$ & $0.25^{+0.08}_{-0.09}$ & $4.62^{+0.11}_{-0.12}$ & $5.22^{+0.12}_{-0.12}$ & $-0.60^{+0.14}_{-0.15}$ \\
             &  Combined$^{\rm a}$ & $0.54^{+0.15}_{-0.14}$ & $1.67^{+0.47}_{-0.47}$ & $0.32^{+0.11}_{-0.11}$ & $4.73^{+0.12}_{-0.12}$ & $5.22^{+0.12}_{-0.12}$ & $-0.49^{+0.14}_{-0.14}$ \\
\hline
\end{tabular}
\begin{tablenotes}
 \item $^{\rm a}$ {Exclude the confirmed binary star in NGC~1846. This star is `N1846-1-r079' in \autoref{tab:sample_n1846}, and the individual velocities are summarized in \autoref{sec:combine_sample} and \autoref{fig:common_stars}.}
\end{tablenotes}
\end{threeparttable}
\end{table*}

Our initial estimates for the masses of NGC~419 and NGC~1846 are based on the K66 model using the structural parameters listed in \autoref{tab:basic} and scaled by the PM50 $\sigma_{\rm p,0}$ value listed in \autoref{tab:box}. 
The mass uncertainty is estimated following \citet{Illingworth76b} by referring to the K66 total-mass estimator, $M_{\rm tot}=167 r_0 \mu \sigma_{{\rm p},0}^2$ (with $r_0$ in pc, $\sigma_{{\rm p},0}$ in \kms\ and M in $M_{\sun}$); the final mass uncertainty is derived from the known errors in the squared velocity dispersion and distance modulus (which propagates to the uncertainty in the scale radius, $r_0$). 

The $V$-band luminosities of the clusters are obtained by integrated the K62 profile scaled to aperture photometry profiles to a maximal reference radius, $x$. The resulting relation is 
\begin{equation}
L_{V}(x)=\pi r_0^2 \Sigma_{V,0}\left[ \ln (1+x)-4{(1+x)^{1/2}-1\over (1+x_t)^{1/2}}+{x\over 1+x_t}\right],
  \label{eq:lum}
\end{equation}
where $x=(r/r_c)^2$, $x_t=(r_t/r_c)^2$ and $\Sigma_{V,0}$ is the central surface brightness in $V$-band.
To estimate $\Sigma_{V,0}$, we compared the ground-base aperture photometry $V_{\rm ap}$ listed in \autoref{tab:basic} with the results of \autoref{eq:lum} if $x=r_{\rm ap}/r_0$. The total luminosity $L_{V,\,\rm tot}$ can then be obtained by setting $x=x_t$ in \autoref{eq:lum}. When transforming magnitudes to luminosities, we used $M_{V,{\sun}}=4.85$ in addition to the distance moduli and extinction values listed in \autoref{tab:basic}. According to \autoref{eq:lum}, the luminosity uncertainty comes from the errors in the central surface brightness, the squared scaled radius in arc and the squared distance.

The empirical $M/L_V$ of a cluster can be derived by comparing the total mass to the total $V$-band luminosity determined above. The uncertainty in $M/L_V$ is estimated from the errors in the squared velocity dispersion, the central surface brightness, the scaled radius in arc and the distance. \autoref{tab:MLv} lists the masses, luminosities and $M/L$ ratios.

\subsection{Rotation}
\label{sec:rot}

\begin{figure*}
   \centering
  \includegraphics[width=0.50\textwidth]{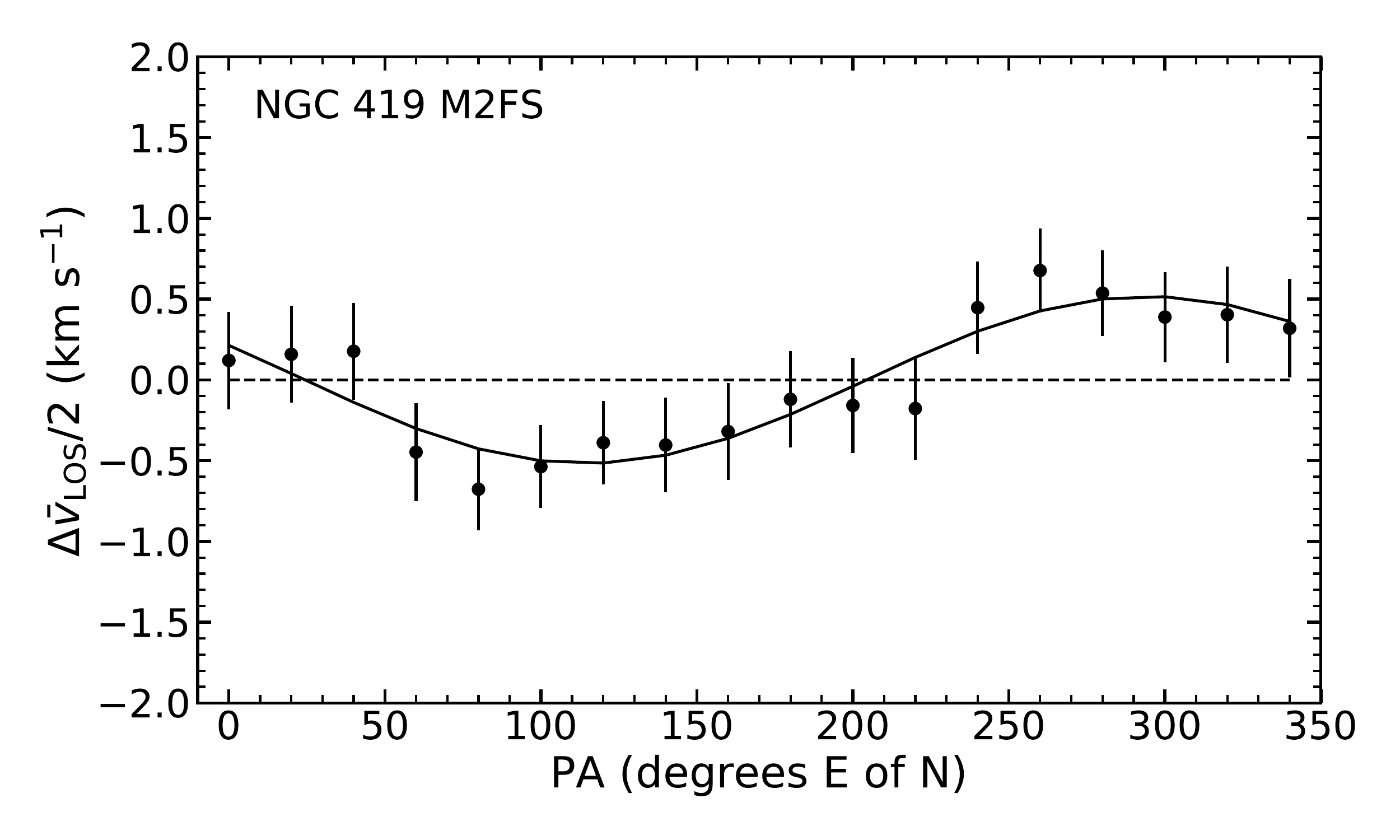}
  \includegraphics[width=0.40\textwidth]{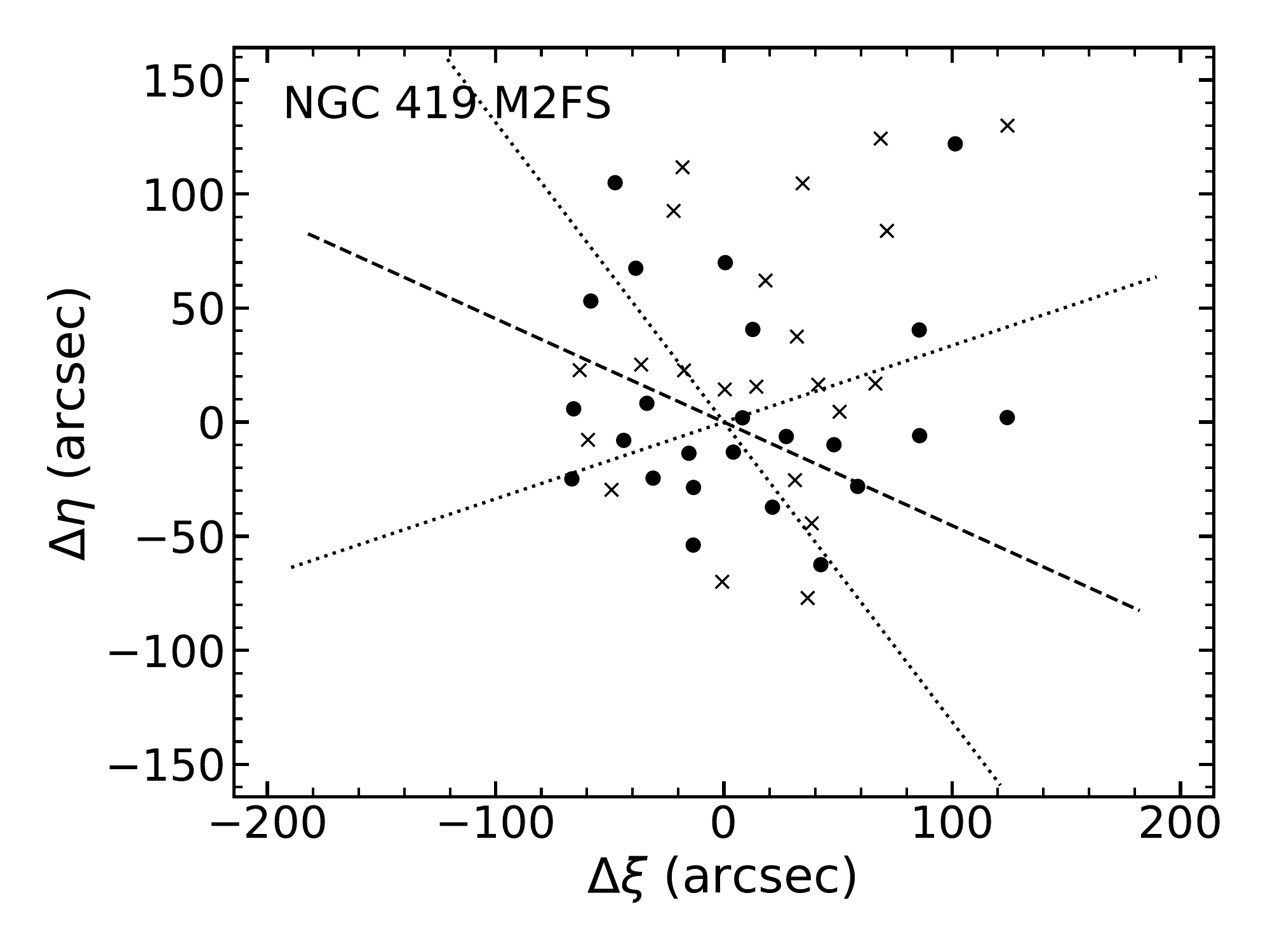}        
  \includegraphics[width=0.50\textwidth]{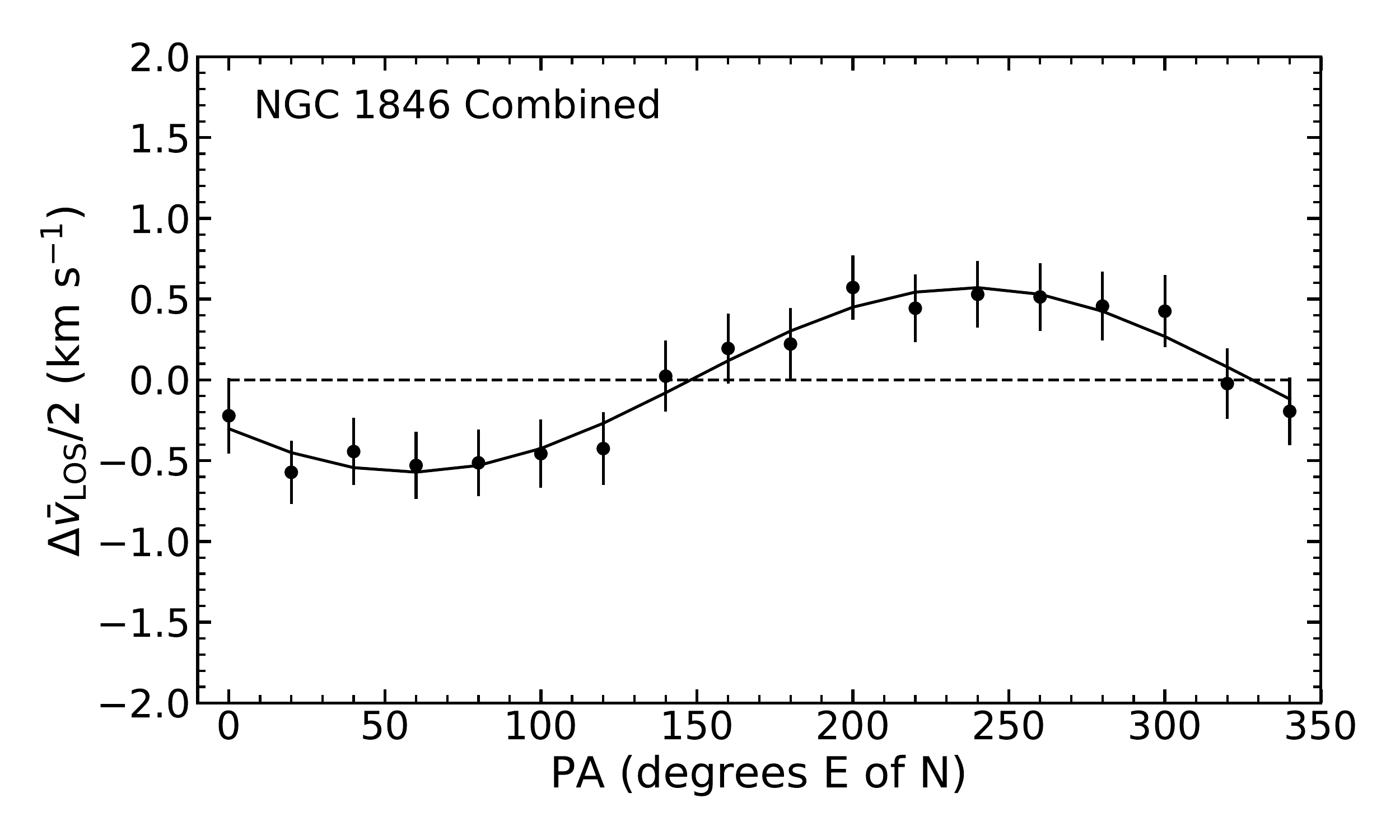}
  \includegraphics[width=0.40\textwidth]{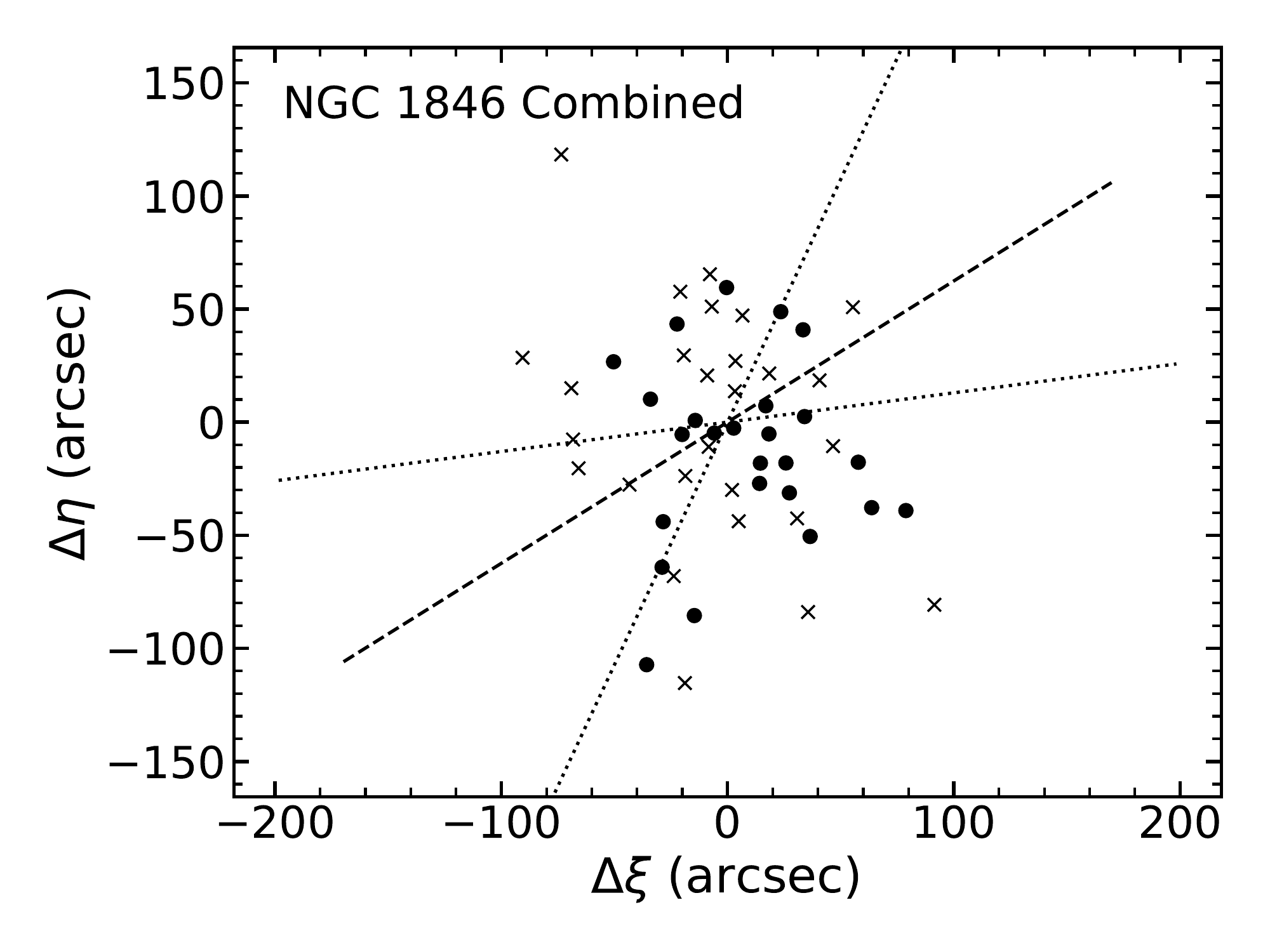}
   \caption{Simple rotation analysis for the stars with $P_{\rm M}>0.5$ in NGC~419 (top) and NGC~1846 (bottom) (see \autoref{sec:rot} and \autoref{tab:rot}). The adopted cluster centers are listed in \autoref{tab:obs}. The left panels show $\Delta \bar{V}_{\rm los}/2$ as a function of the bisector position angle, together with the best-fit sinusoid model. The best-fit parameters are listed in \autoref{tab:rot}. In the right panels, crosses (dots) indicate stars with velocities greater (less) than the systemic velocity. The best-fit rotation axis from the left panel is marked as a dash line in each panel, and two dotted lines denote the 1$\sigma$ uncertainties.}
   \label{fig:rot}
\end{figure*}

\begin{table}
\caption{Rotation Analysis.}
\label{tab:rot}
\begin{threeparttable}
\begin{tabular}{ccccc}
\hline
    {Cluster} & 
    {Dataset} &  
    {$N_{\rm mem}$} &  
    {$A_{\rm rot}$} &  
    {$\phi$}   
\\
    {} & 
    {} &  
    {} &  
    {$\rm (km\ s^{-1})$} &  
    {$(^\circ)$}   \\
\hline
    NGC 419 & M2FS &  46 & $-0.5^{+0.2}_{-0.3}$ & $-24^{+34}_{-43}$ \\
    NGC 1846 & Combined$^{\rm a}$ & 52 & $-0.6^{+0.3}_{-0.2}$ & $32^{+25}_{-37}$ \\
         & Combined & 53 & $-0.5^{+0.2}_{-0.2}$ & $31^{+31}_{-37}$ \\
         & M2FS & 41 & $-0.4^{+0.3}_{-0.2}$ & $23^{+35}_{-38}$ \\    	 
         & Ma13$^{\rm a}$  & 21 & $-1.0^{+0.5}_{-0.3}$ & $37^{+18}_{-46}$ \\
    	 & Ma13 & 22 & $-1.2^{+0.5}_{-0.4}$ & $38^{+8}_{-39}$ \\
\hline
\end{tabular}
\begin{tablenotes}
\item $^{\rm a}$ {Exclude the confirmed binary star in NGC~1846. This star is `N1846-1-r079' in \autoref{tab:sample_n1846}, and the individual velocities are summarized in \autoref{sec:combine_sample} and \autoref{fig:common_stars}.}
\end{tablenotes}
\end{threeparttable}
\end{table}

Up to now, our analysis has assumed both NGC~419 and NGC~1846 are exclusively pressure-supported systems.  However, in the case of NGC~1846, \citetalias{Mackey13} suggested that the cluster may exhibit some coherent rotation.  Since rotation can partly dynamically support the clusters, this effect could alter our estimates of their masses.  In this subsection we explore the evidence for rotation in both clusters and comment on the magnitude of the effects of rotation on our final mass estimates for both systems.

We examined the observations for evidence of internal rotation in each cluster by comparing the mean velocity differences on opposite sides of a line passing through the projected cluster centers as the line's orientation is advanced in position angle (PA, east of north; see, e.g., \citealt{Cote95, Lane09, Lane10a, Lane10b, Bellazzini12, Mackey13}). For this analysis we used the PM50 sample for NGC~419 and the PM50 subsample derived from the Combined Sample for NGC~1846 (see \autoref{sec:box} and \autoref{tab:box}).   If any internal rotation exists that has a significant line-of-sight component, the mean velocity differences between the mean velocities on the two sides of the fiducial line ($\Delta \langle {V_{\rm los}}\rangle$) should exhibit a sinusoidal pattern (see also, e.g. Eq. 8 from \citealt{Kimmig15}) with a statistically significant amplitude.  We use the following relation to parameterize this behavior:
\begin{equation}
{\Delta \langle {V_{\rm los}}\rangle \over 2}=A_{\rm rot}\sin{({\rm PA}+\phi)},
  \label{eq:rotation_sine}
\end{equation}
where $A_{\rm rot}$ is the maximum amplitude of this relation and $\phi$ is related to the rotation axis ${\rm PA}_0$. 
Due to projection, the measured $A_{\rm rot}$ is clearly a lower limit to the true rotation amplitude measured using this parameterization. 

The uncertainties in the fitting parameters for the results for NGC~419 and NGC~1846 were estimated in the same manner as described in \citetalias{Mackey13} (see their Section 4.1) and using the same projected cylindrical rotation curve used in that paper.  In summary, we Monte-Carlo new velocities to each star at their known positions given their velocity measurement errors and the observed cluster dispersion profile consistent with the adopted K66 model for each cluster (\autoref{tab:box}), then re-determined the parameters in \autoref{eq:rotation_sine} 1000 times (see \autoref{tab:rot}).   From this procedure, we estimate rotation amplitudes of $0.5^{+0.3}_{-0.2}$ and $0.6^{+0.2}_{-0.3}$ \kms\ for NGC~419 and NGC~1846, respectively, with rotation axis PAs of $114^{+43}_{-34}$ and $58^{+37}_{-25}$ degrees east of north.  We note here that \citetalias{Mackey13} used a very similar approach to estimate the rotation position angle of NGC~1846 to be $60^\circ \pm 20^\circ$, in good agreement with our result based on the PM50 Combined Sample (see \autoref{tab:rot}).

Based on the results from this section, we can estimate the ratios of the rotational amplitude and observed central dispersion for NGC~419 and NGC~1846 to be $A_{\rm rot}/\sigma_{\rm p,0}\approx0.2^{+0.1}_{-0.1}$ and $0.3^{+0.1}_{-0.2}$, respectively.  Both values are marginally significantly different from zero.  However, if we take these ratios at face value, we can estimate the systematic effects on the masses we derive for both clusters as follows.   First, we have removed the rotational component of the LOS velocities of every star using the projected rotational velocities our cylindrical rotation model predicts.  We then applied our EM estimator to the adjusted samples to obtain a new estimate of the projected central velocity dispersion and estimated the masses of the clusters using the techniques in \autoref{sec:MLv}; these masses tend to be $\sim$9\% lower than the masses based on the uncorrected central velocity dispersions that are listed in \autoref{tab:box}. However, these mass estimates---by design---neglect the mass being supported by the rotational component of the cluster.  Without a better rotational model--not to mention one that is more statistically significant--it is difficult to make a more precise rotational correction (see e.g. \citealt{Fischer92a}).

Another complication has to do with the unknown inclination of any rotation with the plane of the sky.  However, even if the rotations of both clusters are fully in the plane of the sky (strictly not possible to the extent that we see a rotation signal), then the observed central dispersion would be about $\sqrt{2}$ smaller than in the no-rotation case, implying an underestimate of order 30\% in the true masses of the clusters.    We conclude that rotation likely has a negligible impact on the masses we derive for NGC~419 and NGC~1846, and is unlikely to affect our results at a level significantly higher than implied by the error bars on the derived masses (\autoref{tab:MLv}) that are based on the measurement and statistical uncertainties in the kinematic and structural properties of the clusters.  

\section{Comparison with previous studies}
\label{sec:comp}

\begin{table*}
\caption{Comparison of kinematic results with previous studies.}
\label{tab:comp}
\begin{threeparttable}
\begin{tabular}{ccccccccc}
\hline
    {Cluster} & 
    {Dataset} & 
    {RC$^{\rm a}$} & 
    {${V_{\rm rot}}$} & 
    {${\rm PA_{\rm 0}}$} & 
    {${V_{\rm sys}}$} & 
    {${\sigma_{\rm p,\,0}}$} & 
    {${M_{\rm tot}}$} & 
    {$M/L_{\rm V}$}  
    \\
    {} & 
    {} & 
    {} & 
    {$(\rm km\,s^{-1})$} & 
    {$(^{\circ})$} & 
    {$(\rm km\,s^{-1})$} & 
    {$(\rm km\,s^{-1})$} & 
    {$(\times 10^5\ {\rm M}_{\sun})$} & 
    {$({\rm M}_{\sun}\ {\rm L}_{\sun}^{-1})$}  \\
\hline
    NGC 419  & M2FS                      & N & $0.5^{+0.3}_{-0.2}$ & $114^{+43}_{-34}$ & $189.5^{+0.3}_{-0.3}$ & $2.44^{+0.37}_{-0.21}$ & $0.76^{+0.25}_{-0.13}$ & $0.22^{+0.08}_{-0.05}$ \\
             & MUSE$^{\rm b}$     & N & $0.7\pm0.2$ & $13\pm17$ & $190.5\pm0.2$ & $3.3\pm0.2$ & $1.4\pm0.2$ & $0.67\pm0.08$ \\   
    NGC 1846 & Combined$^{\rm c}$ & N & $0.6^{+0.2}_{-0.3}$ & $58^{+37}_{-25}$ & $239.3^{+0.2}_{-0.2}$ & $2.04^{+0.28}_{-0.24}$ & $0.54^{+0.15}_{-0.14}$  & $0.32^{+0.11}_{-0.11}$  \\
             & M2FS                      & N & $0.4^{+0.2}_{-0.3}$ & $67^{+38}_{-35}$ & $239.4^{+0.2}_{-0.2}$ & $1.80^{+0.23}_{-0.24}$ & $0.42^{+0.11}_{-0.12}$ & $0.25^{+0.08}_{-0.09}$ \\
             & Ma13$^{\rm d}$      & N & $1.2^{+0.4}_{-0.5}$ & $52^{+39}_{-8}$ & $239.0^{+0.5}_{-0.5}$ & $2.64^{+0.47}_{-0.50}$ & $0.93^{+0.35}_{-0.37}$ & $0.65^{+0.25}_{-0.26}$\\
             & Ma13$^{\rm d}$      & Y & $1.2^{+0.4}_{-0.5}$ & $52^{+39}_{-8}$ & $239.2^{+0.4}_{-0.5}$ & $2.16^{+0.33}_{-0.25}$ & $0.63^{+0.20}_{-0.16}$ & $0.44^{+0.15}_{-0.12}$\\
             & Ma13$^{\rm e}$     & Y & $1.1\pm0.4$ & $60\pm20$ & $239.1\pm0.4$ & $2.52^{+0.26}_{-0.18}$ & $0.84^{+0.17}_{-0.12}$ & $0.59^{+0.13}_{-0.10}$\\
\hline
\end{tabular}
\begin{tablenotes}
 \item $^{\rm a}$ {Rotation Correction.}
 \item $^{\rm b}$ {The results of \citetalias{Kamann18}. $\sigma_{\rm p,\,0}$ was estimated using the K66 model for NGC~419 listed in \autoref{tab:basic} and the measured $M_{\rm tot}$ by \citetalias{Kamann18}.}
 \item $^{\rm c}$ {Exclude the confirmed binary star in NGC~1846. This star is `N1846-1-r079' in \autoref{tab:sample_n1846}, and the individual velocities are summarized in \autoref{sec:combine_sample} and \autoref{fig:common_stars}.}
 \item $^{\rm d}$ {The results of our analysis using the full dataset of \citetalias{Mackey13}.}
 \item $^{\rm e}$ {The results of \citetalias{Mackey13}.}
\end{tablenotes}
\end{threeparttable}
\end{table*}

As noted in \autoref{sec:prev_samples}, there are previously published kinematic studies of NGC~419 and NGC~1846. For NGC~419, \citetalias{Kamann18} used adaptive optics assisted integral-field spectroscopy using MUSE at the VLT.  For NGC~1846, \citetalias{Mackey13} obtained individual spectra using VLT/FLAMES.  In this section we critically compare the results of the present paper with the findings of these earlier studies.

\subsection{NGC~419}
\label{sec:comp_N419}

\begin{figure}
   \centering
   \includegraphics[width=0.45\textwidth]{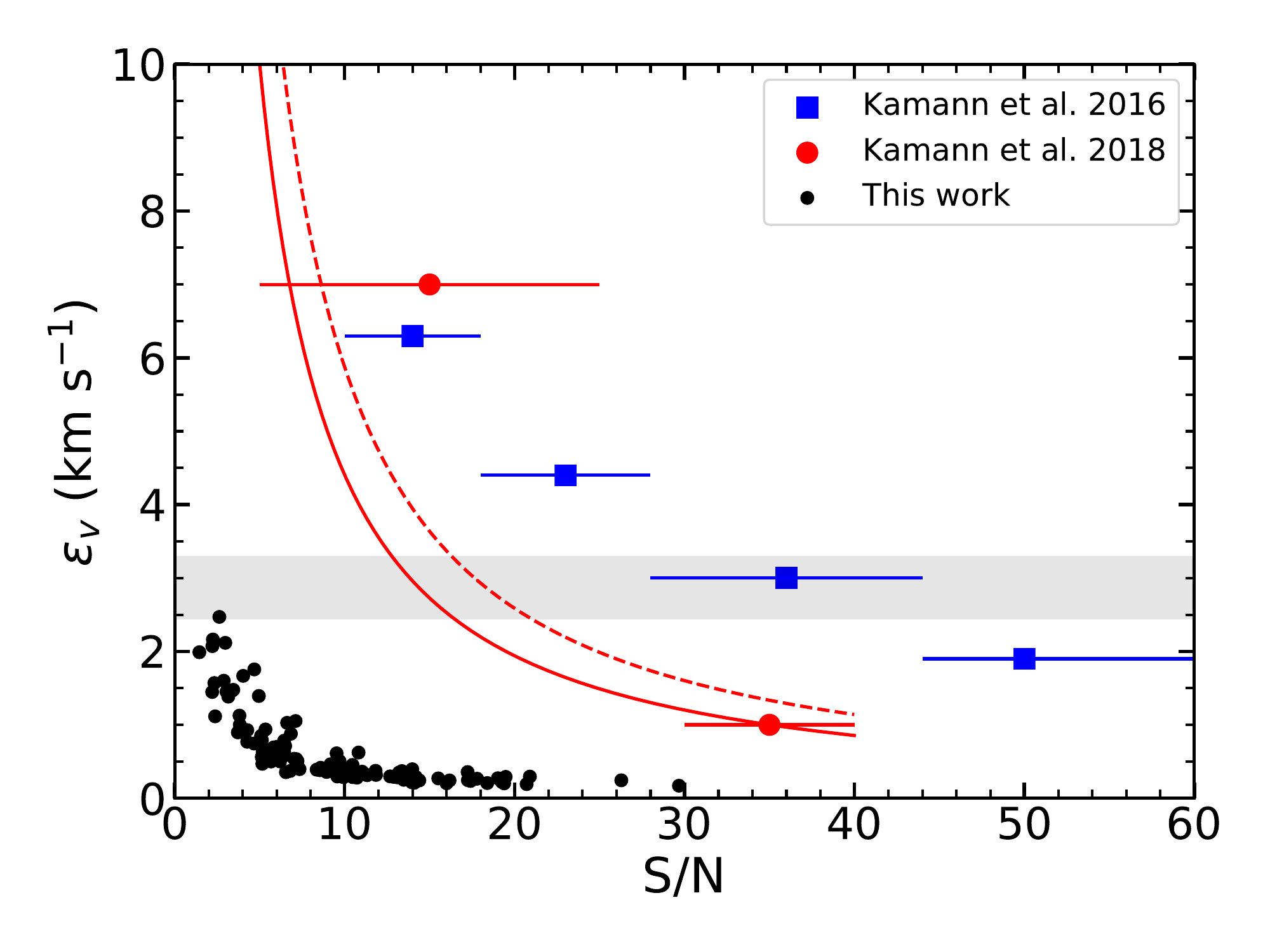} 
   \caption{Velocity uncertainties for kinematic measurements of NGC~419. The black dots are taken from our sample in \autoref{fig:e_vlos}. The red dots show our estimate of $\varepsilon_V$-S/N relation for NGC~419 at fiducial S/N values based on the discussion in \citetalias{Kamann18} (see their Figure 3 and Section 4). The blue squares show the error distribution for stars in NGC 6397 also observed with MUSE but without AO \citep[][and based on their Figure 3]{Kamann16}. The gray shaded horizontal bar spans the range of the central velocity dispersion measurements for NGC~419. The lower bound of this region is $2.44$ \kms\ (the PM50 Combined sample) and the upper bound is $3.3$ \kms\ \citepalias{Kamann18}. The solid red line denotes the parametric form of the error distribution for NGC~419 that we adopted for the \citetalias{Kamann18} observations (see \autoref{sec:comp_N419}). The dashed red line denotes the same error distribution but multiplied by a factor of $1/0.75 = 1.33$ (see \autoref{sec:comp_N419}). Note that both red curves lie well below the error distribution observed in the NGC~6397 study \citep{Kamann16}.}
   \label{fig:e_vlos_comp}
\end{figure}

Using MUSE, \citetalias{Kamann18} obtained radial velocity measurements of 1049 individual sources within the central $1 \times 1$ {\it arcmin}$^{2}$ field of NGC~419 using the AO system to improve spatial resolution in the core region of the cluster.  The spectral resolution of $\mathcal{R} \sim 2800$ yielded radial velocity uncertainties $\lesssim 10\ \rm km\,s^{-1}$ (more on this below).  They measured or constrained many of the same dynamical parameters that we have obtained for the cluster.    A direct comparison of their results and the results from our PM50 sample is provided in \autoref{tab:box}.  In the case of the central velocity dispersion of NGC~419, the value in \autoref{tab:comp} was obtained by adopting their preferred dynamical model (see below) and extrapolating the corresponding dispersion profile to the cluster center (see their Figure 4).
The $M/L$ ratio for NGC~419 was determined by \citetalias{Kamann18} using spherical isotropic Jeans models with different (constant) $M/L$ ratios, and then a maximum likelihood approach was used to sum up the likelihoods for observed radial velocities, given their measurement uncertainties, and the predicted radial velocities for each model at the corresponding positions of each star.  This process yields directly the $V$-band $M/L$ of the cluster from which \citetalias{Kamann18} then estimated a total cluster mass.  These values are also listed in \autoref{tab:comp} for ease of comparison with our results.

The systemic radial velocities measured by \citetalias{Kamann18} and our study differ by $1.0\pm 0.4$ \kms\ (see \autoref{tab:comp}).   Given the possible zero-point errors we have identified in our M2FS data (see \autoref{tab:comp}), these values are in reasonable agreement\footnote{There could of course be a zero-point shift in the MUSE results, but there is no assessment of this in \citetalias{Kamann18}.}  Of course, the systemic velocity is ultimately immaterial to any of the conclusions regarding the mass or $M/L$ ratio for NGC~419 from either paper.  Unlike the case for NGC~1846 (see \autoref{sec:comp_N1846}), we could not carry out a star-by-star velocity comparison since \citetalias{Kamann18} did not provide the velocities of individual sources extracted from their IFU observations.

A comparison of the $M/L_{V}$ values obtained by \citetalias{Kamann18} and this paper differ by a significant factor: $M/L_{V,\,\rm K18}/M/L_{V,\,\rm M2FS} = 3.0^{+1.1}_{-0.8}$.   This comparison is complicated by the different paths by which the respective M/L values were obtained.  If we compare instead the masses derived by \citetalias{Kamann18} and ourselves, we find a ratio of $M_{\rm K18}/M_{\rm M2FS} = 1.84^{+0.66}_{-0.41}$, a 2-$\sigma$ discrepancy that ultimately arises from the different central velocity dispersions measured by the two studies.

One reason for the differences between the \citetalias{Kamann18} results and our new M2FS results could reflect the very different distributions of the tracers sampled in the cluster by the respective studies.  As already noted, the \citetalias{Kamann18} sample consists exclusively of stars within the MUSE field of view, which corresponds almost precisely within the region inside the core radius of NGC~419.  Our M2FS sample consists of stars spread throughout the cluster from the central core to (and beyond) the tidal radius.   If the $M/L$ ratio of NGC~419 varies radially, modeling our sample with an assumed well-mixed population (therefore with constant $M/L$ over all radii) could result in a different value from observations and modeling restricted to the core.  \citetalias{Kamann18} noted that the observed velocity dispersion profile was slightly steeper in the cluster core than the Jeans models they adopted.   To the extent that this reflects radial variations in $M/L$ in NGC~419, it would imply that neither of the models used in either paper is strictly correct and may result in biased mass and $M/L_V$ ratio estimates.  Both studies have to extrapolate to a central dispersion value and the methods by which that was done differ in detail and used different dynamical models.


Another possibility is that the discrepancy in the MUSE and M2FS dynamical results (\autoref{tab:comp}) reflects some sort of observational issues.   Apart from sample size, the key difference in the \citetalias{Kamann18} MUSE and our M2FS results lies in the relative velocity uncertainties.  In one case, \citetalias{Kamann18}/MUSE, the kinematic errors appear to be comparable and often larger than the intrinsic cluster dispersion (this remains true regardless of which value for the central dispersion is adopted).  We can see this in \autoref{fig:e_vlos_comp} where we have plotted, as a function of mean spectral S/N, our estimate of the MUSE errors based on \citetalias{Kamann18}'s description (red dots) and \citet{Kamann16} (blue squares). The same figure shows our measured M2FS velocity uncertainties (black dots) based on repeat measurements as described in \autoref{sec:e_vlos}.   The gray shaded horizontal bar in Figure 15 denotes the range of central dispersion values for NGC~419 based on the \citetalias{Kamann18} and M2FS results and shows graphically how each dataset's precision relates to the likely intrinsic cluster dispersion.

We have carried out simulations to determine how the different error distributions may affect inferred central velocity dispersion estimates for NGC~419.   In the case of the MUSE data, we adopted a specific $\varepsilon_V$-S/N relation shown as a sequence of solid red line in \autoref{fig:e_vlos_comp}.  We then used LIMEPY.SAMPLE to produce a dataset of 1000 targets (roughly comparable to the MUSE sample) from a K66 model with a central dispersion of 2.44 \kms\ (the M2FS value; see \autoref{tab:comp}) and then applied Gaussian deviates to each star according to its expected error as given by the relation in \autoref{fig:e_vlos_comp}.  The distribution of S/N values adopted for individual targets was chosen based on a luminosity function (translated to S/N) that matches the slope of the LF along the cluster's RGB \citep[e.g.,][]{Paust07, Feuillet14}.  

If the errors are precisely known, the EM algorithm does an excellent job of returning the correct central dispersion:  Out of 1000 samples, the mean dispersion was found to be $2.44\pm0.2$ \kms, in essentially perfect agreement with the input value.  Tellingly, this simulation found zero cases out of the 1000 trials where the dispersion was as high as 3.3 \kms, the inferred central dispersion according to \citetalias{Kamann18}.

A problem arises, however, if we assume the velocity uncertainties are not precisely known.  There are clearly reasons to expect that they may not be.  In our M2FS data, we found that the velocity uncertainties returned by the Bayesian spectra fitting underestimate the true errors by about 23\% (see \autoref{sec:e_vlos}).  In the case of the MUSE, the implied errors for the NGC~419 data (red dots in \autoref{fig:e_vlos_comp}) are considerably lower, at a given S/N value, to the results presented by \citet{Kamann16} for NGC~6397. The use of AO for the NGC~419 observations would not obviously improve spectral resolution (the effective slit widths are imposed by the MUSE image slicers); this suggests that even after correction (see \citetalias{Kamann18}), the velocity uncertainties claimed for NGC~419 (\autoref{fig:e_vlos_comp}) may still be significantly underestimated.  

To demonstrate how the precision of the velocity uncertainties can affect dynamical results in a system like NGC~419, we carried out the same simulations as above, but this time we adopted a scaling factor $F$ by which we modified the claimed observational errors (solid red line in \autoref{fig:e_vlos_comp}) compared to the actual kinematic uncertainties used to assign the simulation velocities (the dashed red line in \autoref{fig:e_vlos_comp}).   Our results indicate that for $F\sim0.75$, half of the simulations of the MUSE data drawn from a model with a central dispersion of 2.44 \kms\ yield a dispersion of 3.3 \kms\ or greater.  Thus, a 25\% mean underestimate in the velocity uncertainties can alter the inferred dispersion at a level comparable to the differences between the MUSE and M2FS results.  The magnitude and direction of this error is comparable to the value we found from our analysis of the kinematic errors in the M2FS data, but the quantitative effect on the central dispersion in this case is very different because the M2FS errors are so much smaller than those in the MUSE data.  To quantify this, we ran the same set of simulation on the M2FS dataset including the $F$ factor on the kinematic errors.  In this case, when we assume that we have underestimated the true kinematic errors by a factor of two, we find that only 4 out of 1000 trials produce a dispersion as large as 3.3 \kms.  Alternatively, if we assume the actual dispersion is 3.3 \kms, only 8 of 1000 trials results in an inferred dispersion as low as 2.44 \kms\ for a factor of two underestimate of the velocity uncertainties.

We draw from this the well-known conclusion that when the velocity uncertainties are comparable to the velocity dispersion of a system, the uncertainties must be known to very high precision.  In this respect, a sample such as ours for NGC~419 is far more robust to inaccurate estimates of the velocity uncertainties than the far larger but less precise MUSE dataset.    A similar conclusion would apply to many integrated-light studies of clusters where the instrumental resolution is often much larger than the intrinsic dispersion of the systems being studied.  In such cases one can in principle derive a reliable dispersion, but the instrumental line profile must be shown to have been determined to exceptional precision and be free of any systematics due to, for example, temperature changes, focus drift, optical alignment, {\it etc}.  


\subsection{NGC~1846}
\label{sec:comp_N1846}
The full dataset for NGC~1846 published by \citetalias{Mackey13} consists of radial velocities for 106 targets, 22 of which were deemed to be probable members (including the planetary nebula Mo-17). The measured radial velocities of the remaining 84 stars---considered to be field stars---were not published in \citetalias{Mackey13} but are included here in \autoref{tab:sample_n1846}.   \citetalias{Mackey13} fitted a three-point velocity dispersion profile using the velocities of the 22 likely members using a projected Plummer model scaled by a derived central velocity dispersion of $2.52^{+0.26}_{-0.18}$ \kms\ (see \autoref{tab:comp}).  They separately measured a total K62 luminosity of $1.44\pm0.14\times10^5\ {\rm L}_{\sun}$ in $V$-band. As a result they obtained $M_{\rm tot}=8.4^{+1.7}_{-1.2}\times10^4\ {\rm M}_{\sun}$ and $M/L_V=0.59^{+0.13}_{-0.10}$ (\autoref{tab:comp}).  Note that before fitting their data with the model dispersion profile, \citetalias{Mackey13} corrected the observed velocities by a rotation amplitude $1.1\pm0.4$ \kms\ with a position angle of $60\pm20$ degrees east of north (see \autoref{tab:comp}).   

As a check on the dynamical analysis in the present paper, we have applied our EM analysis (\autoref{sec:EM}) using the published kinematic results from \citetalias{Mackey13} along with the previously unpublished results for the non-members (\autoref{tab:sample_n1846}).  The results we compare here do not include any sort of rotation correction, nor do they exclude the binary identified in \autoref{sec:combine_sample}.  The results (listed in \autoref{tab:comp}) indicate that our analysis of the \citetalias{Mackey13} data---including the associated but heretofore unpublished field-star data---results in dynamical parameters that agree well (to within 1$\sigma$) with \citetalias{Mackey13}'s results obtained using a different analysis technique and different dynamical models. Certainly, at this stage we cannot disentangle any statistically significant systematic offsets that may exist between the \citetalias{Mackey13} and M2FS-only due to analysis differences with the expected statistical noise due to the limited samples sizes in the respective studies.  We will explore the systematic role of dynamical models in greater detail in later papers as we analysis a larger sample of MC clusters.

\section{Summary and Conlusions}
\label{sec:summary}
In this paper, we present a pair of {\it Magellan}/M2FS observations of red giants in and around the intermediate-age Magellanic Cloud star clusters NGC~419 and NGC~1846, respectively. We implement a pipeline to these data that extracts stellar spectra from the raw observational data, and apply a Bayesian method to measure the radial velocities and several physical parameters for the individual target stars in our datasets.  We estimate the projected central velocity dispersion of each cluster using an Expectation-Maximization (EM) algorithm \citep{Walker15mn, Walker15apj} with the assumption that cluster members are spatially and kinematically distributed as expected for a single-mass K66 model and also assuming that superimposed on the cluster is a spatially uniform field population that follows a kinematically much broader Gaussian distribution. We use a number of different approaches to estimate cluster membership probabilities for individual targets in order to properly account for the influence of likely non-members. We find that allowing the EM algorithm to assign probabilities to all targets gives essentially the same results as assigning all stars with membership probabilities $> 50$\% as certain members and all others as certain non-members.

The primary results of both clusters are as follows:
\begin{enumerate}
\item The median velocity uncertainties for the samples of NGC~419 and NGC~1846 are 0.38 and 0.22 \kms, respectively.  These are suitably small for recovering velocity dispersion as low as $2$ \kms\ with high precision even if individual velocity uncertainties are mis-estimated by up to a significant (and unlikely) factor.
\item Our individual velocity measurements of NGC~1846 are in good agreement with those of \citetalias{Mackey13} for 17 targets in common (see \autoref{fig:common_stars}).  This comparison reveals one target that is a likely binary with a velocity difference of $5.6$ \kms\ (defined as \citetalias{Mackey13} {\it minus} M2FS).  A comparison of the velocities for the remaining in-common stars reveals a systematic velocity offset of $-0.54\pm0.15$ \kms\ (again, defined as \citetalias{Mackey13} {\it minus} M2FS) and comparable single-star velocity uncertainties from both studies.  This small offset has been applied to the \citetalias{Mackey13} sample to produce, in combination with our dataset, a larger `Combined Sample' for this cluster.
\item For NGC~419 we measure a systemic velocity $V_{\rm sys}=189.5^{+0.3}_{-0.3}$ \kms\ and a projected central velocity dispersion $\sigma_{\rm p,0}=2.44^{+0.37}_{-0.21}$ \kms\ based on 46 likely members out of an initial sample of 111 targets.  For NGC~1846 we obtain $V_{\rm sys}=239.3^{+0.2}_{-0.2}$ \kms\ and $\sigma_{\rm p,0}=2.04^{+0.28}_{-0.24}$ \kms\ from 52 likely members from our Combined Sample consisting of 195 targets (108 targets from the present study). Details related to these results and the other methodologies used to obtain $V_{\rm sys}$ and $\sigma_{\rm p,0}$ are provided in Table \autoref{tab:box}.  These results are based on an assumption that both clusters' velocity dispersion profiles are adequately described by the appropriate single-mass King (K66) model that fites their respective surface brightness profiles.
\item The total masses of NGC~419 and NGC~1846 are $7.6^{+2.5}_{-1.3}\times10^4\ {\rm M}_{\sun}$ and $5.4^{+1.5}_{-1.4}\times10^4\ {\rm M}_{\sun}$, respectively. We estimate the total masses by scaling the dynamical K66 models with our best-fit $\sigma_{\rm p,0}$ values. The structural parameters of a K66 model are transformed from those of a corresponding empirical surface density profile (K62 profile).  These results and variants are listed in \autoref{tab:MLv}. 
\item Both clusters show marginal signals of systemic rotation. The amplitudes in the plane of sky are $0.5^{+0.3}_{-0.2}$ (NGC~419) and $0.6^{+0.2}_{-0.3}$ \kms\ (NGC~1846), respectively. The rotation signals have negligible impact on the masses we derive for both clusters and do not affect our results at a level significantly higher than implied by the error bars on the derived masses (see above and \autoref{tab:MLv}). 
\item The $V$-band $M/L$ ratios are $0.22^{+0.08}_{-0.05}$ and $0.32^{+0.11}_{-0.11}$ in solar units. In this calculation, the luminosities of the clusters are obtained by integrating the best-fitting K62 profile scaled to agree with published aperture photometry. 
\item The mean metallicities of NGC~419 and NGC~1846 are estimated from out spectra as $\rm [Fe/H]=-0.84\pm0.19$ and $-0.70\pm0.08$, respectively. These results may suffer from systematic offsets due to uncertainties in background correction (\autoref{sec:reduction}) and uncertainties in effective temperatures.
\end{enumerate}

As we have tried to demonstrate in this paper, our current cluster sample---NGC~419 and NGC~1846---is largely immune to common observational uncertainties.  For example, the data on which our analyses are based reside exclusive in the parameter space in which the cluster central dispersions are considerably larger than typical velocity errors for individual targets.   This makes the resulting kinematic parameters largely immune to reasonable mis-estimates of velocity uncertainties and cluster structural parameters.  By obtaining data on individual stars, we also avoid luminosity biases that arise from, say, integrated-light kinematics measurements or results exclusively from cluster cores where blending and scattered light can present challenges.

In future papers of this series, we will apply the techniques described in this paper to a large sample of homogeneously observed MC clusters.  The final aim will be to critically compare our derived M/L ratios for simple stellar systems with those expected from stellar population models.   Of course, clusters are not quite so 'simple' as one might like for this comparison, so we will also be exploring ways in which the M/L ratios of clusters can evolve due to internal dynamical processes and not just due to population evolution.  Our ultimate aim is to provide a strong empirical test of M/L models that will improve the systematic uncertainties when such models are applied to distant, unresolved systems.

\section*{Acknowledgements}
We thank the anonymous referee for helpful comments.  
M.M. and Y.-Y.S. are supported by National Science Foundation (NSF) grants AST-1312997 and AST-1815403.  
E.O. is partially supported by NSF grant AST-1815767.
I.U.R. acknowledges support from NSF grants AST-1613536, AST-1815403, and PHY 14-30152 (Physics Frontier Center/ JINA-CEE).
M.G.W. is partially supported by NSF grant AST-1813881.
We thank Jeff Crane, Steve Shectman and Ian Thompson for invaluable contributions to the design, construction and support of M2FS.  
We thank the M2FS Team for obtaining the spectroscopic data at the {\it Magellan}/Clay telescope, especially Anthony Kremin and Meghin Spencer for the data of NGC~419, and Daniela Barrientos, Valentino Gonzalez and Terese Hansen for the data of NGC~1846.  

\bibliographystyle{mnras}
\bibliography{sc_bib} 



\bsp	
\label{lastpage}
\end{document}